\input lanlmac
\input epsf.tex
\input mssymb.tex
\overfullrule=0pt

\newcount\figno
\figno=0
\def\fig#1#2#3{
\par\begingroup\parindent=0pt\leftskip=1cm\rightskip=1cm\parindent=0pt
\baselineskip=11pt
\global\advance\figno by 1
\midinsert
\epsfxsize=#3
\centerline{\epsfbox{#2}}
\vskip 12pt
{\bf Fig.\ \the\figno:} #1\par
\endinsert\endgroup\par
}
\def\figlabel#1{\xdef#1{\the\figno}%
\writedef{#1\leftbracket \the\figno}%
}
\def\omit#1{}

\def\pre#1{{\tt
#1}}

\def\pf{{\rm Pf}}
\def\der{\partial}
\def\aff#1{V_{#1}}
\def\qed{\nobreak\hfill\vbox{\hrule height.4pt%
\hbox{\vrule width.4pt height3pt \kern3pt\vrule width.4pt}\hrule height.4pt}\medskip\goodbreak}
\lref\Kn{A.~Knutson, {\sl Some schemes related to the commuting variety},
\pre{math.AG/0306275}.}
\lref\KnM{A.~Knutson and E.~Miller, {\sl Gr\"obner geometry of Schubert polynomials},
{\it Annals of Mathematics} (2003), \pre{math.AG/0110058}.}
\lref\BRAU{J. De Gier and B. Nienhuis, {\sl Brauer loops and the commuting variety},
\pre{math.AG/0410392}.}
\lref\GAUDIN{M. Gaudin, {\sl La fonction d'onde de Bethe}, Masson (1997), Paris.}
\lref\DFZJ{P.~Di Francesco and P.~Zinn-Justin, {\sl Around the Razumov--Stroganov conjecture:
proof of a multi-parameter sum rule}, \pre{math-ph/0410061}.}
\lref\RS{A. V. Razumov and Yu. G. Stroganov, 
{\sl Combinatorial nature
of ground state vector of $O(1)$ loop model},
{\it Theor. Math. Phys.} 
138 (2004) 333-337; {\it Teor. Mat. Fiz.} 138 (2004) 395-400, \pre{math.CO/0104216}.}
\lref\PRdGN{P. A. Pearce, V. Rittenberg, J. de Gier and B.~Nienhuis, 
{\sl Temperley--Lieb Stochastic Processes},
{\it J. Phys.} A35 (2002), L661--L668,
\pre{math-ph/0209017}.}
\lref\MR{M. J.~Martins and P. B.~Ramos, {\sl The Algebraic Bethe Ansatz for rational
braid-monoid lattice models}, {\it Nucl. Phys.} B500 (1997) 579--620,
\pre{hep-th/9703023}.}
\lref\MNR{M. J.~Martins, B.~Nienhuis and R.~Rietman, {\sl An Intersecting Loop Model as a Solvable Super Spin Chain},
{\it Phys. Rev. Lett.} 81 (1998) 504-507,
\pre{cond-mat/9709051}.}
\lref\BdGN{M. T. Batchelor, J. de Gier and B. Nienhuis,
{\sl The quantum symmetric XXZ chain at $\Delta=-1/2$, alternating sign matrices and 
plane partitions},
{\it J. Phys.} A34 (2001) L265--L270,
\pre{cond-mat/0101385}.}
\lref\RSa{A. V. Razumov and Yu. G. Stroganov, {\sl Spin chains and combinatorics},
{\it J. Phys.} A34 (2001) 3185, \pre{cond-mat/0012141}\semi
{\sl Spin chains and combinatorics: twisted boundary conditions},
{\it J. Phys.} A34 (2001) 5335--5340, \pre{cond-mat/0102247}.}
\lref\PDFone{P. Di Francesco, {\sl A refined Razumov--Stroganov conjecture}, 
{\it JSTAT} P08009 (2004),
\pre{cond-mat/0407477}.}
\lref\PDFtwo{P. Di Francesco, {\sl A refined Razumov--Stroganov conjecture II},
{\it JSTAT} P11004 (2004),
\pre{cond-mat/0409576}.}
\lref\LLT{
A.~Lascoux, B.~Leclerc, J.-Y.~Thibon,
{\sl Twisted action of the symmetric group on the cohomology of the flag
manifold}, {\it Banach Center Publications} vol. 36 (1996),  111--124.
}
\lref\LGV{B. Lindstr\"om, {\sl On the vector representations of
induced matroids}, {\it Bull. London Math. Soc.} 5 (1973)
85--90\semi
I. M. Gessel and X. Viennot, {\sl Binomial determinants, paths and
hook formulae}, {\it Adv. Math.} 58 (1985) 300--321. }
\lref\Che{I. Cherednik, {\sl Double affine Hecke algebras, Knizhnik--Zamolodchikov
equations, and McDonald's operators}, {\it IMRN (Duke Math. Jour.)} 9 (1992) 171--180.
}
\lref\FK{S. Fomin and A.N. Kirillov, {\sl The Yang--Baxter equation, symmetric functions
and Schubert polynomials}, {\it Discrete Mathematics} 153 (1996), 123--143;
{\sl The Yang--Baxter equation, symmetric functions
and Grothendieck polynomials}, \pre{hep-th/9306005}.}
%
%
%
%
\Title{SPhT-T04/158}
{\vbox{
\centerline{Inhomogenous model of crossing loops}
\medskip
\centerline{and multidegrees of some algebraic varieties}
}}
\bigskip
\centerline{P.~Di~Francesco \footnote{${}^\#$}
{Service de Physique Th\'eorique de Saclay,
CEA/DSM/SPhT, URA 2306 du CNRS,
C.E.A.-Saclay, F-91191 Gif sur Yvette Cedex, France}
and P. Zinn-Justin \footnote{${}^\star$}
{LIFR--MIIP, Independent University, 
119002, Bolshoy Vlasyevskiy Pereulok 11, Moscow, Russia   and
Laboratoire de Physique Th\'eorique et Mod\`eles Statistiques, UMR 8626 du CNRS,
Universit\'e Paris-Sud, B\^atiment 100,  F-91405 Orsay Cedex, France}}
\medskip

\vskip0.5cm

\noindent

We consider a quantum integrable inhomogeneous model based on the Brauer algebra $B(1)$
and discuss the properties of its ground state eigenvector. In particular we derive various sum rules,
and show how some of its entries are related to multidegrees of algebraic varieties.

\Date{12/04}
%
%
\newsec{Introduction}
Recently, a new connection between quantum integrable models and combinatorics has emerged.
This relation can be traced back to the 
idea, as expressed e.g.\ in \PRdGN, that in stochastic integrable processes,
due to the existence of a simple ground state eigenvalue
(without any finite-size corrections), the entries of the ground state are
integers and must have some combinatorial significance. This idea was based 
on experience with a particularly succesful case: the model of non-crossing loops
related to the Temperley--Lieb algebra $TL(1)$, whose special properties \refs{\RSa,\BdGN}
led Razumov and Stroganov to conjecture the combinatorial significance of {\it each}\/ entry of
the ground state \RS. This conjecture has generated a lot of activity (see for example references
in \DFZJ) but has not been proved yet in its full generality.

The latest model that falls into the framework described above is the model of crossing loops
proposed by de Gier and Nienhuis in \BRAU, which is related to the Brauer algebra $B(1)$
and to standard integrable models with
symmetry $OSp(p|2m)$ \refs{\MR,\MNR}, $p-2m=1$.
By abuse of language, as in the non-crossing case, we shall call this model the ``$O(1)$'' crossing
loop model. The novelty in the work \BRAU\ is that the entries of the ground state
are integers that do not appear to be obviously related to statistical mechanics, 
but rather belong to the realm of enumerative geometry.
Indeed some of them are conjectured to be degrees of algebraic varieties that appear in work
of Knutson \Kn\ revolving around the commuting variety. 
The present article tries to shed some light on the origin of these numbers in the model. In
particular we shall see how the algebra of the model naturally leads to
an action of the symmetric group as divided difference operators, which
have well-known meaning in the context of Schubert calculus.

Our work is motivated by
recent progress in understanding the model of non-crossing loops \DFZJ\ for the similar
Razumov--Stroganov conjecture. The idea of \DFZJ\ is
to make better use of the integrability of the model. It involves in particular
the introduction of inhomogeneities (spectral parameters), which give a much
more powerful tool to study
the ground state, whose coefficients become polynomials in these variables.
Here, we shall try to do the same to the $O(1)$ crossing loop model.
As in \DFZJ, our results include multi-parameter sum rules for the entries of the ground state vector;
we find in fact two different sum rules, one for the sum of all entries,
and one for the sum in the so-called permutation sector, in which the entries clearly play a special role:
these are precisely the coefficients which are conjecturally related to degrees of varieties.
In fact we show that this connection is much deeper and that the full polynomial entries are related
to so-called multidegrees.
We also prove some conjectured properties formulated in \BRAU, involving factorizability of the ground state
vector entries.

The paper is organized as follows. 
In Section 2 we introduce the model and its ground state eigenvector.
Section 3 contains general factorization properties of the entries
of the ground state, as well as their construction in terms of divided difference operators
in the space of polynomials.
Section 4 analyzes in detail the special case of so-called ``permutation patterns'',
which is also the focus of \BRAU;
we formulate a conjecture that relates some of its entries to (multi)degrees of some algebraic varieties,
prove some results including a sum rule, and give a sketch of proof of this generalized
de Gier--Nienhuis conjecture.
Section 5 concerns recursion relations and the sum rule for all entries.
A few concluding remarks are gathered in section 6.
The appendices contain some explicit data for $n=2,3,4$.

\newsec{The inhomogeneous $O(1)$ crossing loop model: transfer matrix and ground state vector} 
The $O(1)$ crossing loop model is based on the following solution to the Yang--Baxter equation,
expressed as a linear combination of generators of the Brauer algebra $B_{2n}(1)$. 
These are the identity $I$, the ``crossing" operators $f_i$, and the generators $e_i$ of the
Temperley--Lieb algebra $TL_n(1)$, 
$i=1,2,...,2n$, with the pictorial representations
$$I=\vcenter{\hbox{\epsfbox{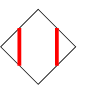}}},\qquad f_i=\vcenter{\hbox{\epsfbox{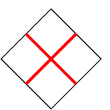}}}, 
\qquad e_i=\vcenter{\hbox{\epsfbox{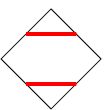}}}$$
and acting vertically on the vector space generated by 
crossing link patterns, that is chord diagrams of $2n$ labeled points
around a circle, connected by pairs via straight lines across the inner disk.
We denote by $CP_n$ the set of these (crossing) link patterns on $2n$ points, with
cardinality
$|CP_n|=(2n-1)!!$. A simple way of indexing these link patterns is via
permutations of ${\cal S}_{2n}$ with only 2-cycles (fixed-point free involutions), 
each cycle being made of the labels of 
the two points connected via a chord. The pictorial representation above makes it
straightforward to derive the $B_{2n}(1)$ Brauer algebra relations:
\eqn\bbralg{\eqalign{ e_i^2&=e_i, \qquad f_i^2=I,\qquad e_ie_{i\pm 1} e_i=e_i, \qquad
f_if_{i+1}f_i=f_{i+1}f_if_{i+1}, \cr
[e_i,e_j]&=[e_i,f_j]=[f_i,f_j]=0\ {\rm if}\ |i-j|>1, \qquad f_ie_i=e_if_i=e_i\cr}}

Looking for a solution for a face transfer matrix operator
$X_i(u)=a(u)I+b(u) f_i+c(u) e_i$ to
the Yang--Baxter equation
\eqn\ybe{ X_i(u)X_{i+1}(u+v)X_i(v)=X_{i+1}(v)X_i(u+v)X_{i+1}(u)}
further fixed by the normalization $X_i(0)=I$,
we find the solution
\eqn\defx{X_i(u)=(1-u)I+{u\over 2}(1-u)f_i+u e_i \ ,}
unique up to scaling of $u$, 
as a direct consequence of the relations \bbralg. The solution \defx\
also satisfies the unitarity relation
\eqn\unita{ X_i(u)X_i(-u)=(1-u^2)(1-{u^2/ 4})I }
This solution appeared first in \GAUDIN, and was further studied in \MR, and shown 
to be related to vertex models based on orthosimplectic groups.

We now introduce an inhomogeneous integrable model based on the above solution 
of the Yang--Baxter equation.
It is defined on an infinite cylinder of square lattice of perimeter 
$2n$ represented as an infinite strip of width $2n$ glued along its two 
borders.
A configuration of the model is defined by assigning the plaquettes
$\vcenter{\hbox{\epsfbox{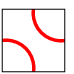}}}$,
$\vcenter{\hbox{\epsfbox{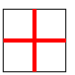}}}$,
or
$\vcenter{\hbox{\epsfbox{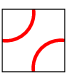}}}$ 
to each elementary face of the cylinder, with certain weights. 

\fig{A typical configuration of the crossing
loop model on a semi-infinite cylinder of square lattice 
with perimeter $2n$.}{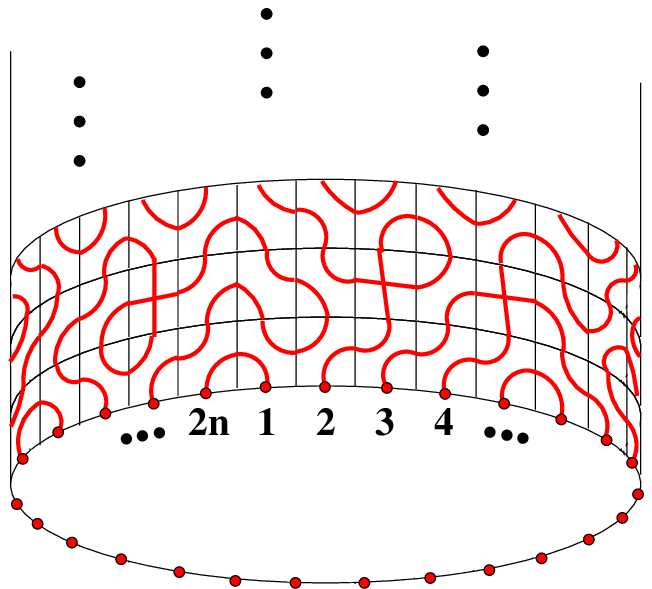}{9.cm}
\figlabel\cylinder

In the transfer matrix approach, one considers a semi-infinite cylinder (see Fig.~\cylinder).
The space of states then represents
the pattern of pair connectivity of the $2n$ labeled midpoints of the boundary edges
of the semi-infinite cylinder via plaquette configurations of the model. Finally, the transfer
matrix represents the addition of one row of plaquettes to the semi-infinite cylinder:
\eqn\transmat{ T_n(t\vert z_1,\ldots,z_{2n})=\prod_{i=1}^{2n}
\Big((1-t+z_i) 
\vcenter{\hbox{\epsfbox{mov1.eps}}}
+{(t-z_i)(1-t+z_i)\over2}
\vcenter{\hbox{\epsfbox{mov3.eps}}}
+(t-z_i)
\vcenter{\hbox{\epsfbox{mov2.eps}}}
\Big)}
where the weights depend on the label $i$ of the site, and correspond to a tilted version
of the operators $X_i$ of Eq.~\defx.
The parameter $t$, which is independent of the row, plays no role in what follows due
to the commutativity property
\eqn\comm{
[T_n(t),T_n(t')]=0}
itself a direct consequence of the Yang--Baxter equation.

For values of $z_i$ and $t$ such that $0<t-z_i<1$, the weights are strictly positive
and can be interpreted as unnormalized probabilities, and
the transfer matrix as an unnormalized matrix of transition probabilities. 
Conservation of probability can be expressed in the following way: 
define the linear form $v_n$ with entries in the canonical basis
$v_{\pi}=1$ for all $\pi\in CP_n$. Then summing the weights in Eq.~\transmat, we obtain
\eqn\eigtransmat{
v_n T_n(t|z_1,\ldots,z_{2n})=v_n\prod_{i=1}^{2n} (1-{1\over2}(t-z_i))(1+t-z_i)
}
This means that $\prod_{i=1}^{2n} (1-{1\over2}(t-z_i))(1+t-z_i)$ is an eigenvalue of $T_n$ (with left
eigenvector $v_n$), and there
must exist a right eigenvector:
\eqn\gsvec{ \left( T_n(t\vert z_1,\ldots,z_{2n})-\prod_{i=1}^{2n} (1-{1\over2}(t-z_i))(1+t-z_i) I\right)
\Psi_n(z_1,\ldots,z_{2n})=0}
In the aforementioned range, Eqs.~\eigtransmat\ and \gsvec\ are nothing but Perron--Frobenius eigenvector
equations for the transpose of $T_n$ and for $T_n$, and the entries $\Psi_{\pi}$ of $\Psi_n$ are
interpreted, up to normalization, as the equilibrium 
probabilities, in random configurations of the model on a semi-infinite cylinder, that the 
boundary vertices be connected according to $\pi$.

As $T_n$ is polynomial, we may assume that $\Psi_n$ is also a polynomial of the $z_i$ (whose
entries are non-identically-zero due to the Perron--Frobenius property).
Since we can always factor out the GCD of the entries $\Psi_{\pi}$,
we assume that they are coprime.
The main purpose of the present article is the investigation of these entries. A special case,
extensively studied in \BRAU, corresponds to choosing the $z_i$ to be all equal.
In this ``homogeneous'' case,
$T_n(t)$ commutes with the Hamiltonian $H_n=\sum_{i=1}^{2n}(3-2e_i-f_i)$, and $\Psi_n$
is the null eigenvector of $H_n$. It was conjectured in \BRAU\ that with proper normalization,
the entries of $\Psi_n$ may be chosen to be all non-negative integers, 
the smallest of which is $1$. Here we use the latter condition to fix the 
remaining arbitrary numerical factor in the 
normalization of the entries, so that it coincides in the homogeneous case with that of \BRAU.

Before going into specifics, let us mention a preliminary property satisfied by the entries of $\Psi_n$.
Our semi-infinite cylinder problem is clearly invariant under rotation by one lattice step. Denoting 
by $\rho=f_{2n-1}f_{2n-2}\ldots f_{1}$ the corresponding rotation operator acting on the crossing 
link patterns by cyclically shifting the labels $i\to i+1$,
we have the relation $T_n(t\vert z_2,\ldots,z_{2n},z_{1})\rho=
\rho T_n(t\vert z_1,\ldots,z_{2n})$,
from which we deduce that $\rho \Psi_n(z_1,\ldots,z_{2n})=\lambda \Psi_n(z_2,\ldots,z_{2n},z_1)$.
Noting that $\Psi_n$ is generically non-zero due to the Perron--Frobenius property, and that
$\lambda$ takes discretes values $\lambda^{2n}=1$ and must therefore be independent of the $z_i$,
we immediately get that $\lambda=1$ in the range where $\Psi_{\pi}>0$, 
henceforth the entries of $\Psi_n$ satisfy the following 
cyclic covariance relation:
\eqn\cyclicov{ \Psi_{\rho\cdot\pi}(z_2,z_3,\ldots z_{2n},z_1)=\Psi_{\pi}(z_1,z_2,\ldots, z_{2n})}
Similarly, one can prove a reflection relation: if $r$ exchanges $i$ and $2n+1-i$,
\eqn\refl{
\Psi_{r\cdot\pi}(-z_{2n},-z_{2n-1},\ldots,-z_1)=\Psi_{\pi}(z_1,z_2,\ldots, z_{2n})
}

\newsec{Factorization and degree}
We now establish factorization properties of the transfer matrix $T_n$ and of its eigenvector $\Psi_n$. 
Note that this section (as well as Sect.~5 below)
possesses some strong similarities with Sect.~3 of \DFZJ, though the 
model under consideration is different. 
It is sometimes convenient
to use the following pictorial representations for the matrix $X_i(t-z)$ and for
the transfer matrix:
\eqn\tma{X_i(u)=\vcenter{\hbox{\epsfbox{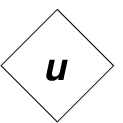}}},\quad 
T_n(t|z_1,\ldots,z_{2n})= \vcenter{\hbox{\epsfxsize=3cm\epsfbox{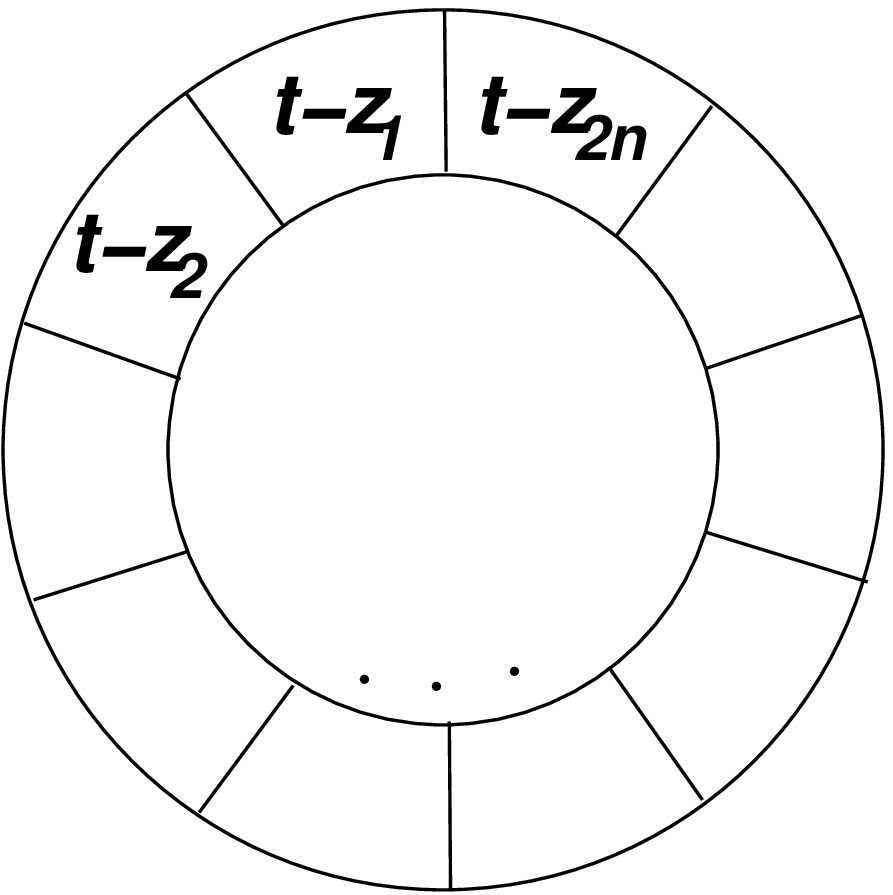}}}}
In this language, the Yang--Baxter and unitarity relations
read respectively:
\eqn\ybuni{\vcenter{\hbox{\epsfbox{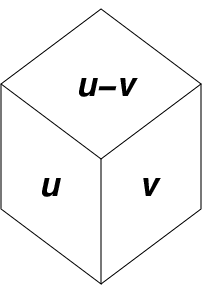}}}\ =\ \vcenter{\hbox{\epsfbox{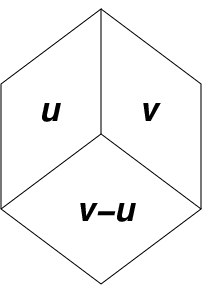}}}\ \ \ \ \ 
{\rm and}\ \ \ \ \  \vcenter{\hbox{\epsfbox{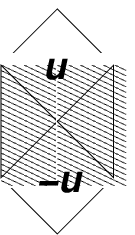}}}\ =(1-u^2)(1-{u^2\over 4}) }

In all that follows, due to periodic boundary conditions indices are meant modulo $2n$ ($2n+1\equiv 1$).

\subsec{Vanishings and factorizations}
Let us show a first intertwining property:

\proclaim Lemma 1. The matrices $T_n(t\vert z_1,\ldots,z_i,z_{i+1},\ldots,z_{2n})$ and
$T_n(t\vert z_1,\ldots,z_{i+1},z_i,,\ldots,z_{2n})$ are intertwined by $X_i(z_{i+1}-z_i)$, namely
\eqn\interone{\eqalign{
T_n(t\vert z_1,\ldots,z_i,&z_{i+1},\ldots,z_{2n})X_i(z_{i+1}-z_i)\cr
&=X_i(z_{i+1}-z_i) T_n(t\vert z_1,\ldots,z_{i+1},z_i,\ldots,z_{2n})\cr}}

Proof. This is a direct consequence of the Yang--Baxter relation and reads pictorially:
\eqn\pictoyb{ \vcenter{\hbox{\epsfxsize=6cm\epsfbox{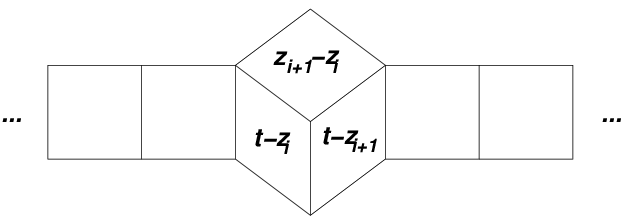}}}\ =\ 
\vcenter{\hbox{\epsfxsize=6cm\epsfbox{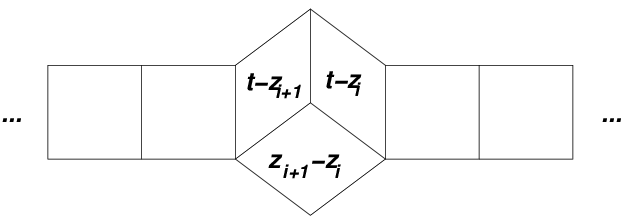}}}}
\qed

We now remark that at the value $1$ of the parameter, 
the face transfer matrix reduces to $X_i(1)=e_i$.
This means that for $z_{i+1}=z_i+1$, the above transfer matrices say $T$ and $\tilde T$
satisfy $Te_i=e_i{\tilde T}$. When acting on ${\tilde \Psi}_n\equiv \Psi_n(\ldots z_{i+1},z_i,\ldots)$
at $z_{i+1}=z_i+1$, we get: $Te_i{\tilde \Psi}_n=\Lambda e_i{\tilde \Psi}_n$, with
$\Lambda=\prod_{i=1}^{2n} (1-{1\over2}(t-z_i))(1-z_i+t)$. Hence $e_i{\tilde \Psi}_n$ is a non-vanishing vector
proportional to $\Psi_n$, and there exists a rational function $\alpha$, such that
$\Psi_n=\alpha e_i{\tilde \Psi}_n$.
When written in components, this implies that whenever $i$ and $i+1$ are not connected via a 
``little arch" in a link pattern $\pi\in CP_n$, the entry $\Psi_{\pi}$ 
vanishes when $z_{i+1}=z_i+1$. We may extend this remark into a:

\proclaim Proposition 1.
If the link pattern $\pi\in CP_n$ has no arch connecting a pair of points between labels
$i$ and $j$, then the entry $\Psi_{\pi}$ vanishes for $z_j=z_i+1$.

The proof is already done in the case $j=i+1$. For more distant points, we use a generalized intertwining
property $T P =P {\tilde T}$, where $P$ is a suitable product of $X$ matrices. Using again
the fact that $X(1)=e_i$, we see that at $z_j=z_i+1$ the product of $X$ forming $P$ contains a factor
$e_i$ at the intersection between the lines $i$ and $j$.
We deduce that $\Psi_n=\alpha P{\tilde \Psi}_n$ has no non-vanishing
entry with at least an arch linking two points between $i$ and $j$. Indeed, by expanding the
product of $X_i$ that form $P$ as a sum of products of $f$ and $e$, we see that there is always 
at least one $e_k$ in factor, for $i\leq k<j$, which results in the existence of an arch
connecting two points inbetween $i$ and $j$. \qed

This shows that $\Psi_{\pi}$ is divisible by $\prod_{i\le k<l\le j} (1+z_k-z_l)$ 
(with obvious cyclic notations) for all pairs
of points $i$ and $j$ satisfying the hypothesis of Prop.~1.

As a first application, let us consider the link pattern $\pi_0$ without any little arches, and
the maximum number of crossings: $\pi_0(i)=i+n$, $i=1,\ldots,n$. For each variable $z_i$, we have a factor
of $\prod_{j=i+1}^{i+n-1} (1+z_i-z_j)\prod_{k=i+n+1}^{i-1} (1+z_k-z_i)$.
In total, this gives
\eqn\valpio{\Psi_{\pi_0}=\Omega \prod_{1\leq i<j\leq 2n\atop j-i< n} (1+z_i-z_j)
\prod_{1\leq i<j\leq 2n\atop j-i> n} (1+z_j-z_i)}
where $\Omega$ is a polynomial yet to be determined. 
Apart from this, we find a polynomial
of total degree $n(2n-1)-n=2n(n-1)$ and partial degree $2n-2$ in each variable.
We shall prove in the following that $\Omega=1$.

\subsec{Permutation of variables in $\Psi_n$ and degree}
Let us introduce a normalized version of $X_i$, which we denote by $\check R_i$:
\eqn\goodr{{\check R}_{i}(z,w)={(1-w+z)I+{1\over2}(w-z)(1-w+z)f_i+(w-z)e_i\over (1-{1\over2}(w-z))(1+w-z)}  }
This matrix satisfies the usual unitarity  relation $\check R_{i}(z,w)\check R_{i}(w,z)=I$.

\proclaim Theorem 1. The transposition of any two consecutive spectral parameters 
in $\Psi_n$ is generated by the action of ${\check R}$:
\eqn\vecrel{ \Psi_n(\ldots,z_i,z_{i+1},\ldots)={\check R}_{i}(z_i,z_{i+1})
\Psi_n(\ldots,z_{i+1},z_i,\ldots)} 

Proof.
To show this, we apply Lemma 1 to the vector $\Psi_n(z_1,\ldots,z_{i+1},z_i,\ldots,z_{2n})$.
We find that $\Psi_n(z_1,\ldots,z_{2n})=\alpha_{n,i}(z_1,\ldots,z_{2n})X_i(z_{i+1}-z_i)
\Psi_n(z_1,\ldots,z_{i+1},z_i,\ldots,z_{2n})$, for some rational function $\alpha_{n,i}$.
By the coprimarity assumption for the entries of $\Psi_n$, we deduce that $\alpha_{n,i}$
may have no zero, hence it reads $\alpha_{n,i}=1/\beta_{n,i}$, for some polynomial $\beta_{n,i}$.

Moreover, iterating the above once more, we find that
\eqn\itermore{
\Psi_n=\check R_i(z_{i+1},z_{i})\check R_i(z_{i},z_{i+1}) \Psi_n={\beta_{n,i}(z_1,\ldots,z_{2n})
\beta_{n,i}(z_1,\ldots,z_{i+1},z_i,\ldots,z_{2n})\over
(1-{1\over4}(z_i-z_{i+1})^2)(1-(z_i-z_{i+1})^2)}\Psi_n}
The only polynomials that satisfy this relation are
\eqn\valbet{\beta_{n,i}(z_1,\ldots,z_{2n})=\Big(1+\epsilon_i(z_{i+1}-z_{i})\Big)\Big(1+
\epsilon'_i{1\over2}(z_{i}-z_{i+1})\Big)\epsilon_i''}
for $\epsilon_i,\epsilon_i',\epsilon_i''=\pm 1$. 
These signs are further all fixed to be $+1$ by (i) expressing \valbet\ when all $z_j=0$
($\epsilon_i''=1$), (ii) expressing it when all $z_j\to\infty$ ($\epsilon_i\epsilon_i'=1$),
and (iii) by applying the Lemma 1 ($\epsilon_i=1$).
This yields Eq.~\vecrel. \qed

More explicitly, Eq.~\vecrel\ reads in components:
\eqnn\recpolone
$$\eqalignno{
(1+{1\over2}(z_{i}-z_{i+1}))&(1-z_i+z_{i+1})
\Psi_{\pi}(z_1,\ldots,z_{2n})\cr
&=(1+z_i-z_{i+1}) \Psi_{\pi}(z_1,\ldots,z_{i+1},z_i,\ldots,z_{2n})\cr
&+{1\over2}(z_{i+1}-z_{i})(1+z_i-z_{i+1}))\Psi_{f_i\cdot\pi}(z_1,\ldots,z_{i+1},z_i,\ldots,z_{2n})\cr
&+(z_{i+1}-z_{i})\sum_{\pi', \ e_i\cdot\pi'=\pi} \Psi_{\pi'}(z_1,\ldots,z_{i+1},z_i,\ldots,z_{2n})
&\recpolone\cr}
$$
This is a very efficient recursion relation, allowing to
express all entries of $\Psi_n$ in terms of the maximally crossing pattern
entry. Indeed, two situations may occur for $\pi$:
\item{(i)} $\pi$ has no little arch joining $(i,i+1)$. Then
Eq.~\recpolone\ translates into
\eqn\translat{ \Psi_{f_i\cdot\pi}(z_1,\ldots,z_{2n})=\Theta_{i}\Psi_{\pi}(z_1,\ldots,z_{2n}) }
where the linear operator $\Theta_{i}$ acts on functions $F(z_1,\ldots,z_{2n})$ as
\eqn\deftheta{\eqalign{ \Theta_{i}F(z_1,\ldots,z_{2n})&=2 {
(1+z_i-z_{i+1})F(z_{i+1},z_i)-(1-z_{i}+z_{i+1})F(z_i,z_{i+1})\over
(z_{i}-z_{i+1})(1-z_{i}+z_{i+1})}\cr
& - {1+z_{i}-z_{i+1}\over 1-z_{i}+z_{i+1}}
F(z_{i+1},z_i) \cr}}
where for simplicity we have only represented the arguments $i$ and $i+1$ of $F$ (recall that periodic
boundary conditions for indices are implied: $2n+1\equiv 1$).
Note here that $\Theta_{i}\circ \Theta_{i}=I$, in agreement with $f_i^2=I$, a simple consequence of the
``gauge formula"
\eqn\gauge{ 
\Theta_i=(1+z_i-z_{i+1}) 
\left(2\partial_i-\tau_i\right){1\over 1+z_i-z_{i+1}}}
where $\tau_i$ and $\partial_i$ are respectively the transposition
and divided difference operators,\foot{Note the unusual sign convention for $\der_i$,
which will be fixed in Sect.~4 by renumbering variables in the opposite order.}
acting 
as
\eqn\actaupart{ \eqalign{
\tau_i F(z_i,z_{i+1})&= F(z_{i+1},z_i)\cr
\partial_i F(z_i,z_{i+1})&={F(z_{i+1},z_i)-F(z_i,z_{i+1})\over z_i-z_{i+1}}\cr}}
with the obvious relations
\eqn\obtaupart{ \tau_i^2=1,\quad \partial_i^2=0,
\quad \partial_i \tau_i=-\tau_i\partial_i}

\item{(ii)} $\pi$ has a little arch joining $(i,i+1)$. Then Eq.~\recpolone\ translates into
\eqn\otrans{ \sum_{\pi'\in CP_n\atop \pi'\neq \pi,\ e_i\pi'=\pi} \Psi_{\pi'}(z_1,\ldots,z_{2n})
=\Delta_{i}\Psi_{\pi}(z_1,\ldots,z_{2n})}
where the linear operator $\Delta_{i}$ acts as
\eqn\defdelta{\Delta_{i}F(z_1,\ldots,z_{2n})
=(1+z_i-z_{i+1})(1+{1\over2}(z_{i+1}-z_{i})){F(z_{i+1},z_i)-F(z_i,z_{i+1})
\over z_{i}-z_{i+1}}}
Note also that $\Delta_{i}\circ \Delta_{i}=-\Delta_{i}$, in 
agreement with $(e_i-I)^2=-(e_i-I)$.

Some remarks are in order. From the explicit form of $\Delta_i$ \defdelta, 
one may think that its action on a polynomial increases the degree by one.
However this is not the case if the largest total degree piece of the polynomial is
symmetric under $z_i\leftrightarrow z_{i+1}$. Such a property will be found below (Lemma 2).
Also, it is clear that the set of relations \translat\ and \otrans\ is overdetermined. The compatibility
between these equations is granted by the Yang--Baxter equation, that translates into relations
between the $\Theta_i$ and $\Delta_i$.
Finally, we note that the system of equations \translat\ and \otrans\ is equivalent to the eigenvector
condition \gsvec. To see that this system is sufficient to ensure \gsvec, we note that in analogy
with the Temperley--Lieb case of \DFZJ, we may construct yet another transfer matrix $T'_n$
that commutes with $T_n$, and made of a particular product of matrices of the form ${\check R}_i(z_k,z_\ell)$.
More precisely, we have $T'_n=\prod_{i=1}^n \prod_{j=1}^{n} {\check R}_{i+2j-2}(z_{2j-1},z_{2i+2j-2})$,
with products taken in the order indicated. As a direct consequence of the Yang--Baxter equation
\pictoyb,
we have the commutation relation $[T_n,T'_n]=0$, hence if $\Psi_n$ is 
eigenvector of $T'_n$ with eigenvalue $1$ (that is Perron--Frobenius eigenvector
in appropriate ranges of the $z_i$), then it is also that of $T_n$. The set of conditions
\translat\ and \otrans\ precisely guarantee that $\Psi_n$ is such an eigenvector
of $T'_n$, hence they are sufficient to ensure \gsvec.

As it turns out, we may generate all the entries $\Psi_{\pi}$ for $\pi\in CP_n$ by 
acting with a number of $\Theta_{i}$ on $\Psi_{\pi_0}$. 
This is best seen by recalling that the link patterns $\pi\in CP_n$ can be considered
as permutations of ${\cal S}_{2n}$ with only 2-cycles. As such, $f_i$ acts on $\pi$ as conjugation
of $\pi$ by the elementary transposition $i\leftrightarrow i+1$, and generates
the action of the whole symmetric group ${\cal S}_{2n}$. The well-known property
that two permutations are conjugate if (and only if) they have the same cycle lengths implies that
any $\pi$ can be obtained from $\pi_0$ as
$\pi= f_{i_1}\cdots f_{i_k}\cdot \pi_0$. We assume that $f_{i_{l+1}}\cdots f_{i_k}\cdot \pi_0$ does
not have a little arch $(i_l,i_l+1)$ for $1\le l\le k$, i.e.\ exclude in such a decomposition
any $f_i$ that would act on a pattern with a little arch $(i,i+1)$ since such an action is trivial.
We can therefore apply (i) above repeatedly,
and express the corresponding entry of $\Psi_n$:
\eqn\expraco{ \Psi_{\pi}=\Theta_{i_1}\cdots\Theta_{i_k} \ \Psi_{\pi_0} }

The procedure is illustrated in appendix B in the case $n=3$.

The property \expraco\ has an important consequence: by constructing explicitly
the entries of $\Psi$, we fix their degree and prove that $\Omega=1$ in Eq.~\valpio:

\proclaim Theorem 2. One has:
\eqn\valpiob{
\Psi_{\pi_0}=\prod_{1\leq i<j\leq 2n\atop j-i< n} (1+z_i-z_j)
\prod_{1\leq i<j\leq 2n\atop j-i> n} (1+z_j-z_i)}
and all the entries
of $\Psi_n$ are polynomials of total degree $2n(n-1)$, 
and partial degree $2(n-1)$ in each $z_i$.

The proof goes as follows. Starting with the ``minimal" candidate \valpiob\ for $\Psi_{\pi_0}$
subject to the vanishing conditions of proposition 1 above, we must first show that we can
construct all other
entries $\Psi_\pi$ via Eqs.~\expraco\ {\it independently of the choice of words}\/ $f_{i_1}\cdots f_{i_k}$,
and then check that these moreover satisfy Eqs.~\otrans, and are coprime.

We first use Eq.~\expraco\ to express all the $\Psi_{\pi}$ in terms of the minimal
choice \valpiob\ for $\Psi_{\pi_0}$.
It is easy to check that the factorization properties of proposition 1 are satisfied by
all the $\Psi_\pi$ thus constructed.
As a consequence, all denominators in Eq.~\deftheta\ are cancelled
(see the reformulation \gauge), and 
the action of $\Theta_{i}$ preserves the polynomial character, and the degree.
The fact that the values of $\Psi_\pi$ are independent of the choice of decomposition
$\pi=f_{i_1}\cdots f_{i_k}\cdot \pi_0$
which appears in \expraco\ is a consequence of the following properties:
\item{(i)} The affine symmetric group relations satisfied by the $\Theta$'s,
namely $\Theta_i^2=I$, $\Theta_i\Theta_{i+1}\Theta_i=\Theta_{i+1}\Theta_i\Theta_{i+1}$, and
$\Theta_i\Theta_j=\Theta_j\Theta_i$ for $|i-j|>1$, for all $i=1,2,\ldots,2n$ with cyclic boundary conditions,
ensure that two words corresponding to the same permutation give rise to the same $\Psi_\pi$.
\item{(ii)} As the $\Theta_i$ correspond only to  
the action of the $f_i$ on permutations $\sigma$ with $n$ cycles of length $2$ (the class $[2^n]$ of $S_{2n}$),
while forbidding the action of $f_i$ on permutations $\sigma$ such that $\sigma(i)=i+1$,
we must also enforce the conditions that 
\eqn\stabi{\Theta_i\Theta_{n+i}\Psi_{\pi_0}=\Psi_{\pi_0}} 
for $i=1,2,...,n$. 
\par 
\noindent The corresponding elements $f_if_{i+n}$ of $S_{2n}$
are indeed nothing but the remaining generators of the stabilizer subgroup of the class $[2^n]$, once
the actions of $f_i$ on permutations $\sigma$ such that $\sigma(i)=i+1$, $i=1,2,\ldots ,2n$, have been 
forbidden.\foot{This action is exactly the ``flat first Reidemeister move''. In other words to
obtain a true representation of the $\Theta_i$ without restrictions one should consider link patterns
decorated with extra ``tadpoles''.}
Note that the latter
condition has the net effect of wiping out any affine symmetric group relation that ``winds"
around the circle, such as for instance $f_2f_3\ldots f_{2n}=I$, as its action on link patterns
would involve at least one of the forbidden moves.
The relations \stabi\ are readily checked by using the expression
\valpiob. This defines the vector $\Psi_n$ unambiguously.

To ensure that $\Psi_n$ satisfies the eigenvector equation \gsvec, we must still check that it satisfies
the properties \otrans, which have not been used so far. This is done by induction on the 
number of crossings in $\pi$. Assume a link pattern $\pi$ has a little arch in position $(i,i+1)$,
and that it satisfies Eq.~\otrans. As $[\Delta_i,\Theta_j]=0$ for $j\neq i-1,i,i+1$, we may act
upon Eq.~\otrans\ with any such $\Theta_j$, also such that $\pi$ has no little arch at
positions $(j,j+1)$. This results in a new equation, in which $\pi,\pi'\to f_i\cdot\pi,f_i\cdot\pi'$.
This allows to increase the number of crossings of $\pi$, hence the induction.
We are left with the task of proving Eq.~\otrans\ for $\pi$ with maximal number of crossings
and with a little arch $(i,i+1)$. We first note that $\Psi_{\pi_0}$ as
given by \valpiob\ is cyclically invariant (under $z_i\to z_{i+1}$ cyclically), henceforth we 
need only check this property for $i=2n$, say.
We have to check that
\eqn\propcheck{ \Delta_{2n} \ \Psi
\raise-2.0cm\hbox{\epsfxsize=2.5cm\epsfbox{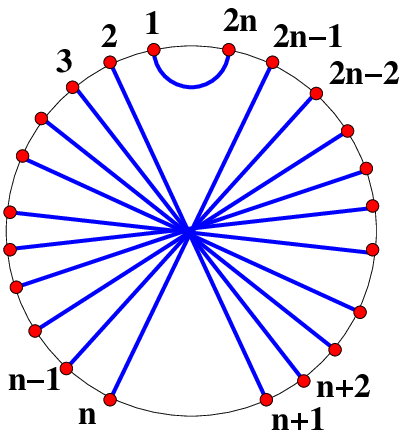}}
=\sum_{j=2}^n 
\Big(\ \Psi
\raise-2.0cm\hbox{\epsfxsize=3cm\epsfbox{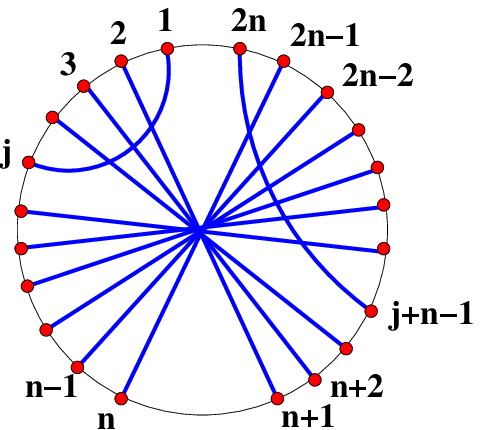}}+
\Psi 
\raise-2.0cm\hbox{\epsfxsize=3cm\epsfbox{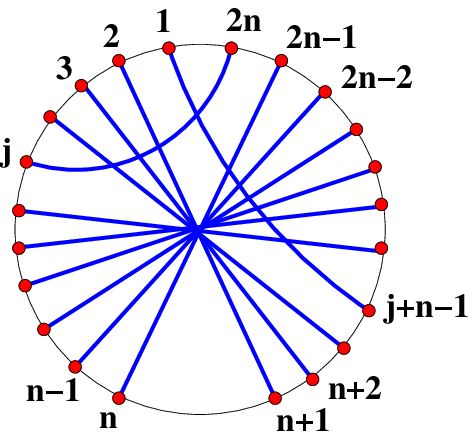}}
\Big)}
or equivalently
\eqn\checprop{ \Delta_{2n} \Theta_1\Theta_2\ldots \Theta_{n-1}\Psi_{\pi_0}=
(1+\Theta_{2n})\sum_{j=2}^n \Theta_1\Theta_2\ldots \Theta_{j-2}\Theta_j\Theta_{j+1}
\ldots \Theta_{n-1} \Psi_{\pi_0} }
where we have used the commutation of the $\Theta_i$ for distant enough indices, and
the fact that $\Theta_i\Psi_{\pi_0}=\Theta_{n+i}\Psi_{\pi_0}$ for all $i$, a direct
consequence of $f_i\cdot\pi_0=f_{i+n}\cdot\pi_0$ for all $i$.
Noting also that $1+\Theta_i =\Delta_i \times 2/(1+z_i-z_{i+1})$, and that $\Delta_i$ is proportional
to $\der_i$, we finally arrive at the condition that the quantity 
\eqn\qtyphi{\Phi_n= \left( \Theta_1\Theta_2\ldots \Theta_{n-1}-{2\over 1+z_{2n}-z_1}\sum_{j=2}^n
\Theta_1\Theta_2\ldots \Theta_{j-2}\Theta_j\Theta_{j+1}
\ldots \Theta_{n-1}\right) \Psi_{\pi_0}}
must be annihilated by $\Delta_{2n}$, hence by $\der_{2n}$, or equivalently that $\Phi_n$
must be symmetric under $z_{2n}\leftrightarrow z_1$.
We have found an explicit expression for $\Phi_n$, displaying this invariance manifestly, namely
\eqn\expliciphi{ 
\Phi_n= \Psi_{\pi_0}^{(n-1)} (z_2,z_3,\ldots z_{2n-1}) \prod_{j=2}^n a_{1,j}a_{2n,j}a_{j+n-1,1}a_{j+n-1,2n}}
where the first term is nothing but the expression for $\Psi_{\pi_0}$ \valpiob\ but for the size $2n-2$,
and with arguments counted cyclically from $2$ to $2n-1$. We have also used the shorthand notation 
$a_{i,j}=1+z_i-z_j$. To prove \expliciphi, we fist note that \checprop\ may be equivalently 
rewritten as
\eqn\rewchec{ \Phi_n= \left( \prod_{i=1}^{n-1} \Theta_i -{2\over 1+z_{2n}-z_i} \right) \Psi_{\pi_0}}
This is easily shown by induction, by noticing that $\Theta_i$ commutes
with $2/a_{2n,j}$ for $j=i+2,i+3,\ldots,n-1$, and that 
$\Theta_i(2/a_{2n,i+1}) =4/(a_{2n,i}a_{2n,i+1})$, from 
the explicit definition of $\Theta_i$ \deftheta. Finally, acting first with $\Theta_{n-1}-2/a_{2n,n-1}$
on $\Psi_{\pi_0}$, we see that the only terms not symmetric under $z_{n-1}\leftrightarrow z_n$,
hence that do not commute with this action, are $a_{n-1,n}a_{n,2n-1} a_{2n,n-1}$, and from
the explicit expression \deftheta, we get 
\eqn\firstep{ \left(\Theta_{n-1}-{2\over a_{2n,n-1}}\right)(a_{n-1,n}a_{n,2n-1} a_{2n,n-1})=
a_{n-1,n} a_{2n,n} a_{2n-1,n-1}}
The same reasoning then applies to the resulting function, whose only term non-symmetric under
$z_{n-2}\leftrightarrow z_{n-1}$ is $a_{n-2,n-1}a_{2n,n-2}a_{n-1,2n-2}$; at the $j^{\rm th}$ step
we act with $\Theta_{n-j}-2/a_{2n,n-j}$ on the product $a_{n-j,n-j+1}a_{2n,n-j}a_{n-j+1,2n-j}$
with the result $a_{n-j,n-j+1}a_{2n,n-j+1}a_{2n-j,n-j}$, and this goes on until $j=n-1$. 
Collecting all the terms, we finally get Eq.~\expliciphi.

We conclude that the $\Psi_n$ constructed above, normalized by \valpiob, satisfies all 
equations \otrans, and therefore also
satisfies \gsvec. Moreover, any GCD of the $\Psi_\pi$ would divide $\Psi_{\pi_0}$ of
\valpiob, but all the factors present in \valpiob\ are required by the proposition 1, hence this GCD
must be a constant and the components of $\Psi_n$ are indeed coprime.
Furthermore, due to the structure of $\Theta_i$, we find that the entries $\Psi_{\pi}$ are 
polynomials with total degree 
and partial degrees in each $z_i$ bounded respectively by the total degree $2n(n-1)$ and partial
degrees $2(n-1)$ of $\Psi_{\pi_0}$. In fact, we have equality of degrees, and an explicit expression
for the highest degree terms in $\Psi_\pi$, as the following lemma shows:

\proclaim Lemma 2. 
The terms of $\Psi_{\pi}$ of maximal degree read 
\eqn\maxdeg{\Psi_{\pi}^{\rm max}(z_1,\ldots,z_{2n})= (-1)^{c(\pi)}
\left(\prod_{1\leq i<j\leq 2n} (z_i-z_j) \right)\prod_{i<j:\pi(i)=j} {1\over z_i-z_j} }
where $c(\pi)$ is the number of crossings of $\pi$.

This is proved by induction on the quantity $n(n-1)/2-c(\pi)$, 
starting from $\pi=\pi_0$, whose leading degree terms read
\eqn\debut{ \Psi_{\pi_0}^{\rm max}(z_1,\ldots,z_{2n})=(-1)^{n(n-1)/2}
\Delta(z) \prod_{i=1}^{n}{1\over z_i-z_{i+n}}}
where we used the standard notation $\Delta(z)=\prod_{1\leq i<j\leq 2n}(z_i-z_j)$ for the 
Vandermonde determinant. This leading term
matches \maxdeg\ upon noting that $c(\pi_0)=n(n-1)/2$. Let us now prove that
if $\pi$ has no little arch linking $(i,i+1)$, then at leading order in the $z$,
the action of $\Theta_{i}$ preserves the form \maxdeg, with $\pi$ replaced by $f_i\cdot\pi$. 
Assuming that $\pi(i)=k$
and $\pi(i+1)=\ell$, we see that the product \maxdeg\ may be rewritten as
$\Psi_{\pi}^{\rm max}=(z_i-z_\ell)(z_{i+1}-z_k)(z_i-z_{i+1})\Phi_{\pi}$, where the polynomial
$\Phi_{\pi}$ is symmetric under the interchange $z_i\leftrightarrow z_{i+1}$. As any polynomial
with such a symmetry may be factored out of the action of $\Theta_{i}$ 
(the latter affects only non-symmetric terms), we are left with the task of finding the 
leading behavior at large $z_i$ of
\eqn\tetact{ \Theta_{i} (z_i-z_\ell)(z_{i+1}-z_k)(z_i-z_{i+1}) \sim -(z_i-z_k)(z_{i+1}-z_\ell)(z_i-z_{i+1})}
obtained as an immediate consequence of \gauge, as $\partial_i$
decreases the degree strictly, and only the $-\tau_i$ term contributes 
at leading order. 
This proves that $\Psi_{f_i\cdot\pi}^{\rm max}$ also has the form
\maxdeg, the overall minus sign accounting for the decrease by 1 of the number of crossings.
This completes the proof of \maxdeg\ for all entries of $\Psi_n$, and of Theorem 2. \qed

As a side remark,
an immediate consequence of Lemma 2 is the property that the maximal degree terms in the ground state
vector entries, $\Psi_{\pi}^{\rm max}$, are invariant under the interchange $z_i\leftrightarrow z_{i+1}$
whenever $\pi$ has a little arch joining $(i,i+1)$. Indeed, as an arch joins $(i,i+1)$, the 
only term involving $z_i$ to be divided out of the Vandermonde determinant in \maxdeg\
is the skew-symmetric term $(z_i-z_{i+1})$. This leaves us with a manifestly invariant product, and
shows that the action of the operators $\Delta_{i}$ on entries of $\Psi_n$ with 
a little arch joining $(i,i+1)$ does not increase the degree, as annouced.

We conclude this section by stressing that we have explicitly constructed the entries of $\Psi$
by repeated action of the $\Theta_i$ on $\Psi_{\pi_0}$ of \valpiob. It is clear that all the entries
of $\Psi$ thus constructed are polynomials of the $z_i$ with {\it integer}\/ coefficients. In particular,
in the homogeneous limit where
all the $z_i$ tend to the same value (say $t$), we see that $\Psi_{\pi_0}\to 1$,
and we end up with a vector whose entries are all integers, moreover positive, as
$\Psi$ then coincides with the Perron--Frobenius eigenvector of the Hamiltonian of \BRAU, with the same 
normalization condition. This yields a proof that, in the homogeneous case, once $\Psi$ is normalized
by $\Psi_{\pi_0}=1$, all its entries are positive integers.

\newsec{Sector of permutations}
\subsec{Definition of the sector $P_n$ and of the associated varieties}
A subset $P_n$ of $CP_n$ consists of the $n!$ ``permutation patterns'' that connect the points
$\{1,2,\ldots,n\}$ to the points $\{n+1,n+2,\ldots 2n\}$. Such patterns $\pi\in P_n$ are in one-to-one
correspondence with permutations $\hat{\pi}\in {\cal S}_n$ 
of $\{1,\ldots,n\}$ via $\pi(i)=\hat{\pi}(n+1-i)+n$, $i=1,\ldots,n$.

\def\myitem{\par\nobreak\indent\hangindent2\parindent\textindent}
To each pattern $\pi\in P_n$ is naturally associated a homogeneous affine variety $\aff{\pi}$:
following Knutson \Kn, we consider the three conditions on
pairs of $n\times n$ complex matrices $X$ and $Y$:

\myitem{(1)} $XY$ lower triangular, $YX$ upper triangular.
\myitem{(2)} $(XY)_{ii}=(YX)_{\hat\pi(i)\hat\pi(i)}$ for $i=1,\ldots,n$.
\myitem{(3)} The matrix Schubert variety conditions: the rank of any upper left (resp.\ lower right)
rectangular submatrix of $X$ (resp.\ $Y$) 
is less or equal to the rank of the corresponding submatrix of the
permutation matrix $\hat\pi$ (resp.\ $\hat\pi^{-1}$).
\smallskip

The $\aff{\pi}$ all have the same dimension $n(n+1)$,
or equivalently codimension $n(n-1)$ in $M_n({\Bbb C})^2$.
They are conjectured (Sect.~3 of \Kn) to be the irreducible components 
of $\aff{}$ which is the set of pairs $(X,Y)$
that satisfy condition (1) only. 
Note that we use the ``lower-upper scheme'' here, as in Sect.~1 of \Kn, where $\aff{}$
is denoted by $D^0$;
however, starting
with Sect.~2, \Kn\ uses the ``upper-upper scheme'', hence slightly differing conventions. 

\subsec{Refined de Gier--Nienhuis conjectures}
We first provide the following interpretation of the entries $\Psi_{\pi}$ with
$\pi\in P_n$ for a special choice of inhomogeneities, which refines a conjecture of Nienhuis and
de Gier \BRAU:

\proclaim Conjecture 1. Set 
$z_i=B/(A+B)$, for $i=1,\ldots,n$ and $z_i=A/(A+B)$ for $i=n+1,\ldots,2n$.
Then we have the following identification for $\pi\in P_n$:
\eqn\bideg{ \left({A+B\over 2}\right)^{n(n-1)}\Psi_{\pi}= d_\pi(A,B) }
where $d_\pi(A,B)$ is the bidegree of $\aff{\pi}$
which is related to the separate scaling of the matrices $X$ and $Y$
(such a bidegree is defined in Sect.~2.2 of
\Kn\foot{Note that we have slightly altered the notation:
in \Kn\ the index $\sigma$ of $d_\sigma(A,B)$
refers to the permutation such that $\pi(i)=n+\sigma(i)$, $i=1,2,\ldots n$.}).

The cases $n=3$ and $n=4$ are given in appendix C.

For $A=B=1$, all $z_i$ are equal, and our conjecture reduces to that of \BRAU.
More precisely, Eq.~\bideg\ then expresses identities between
on the left hand side the entry of the homogeneous problem (i.e.\ of the null vector
of the Hamiltonian $H_n$),
and on the right hand side the usual degrees of the algebraic varieties.

In order to go further, we note that for any $\pi\in P_n$, the entry $\Psi_{\pi}$ has no little arch connecting
pairs $(i,j)$ with $1\le i,j\le n$ or $n+1\le i,j\le 2n$. This motivates the following redefinitions:
set $p_i=z_{n+1-i}$, $q_i=z_{n+i}$, $i=1,\ldots,n$ and
\eqn\defdeg{
\Psi_{\pi}(z_1,\ldots,z_{2n})=
\prod_{1\le i<j\le n} \!\!\! (1+z_i-z_j) 
\prod_{n+1\le k<l\le 2n} \!\!\!\!\!\!\! (1+z_k-z_l)
\ \ \ \delta_\pi(p_1,\ldots,p_n,q_1,\ldots,q_n)
}
where $\delta_\pi$ is some polynomial of its $2n$ variables. 

To interpret $\delta_\pi$,
we also need a more general action of a torus of dimension $2n+2$
that maps $(X,Y)$ to $(a PXQ^{-1},b QYP^{-1})$;
it allows to define multidegrees, i.e.\ torus-equivariant cohomology.
With respect to this action, the weights of the variables are:
\eqn\weideg{
[X_{ij}]=A+p_i - q_j \qquad [Y_{ij}]=B+q_i-p_j
}

We now have the following stronger conjecture:

\proclaim Conjecture 2. $\delta_\pi(p_1,\ldots,p_n,q_1,\ldots,q_n)$ is the $(2n+2)$-multidegree 
of $\aff{\pi}$ associated to the weights \weideg\ in which one sets $A=B=1$.

That Conjecture 2 implies Conjecture 1 is due to the following simple scaling property: if one
sets $A=B=1$ in Eq.~\weideg, and performs the change of variables
\eqn\scalrel{
p_i={A'+2 p'_i\over A'+B'} \qquad q_i={B'+2 q'_i\over A'+B'}
}
one finds $[X_{ij}]={2\over A'+B'}(A'+p'_i-q'_j)B$
and $[Y_{ij}]={2\over A'+B'} (B'+q'_i-p'_j)$, 
which are identical to the original weights up to a factor $2/(A'+B')$.
Since the multidegrees are homogeneous polynomials of degree $n(n-1)$
(the codimension of $\aff{\pi}$), 
the full $(2n+2)$-multidegree can be extracted from $\delta_\pi$
by simply pulling out the factor $(2/(A'+B'))^{n(n-1)}$. 
As a special case, if we set all $p'_i=q'_i=0$,
we recover Eq.~\bideg.

Note also that the terms of highest degree of $\delta_\pi$ are now interpreted, using the same
scaling argument, as the $2n$-multidegree of $V_\pi$ associated to the weights \weideg\ with $A=B=0$.
Explicitly,
\eqn\maxmuldeg{
\delta_\pi^{\rm max}(p_1,\ldots,p_n,q_1,\ldots,q_n)=
(-1)^{c(\pi)} \prod_{\scriptstyle i,j=1\atop\scriptstyle j\ne\hat\pi(i)}^n (p_i-q_j)
}
In Sect.~4.5, we shall give a simple and efficient way to compute
the $\delta_\pi$ for arbitrary $n$.
But first we show some simple properties satisfied by the entries $\Psi_{\pi}$,
$\pi\in P_n$; these properties have clear meaning for the multidegrees (see also Sect.~2.2 of \Kn\ for
analogous statements on (bi)degrees). 
As a simple example, the behavior of $\Psi_n$ under rotation and reflection (Eqs.~\cyclicov--\refl)
implies, modulo Conjecture 1,
the third and fourth statements in Prop.~4 of \Kn; these statements could be easily extended
to multidegrees, and indeed, we find for the 
conjectured multidegrees $\delta_\pi$:
\eqna\refldeg
$$\eqalignno{
\delta_{\rho^n\cdot\pi}(q_n,\ldots,q_1,p_n,\ldots,p_1)
&=\delta_\pi(p_1,\ldots,p_n,q_1,\ldots,q_n)&\refldeg{a}\cr
\delta_{r\cdot\pi}(-q_1,\ldots,-q_n,-p_1,\ldots,-p_n)
&=\delta_\pi(p_1,\ldots,p_n,q_1,\ldots,q_n)&\refldeg{b}\cr
\delta_{r'\cdot\pi}(-p_n,\ldots,-p_1,-q_n,\ldots,-q_1)
&=\delta_\pi(p_1,\ldots,p_n,q_1,\ldots,q_n)&\refldeg{c}\cr
}$$
where $\widehat{\rho^n\cdot\pi}=\hat\pi_0\hat\pi^{-1}\hat\pi_0$, 
$\widehat{r\cdot\pi}=\hat\pi^{-1}$ and $\widehat{r'\cdot\pi}=\hat\pi_0 \hat\pi \hat\pi_0$.

\subsec{Sum rule for the permutation sector $P_n$}
\proclaim Theorem 3.
The sum of entries
of $\Psi_n$ corresponding to permutation patterns has the following factorized form:
\eqnn\permu
$$\eqalignno{ &Y_n(z_1,\ldots,z_{2n})=\sum_{\pi\in P_n} \Psi_{\pi}(z_1,z_2,\ldots z_{2n})&\permu\cr
&=\Bigg(\prod_{1\leq i<j\leq n}(1+z_i-z_j)(2-z_i+z_j) \Bigg)\Bigg(\prod_{n+1\leq k<l\leq 2n}
(1+z_k-z_l)(2-z_k+z_l)\Bigg) \cr}
$$

Proof. Let us introduce
the linear form $b_n$ that is the characteristic function of $P_n$, namely with entries in the canonical basis
\eqn\bnentries{b_{\pi}=\cases{1& if $\pi\in P_n$\cr 0& otherwise\cr}}
$b_n$ satisfies
\eqn\satib{ b_n I=b_n,\qquad b_n f_i=b_n, \qquad b_n e_i=0, \qquad i\neq n,2n}
as no little arch may connect points among $\{1,2,\ldots n\}$ or
$\{n+1,n+2,\ldots,2n\}$ in $\pi\in P_n$. 
We now act with the characteristic function $b_n$ of the permutation sector $P_n$ \bnentries\
on the matrix $\check R$. Using the relations \satib, we immediately find
\eqn\actbx{ b_n \check R_i(z,w)= 
{(1+{1\over2}(w-z))(1+z-w)\over (1-{1\over2}(w-z))(1+w-z)} b_n,\quad i\neq n,2n}
Noting finally that $Y_n=b_n\cdot \Psi_n$, and taking the scalar product of $b_n$ with \vecrel,
we find that the r.h.s.\ of \permu\ divides the sum on the l.h.s.
Moreover, the degree of the r.h.s.\ is $2n(n-1)$, the same as that of the sum
$Y_n$ by application of Theorem 2. The two terms must therefore be proportional by a constant,
further fixed to be $1$ by considering the terms of maximal degree ($2n(n-1)$) in $Y_n$. Indeed, 
the maximal degree term in the r.h.s.\ of \permu\ reads 
\eqn\degmax{\Delta^2(z_1,\ldots z_n)\Delta^2(z_{n+1},\ldots,z_{2n})}
where $\Delta$ stands for the Vandermonde determinant.
On the other hand, resumming all leading behaviors \maxdeg\ over the permutation patterns $\pi\in P_n$,
we get the following quantity
\eqn\folymax{\eqalign{
Y_n^{max}(z_1,\ldots,z_{2n})&=\Delta(z_1,\ldots,z_{2n})\sum_{\pi\in P_n}
(-1)^{c(\pi)} \prod_{{\rm pairs}\ (i,j)\atop j=\pi(i), \ i=1,2,\ldots,n} {1\over z_i-z_j} \cr
&=\Delta(z_1,\ldots,z_{2n}) \det\left({1\over z_i-z_{n+j}}\right)_{1\leq i,j\leq n}\cr} }
where we have identified the sign $(-1)^{c(\pi)}$ with the signature of the underlying permutation,
eventually leading to the determinant.
The identity between \degmax\ and \folymax\ is just the Cauchy determinant formula:
\eqn\cauch{ {\Delta(z_1,\ldots z_n)\Delta(z_{n+1},\ldots,z_{2n})\over
\prod_{i,j=1}^n (z_i-z_{n+j}) }=\det\left({1\over z_i-z_{n+j}}\right)_{1\leq i,j\leq n} }
\qed

Note that the sum rule \permu\ translates, at the special values of inhomogeneities considered 
in Sect.~4.1 for the Conjecture 1, into the following
\eqn\smur{\sum_{\pi\in P_n} d_\pi(A,B)= \left({A+B\over 2}\right)^{n(n-1)} 
Y_n\Big({B\over A+B},\ldots,{A\over A+B},\ldots\Big)
=(A+B)^{n(n-1)} }
which is in agreement with the first statement in 
Prop.~4 of \Kn. Yet another expression, equivalent to Theorem 3, is:
\eqn\sumdeg{
\sum_{\pi\in P_n} \delta_\pi(p_1,\ldots,p_n,q_1,\ldots,q_n)
= \prod_{1\le i<j\le n} (2+p_i-p_j)(2-q_i+q_j)
}
which is clearly the multidegree of the whole variety $\aff{}$ (equations enforcing condition (1)).

{\it Remark}: rotated versions of the sum rule \permu\ are also available, 
upon using the general cyclic covariance property of $\Psi_n$ \cyclicov, namely involving
$Y_n^{(i)}=Y_n( z_k\to z_{k+i})$, with $Y_n^{(0)}\equiv Y_n$. 
These correspond to sums of entries of $\Psi_n$ over the rotated permutation pattern sets
$P_n^{(i)}$ that connect the points $\{i+1,i+2,\ldots,i+n\}$ to $\{n+i+1,n+i+2,\ldots i\}$ 
and will be used in Sect.~5.1 (proof of Theorem~4).

\subsec{Factorization in the permutation sector $P_n$}
We may prove a general factorization property 
for the fully decomposable permutation patterns $\pi\in P_n$ defined as follows.
Assume the points $1,2,\ldots n$ are partitioned into two sets $R_1=\{1,2,\ldots,r\}$,
$R_2=\{r+1,r+2,\ldots,n\}$; define also $S_1=R_1+n$, $S_2=R_2+n$.
A permutation pattern $\pi\in P_n$ is called fully decomposable with respect to the partition $(R_1,R_2)$
if $\pi$ only connects the points of $R_1$ to those of $S_1$ and the points of $R_2$
to those of $S_2$.
We denote by $\pi_i$ the restriction of 
$\pi$ to the set $R_i\cup S_i$.

\proclaim Proposition 2.
For any fully decomposable $\pi$, we have the
factorization property
\eqn\factot{
\Psi_{\pi}(z_1,\ldots,z_{2n})=
\prod_{(i,j)\in X} (1+z_i-z_j)
\ \Psi_{\pi_1}(z_{R_1\cup S_1})\Psi_{\pi_2}(z_{R_2\cup S_2})
}
where $X=R_1\times R_2 \cup R_2\times S_1 \cup S_1\times S_2\cup S_2\times R_1$,
and $z_I$ denotes the sequence of $z_i, i\in I$.

To prove \factot, we proceed by induction on $n(n-1)/2-c(\pi)$.
we start from the link pattern $\pi_0$, viewed as a fully decomposable pattern with respect
to $(R_1,R_2)$. From the explicit expression \valpiob, we immediately
get \factot\ upon noting that the restrictions $\pi_1$ and $\pi_2$ are themselves 
the maximally crossing patterns $\pi_0$ in their respective sets $P_{r}$ and $P_{n-r}$.
Assume some fully decomposable $\pi$ satisfies \factot. We may reduce by 1 the number
of crossings of $\pi$ by acting on $\pi$ with some $f_i$, with either $i\in \{1,2,\ldots,r-1\}$ 
or $i\in \{r+1,r+2,\ldots, n-1\}$. The corresponding
entry $\Psi_{f_i\cdot\pi}$ is obtained by acting with $\Theta_{i}$ on $\Psi_{\pi}$. But this action
only affects the corresponding restriction $\pi_1$ or $\pi_2$, within which the uncrossing is performed.
Hence the form \factot\ is preserved, and we simply have to substitute $\pi\to f_i\cdot\pi$ and 
either $\pi_1\to f_i\cdot\pi_1$ or $\pi_2\to f_i\cdot\pi_2$. This completes the proof of \factot. \qed

As a corollary, when all $z_i$ are taken to zero, Eq.~\factot\ translates into a factorization
property conjectured in \BRAU. The corresponding property for bidegrees is the second statement
of Prop.~4 of \Kn.
More generally, we can rewrite it in terms of our conjectured multidegrees:
\eqnn\factotb
$$\eqalignno{
\delta_\pi(p_1,\ldots,p_n,q_1,\ldots,q_n)
=&\prod_{i=1}^{n-r}\prod_{j=1}^r (1+p_i-q_j)\prod_{i=r+1}^n \prod_{j=n-r+1}^n (1+q_i-p_j)&\factotb\cr
&\delta_{\pi_1}(p_{n-r+1},\ldots,p_n,q_1,\ldots,q_r)
\,\delta_{\pi_2}(p_1,\ldots,p_{n-r},q_{r+1},\ldots,q_n)
}$$
The factors $\prod_{i=1}^{n-r}\prod_{j=1}^r (1+p_i-q_j)\prod_{i=r+1}^n \prod_{j=n-r+1}^n (1+q_i-p_j)$
correspond to equations enforcing the block-triangular shape of the matrices $X$ and $Y$ for
a decomposable $\pi$.

\subsec{Sketch of proof of the multi-parameter de Gier--Nienhuis conjecture}
It is not the purpose of the present paper to give a rigorous proof of Conjecture 2. However,
we shall indicate the main steps of the proof, leaving aside
various technicalities.
Note that no results in this paper depend on proving Conjecture 2.

The property to be proved is written in short as: $\deg\aff{\pi}=\delta_\pi$ (deg denoting the multidegree
as in Conjecture 2, i.e.\ with $A=B=1$).
The proof proceeds as usual by induction on the number of crossings of $\pi$.

$\star$ For the case of the long permutation $\hat{\pi}_0$, 
we have the explicit formula \valpiob, or with our
redefinitions:
\eqn\valpioc{
\delta_{\pi_0}(p_1,\ldots,p_n,q_1,\ldots,q_n)=
\prod_{\scriptstyle 1\le i,j\le n\atop\scriptstyle i+j<n+1} (1+p_i-q_j)
\prod_{\scriptstyle 1\le i,j\le n\atop\scriptstyle i+j>n+1} (1+q_i-p_j)
}
But the corresponding variety $\aff{\pi_0}$ is known explicitly:
\eqn\trivar{
\aff{\pi_0}=\{ (X,Y)\in M_n({\Bbb C})^2 \mid X {\rm\ lower\ right\ triangular}, Y {\rm\ upper\ left\ triangular}\}
}
The equations defining $\aff{\pi_0}$ are simply: $X_{ij}=0$ for $i+j< n+1$ and $Y_{ij}=0$ for
$i+j> n+1$; using Eq.~\weideg\ with $A=B=1$, we immediately 
find that $\deg V_{\pi_0}$ equals the r.h.s.\ of Eq.~\valpioc.

$\star$ Induction.
We want to show the property for a certain permutation pattern, assuming it is true for
all permutation patterns with higher number of crossings.
We can always write this pattern as $f_{\bar\imath}\cdot\pi$, $\bar\imath=1,\ldots, n-1$, such that
$\pi$ has one more crossing than $f_{\bar\imath}\cdot\pi$.
In terms of permutations, $\widehat{f_{\bar\imath}\cdot\pi}=\hat\pi \tau_i$ with $i=n+1-\bar\imath$, 
that is $f_{\bar\imath}$ acts (as elementary transposition $\tau_i$) by multiplication on the 
right.\foot{Note
that one could have used a $f_{\bar\imath}$ with $\bar\imath=n+1,\ldots,2n-1$; this would correspond
to multiplication of the permutation $\hat{\pi}$ on the left. In our multidegree setting (similar to double
Schubert polynomials) there is complete symmetry between left and right multiplication, corresponding
to operators acting on the $p_i$ or on the $q_i$.}

The crucial point is once more to use the formula \translat\ which relates $\Psi_{f_{\bar\imath}\cdot\pi}$ to
$\Psi_{\pi}$ via the operator $\Theta_{j}$ defined by Eq.~\deftheta.
After taking into account our redefinition \defdeg, we find the much simpler expression
\eqn\divdif{
\delta_{f_{\bar\imath}\cdot\pi}=(2\der_i-\tau_i) \,\delta_{\pi}
}
following immediately from \gauge.
Here $\tau_i$ is the natural action of the symmetric group that exchanges variables $p_i$ and
$p_{i+1}$, and $\der_i$ is the {\it divided difference}\/ operator 
$\der_i={1\over p_{i+1}-p_i}(\tau_i-1)$ (with the usual sign convention).
Note that this is a known representation of the symmetric group, studied for example in \LLT.

It is better to rewrite Eq.~\divdif\ as:
\eqn\divdifb{
\delta_{f_{\bar\imath}\cdot\pi}+\delta_\pi=(2+p_i-p_{i+1}) \der_i\,\delta_{\pi}
}
Roughly speaking, $\der_i$ corresponds to the
action of the elementary transposition $\tau_i$ in
the matrix Schubert variety conditions (3), whereas 
$2+p_i-p_{i+1}$ is the additional equation $(XY)_{i\,i+1}=0$ (part of condition (1))
which was lost in the process.

\def\braket(#1|#2){\left< #1 | #2 \right>}
More explicitly, call $X_i$ the row vectors of $X$ and $Y_i$ the column vectors of $Y$.
For any variety $W$ inside $M_n({\Bbb C})^2$ with a torus action,
define $\der_i W \equiv \{(X',Y')|X'_j=X_j\ \forall j\ne i, X'_i=X_i+u X_{i+1}, 
Y'_j=Y_j\ \forall j\ne i+1, Y'_{i+1}=Y_{i+1}-u Y_i\ {\rm for}\ 
(X,Y)\in W, u\in{\Bbb C}\}$. 
One can show that $W\mapsto \der_i W$ translates
into the operator $\der_i$ for multidegrees (indeed it decreases codimension = degree by 1, and
clearly does not act on variables other than $p_i$, $p_{i+1}$ or on symmetric functions
of $p_i$, $p_{i+1}$; 
all these properties characterize $\der_i$ up to normalization, which is easily fixed). 

Now apply this operation $\der_i$ to the variety $\aff{\pi}$. 
By direct inspection,
$\der_i \aff{\pi}$ satisfies condition (3) in which
$\hat\pi$ is replaced with $\hat\pi\tau_i$ (standard reasoning for matrix Schubert varieties),
as well as the set of equations defining condition (1) (noting that $Y'X'=YX$)
except for $(X'Y')_{i\,i+1}=\braket(X'_{i}|Y'_{i+1})=0$. We therefore compute
$
\braket(X'_{i}|Y'_{i+1})
=u(\braket(X_{i+1}|Y_{i+1})-\braket(X_i|Y_i)-u\braket(X_{i+1}|Y_{i}))
$,
and find that the equation $\braket(X'_{i}|Y'_{i+1})=0$ nicely factorizes into $u=0$, which of course
defines $\aff{\pi}$, and a non-trivial linear equation for $u$:
\eqn\eqnu{
\braket(X_{i+1}|Y_{i+1})-\braket(X_i|Y_i)-u\braket(X_{i+1}|Y_{i})=0
}
Now we want to check condition
(2) when this equation is satisfied. We have $(Y'X')_{jj}=(YX)_{jj}$ for all $j$, and 
$(X'Y')_{jj}=(XY)_{jj}$ for $j\ne i,i+1$. Furthermore,
\eqna\compyxb
$$\eqalignno{
(X'Y')_{ii}=\braket(X'_i|Y'_i)=\braket(X_i|Y_i)+u\braket(X_{i+1}|Y_{i})=\braket(X_{i+1}|Y_{i+1})=(XY)_{i+1\,i+1}\qquad&\quad&\compyxb{a}\cr
(X'Y')_{i+1\,i+1}=\braket(X'_{i+1}|Y'_{i+1})=\braket(X_{i+1}|Y_{i+1})-u\braket(X_{i+1}|Y_{i})=\braket(X_i|Y_i)=(XY)_{ii}\quad&\quad&\compyxb{b}\cr
}$$
using Eq.~\eqnu.
We conclude that $(X',Y')$ satisfies condition (2) in which $\hat\pi$ is replaced with $\hat\pi\tau_i$. Since the operation is an involution, all of
$\aff{f_{\bar\imath}\cdot\pi}$ is obtained this way.

What we have found is that the additional relation $(X'Y')_{i\,i+1}=0$ restricts $\der_i \aff{\pi}$
to $\aff{\pi}\cup\aff{f_{\bar\imath}\cdot\pi}$.
This means that it increases codimension by 1, i.e.\ is independent from other
equations and therefore amounts to multiplication by $2+p_i-p_{i+1}$ of
the multidegree (cf Eq.~\weideg\ with $A=B=1$). This proves that $\deg \aff{f_{\bar\imath}\cdot\pi}+\deg\aff{\pi}=
(2+p_i-p_{i+1})\der_i \deg\aff{\pi}$; by the induction hypothesis $\deg\aff{\pi}=\delta_\pi$,
and comparing with Eq.~\divdifb, we conclude that $\deg\aff{f_{\bar\imath}\cdot\pi}=\delta_{f_{\bar\imath}\cdot\pi}$. \qed

As already mentioned at the end of Sect.~3, a corollary of the recursive construction of the $\Psi_\pi$,
or equivalently of the $\delta_\pi$ in the permutation sector, is that they are sums of products
of $1-z_i+z_j$, and in particular polynomials with integer coefficients;
this could also be seen as a consequence,
assuming Conjecture 2, of the general theorem 1.7.1.\ of \KnM. 

\newsec{Recursion relations and full sum rule}
In this section, we will derive recursion relations relating specialized entries
of $\Psi_n$ to entries of $\Psi_{n-1}$, that will allow us to prove
the full sum rule. 

\subsec{Recursion relations for the entries of $\Psi_n$}
Let us examine the case when a little arch connects two neighboring points $i,i+1$ in
some $\pi\in CP_n$. We have the following 

\proclaim Lemma 3.
Let $\varphi_i$ denote the embedding $CP_{n-1}\to CP_n$ that inserts a little arch between positions
$i-1$ and $i$ and relabels the later positions $j\to j+2$ in any $\pi\in CP_{n-1}$; we also denote by
$\varphi_i$ the induced embedding of vector spaces. We have
\eqn\interphi{\eqalign{ T_n&(t,z_1,\ldots,z_i,z_{i+1}=z_i+1,\ldots,z_{2n})\varphi_i\cr
&={1\over4}(t-z_i)(1-z_i+t)(2-t+z_i)(3-t+z_i)
\varphi_i T_{n-1}(z_1,..,z_{i-1},z_{i+2},\ldots,z_{2n})\cr}}

Proof.
This is a direct consequence of the unitarity relation \ybuni\ 
together with the so-called crossing relation
\eqn\cross{ \hbox{\epsfbox{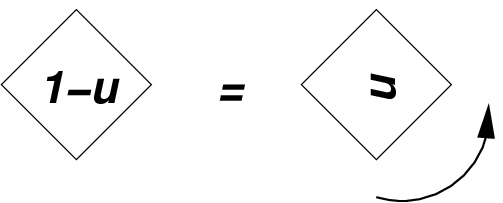}}}
in  which the second picture has been rotated by $+\pi/2$. Indeed, such a rotation
exchanges the roles of $I$ and $e_i$ while $f_i$ is left invariant, and the coefficients
in \defx\ are interchanged accordingly under $u\to 1-u$.
However, it is more instructive to prove \interphi\ directly, by
attaching to the little arch the two plaquettes at sites $i$ and $i+1$
We denote by $u_i=1-t+z_i$, $v_i={1\over2}(t-z_i)(1-t+z_i)$ and $w_i=t-z_i$. 
Pictorially, this gives rise to 9 situations:
\eqnn\picproof
$$
\eqalignno{
&{\raise-5mm\hbox{\epsfxsize=1.35cm\epsfbox{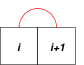}}}\cr
&=w_{i} w_{i+1}{\raise-3.2mm\hbox{\epsfxsize=1.35cm\epsfbox{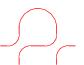}}}
+u_i u_{i+1}{\raise-3.2mm\hbox{\epsfxsize=1.35cm\epsfbox{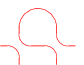}}}
+u_i w_{i+1}{\raise-3.2mm\hbox{\epsfxsize=1.35cm\epsfbox{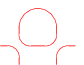}}}
+u_iv_{i+1}{\raise-3.2mm\hbox{\epsfxsize=1.35cm\epsfbox{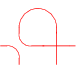}}}
+v_iw_{i+1}{\raise-3.2mm\hbox{\epsfxsize=1.35cm\epsfbox{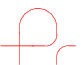}}}\cr
&+w_iv_{i+1}{\raise-3.2mm\hbox{\epsfxsize=1.35cm\epsfbox{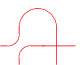}}}
+v_iu_{i+1}{\raise-3.2mm\hbox{\epsfxsize=1.35cm\epsfbox{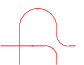}}}\cr
&+w_{i} u_{i+1}{\raise-3.2mm\hbox{\epsfxsize=1.35cm\epsfbox{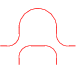}}}
+v_iv_{i+1}{\raise-3.2mm\hbox{\epsfxsize=1.35cm\epsfbox{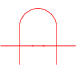}}}&\picproof
}$$
We have displayed on the same line the terms contributing to the
same picture.
We now note that when $z_{i+1}=z_i+1$, the first and second line
both have a global vanishing factor
$w_{i} w_{i+1}+u_i u_{i+1}+u_i w_{i+1}+u_iv_{i+1}+v_iw_{i+1}=
w_iv_{i+1}+v_iu_{i+1}=0$.
We are only left with the contribution where the little arch has safely
gone across the horizontal line, and where in passing the two spaces
$i$ and $i+1$ have been erased. The prefactor is
\eqn\prefa{ w_{i} u_{i+1}+v_iv_{i+1}={1\over4}(t-z_i)(1-z_i+t)(2-t+z_i)(3-t+z_i) }
and yields the prefactor in Eq.~\interphi. \qed

Together with the results of previous sections, this leads to:

\proclaim Theorem 4.  For a given $\pi\in CP_n$,
taking $z_{i+1}=z_i+1$, we have either of the
two following situations:
\item{(i)} There is no little arch $(i,i+1)$ in $\pi$. Then
\eqn\vanish{ \Psi_{\pi}(z_1,\ldots,z_i,z_{i+1}=z_i+1,\ldots,z_{2n})=0}
\item{(ii)} There is a little arch connecting $i$ and $i+1$. Then we have
the recursion relation
\eqn\recure{\eqalign{\Psi_{\pi}(z_1,&\ldots,z_i,z_{i+1}=z_i+1,\ldots,z_{2n})=\cr
&
\left(\prod_{\scriptstyle k=1\atop\scriptstyle k\ne i,i+1}^{2n} (1+z_{i+1}-z_k)(1+z_k-z_i)
\right)
\ \Psi_{\pi'}(z_1,\ldots,z_{i-1},z_{i+2},\ldots,z_{2n})\ .\cr}}
between the entry $\Psi_\pi$ of $\Psi_n$ and the entry $\Psi_{\pi'}$ of $\Psi_{n-1}$, 
where $\pi'$ is the link pattern $\pi$ with the little arch $i,i+1$
removed ($\pi=\varphi_i\pi'$, $\pi'\in CP_{n-1}$).

Proof. The situation (i) is covered by Proposition 1. To prove \recure, we use Lemma 3, and act on
$\Psi_{n-1}$, with the obvious values of the parameters. We have 
$T\varphi_i\Psi_{n-1}=\Lambda/\Lambda'\varphi_i T' \Psi_{n-1}=\Lambda \varphi_i\Psi_{n-1}$, hence
$\varphi_i\Psi_{n-1}$ is proportional to $\Psi_n$, when $z_{i+1}=z_i+1$. This yields Eq.~\recure,
up to a proportionality factor say $\gamma_{n,i}$, rational fraction of the parameters. This factor is
further fixed by applying \vanish--\recure\ to the sum over the suitably rotated set of
permutation patterns $P_n^{(i)}$, that allows for little arches in positions
$(i,i+1)$ or $(i+n,i+n+1)$.  Due to the properties (i-ii), the only non-vanishing contributions
to this sum of entries when $z_{i+1}=z_i+1$ are those for which a little arch connects $(i,i+1)$.
These are the images under $\varphi_i$ of the permutation sector $P_{n-1}^{(i-1)}$, hence we get
\eqn\normay{ Y_n^{(i)}(z_1,\ldots,z_{2n})\big\vert_{z_{i+1}=z_i+1}=
\gamma_{n,i} Y_{n-1}^{(i-1)}(z_1,\ldots,z_{i-1},z_{i+2},\ldots,z_{2n}) }
Applying the suitable rotations to the result of Theorem 3 yields the value of $\gamma_{n,i}$
and \recure\ follows. \qed

\subsec{Sum rule for the entries}
We now compute the sum of all the entries of $\Psi_n$ an express it in
terms of a Pfaffian. We start with the following

\proclaim Lemma 4.
The sum of entries of $\Psi_n$:
\eqn\sumofent{ Z_n(z_1,\ldots,z_{2n})=\sum_{\pi\in CP_n} \Psi_{\pi}(z_1,\ldots,z_{2n})}
is a symmetric polynomial.

Proof.
the linear form $v_n$, with entries $v_{\pi}=1$, clearly satisfies 
$v_nI=v_n$, $v_nf_i=v_n$, $v_ne_i=v_n$, as each of the operators $I,f_i,e_i$ sends each link pattern to
a unique link pattern. Therefore
$v_n \check R_i(z,w)=v_n$. The sum over entries
is nothing but $Z_n=v_n\cdot \Psi_n$, and taking the scalar product of $v_n$ with Eq.~\vecrel,
we deduce that $Z_n$ is invariant under the interchange of $z_i\leftrightarrow z_{i+1}$, hence is
fully symmetric. \qed

By virtue of Theorem 2 and Lemma 4,
$Z_n(z_1,\ldots,z_{2n})$, the sum of entries of $\Psi_n$, is 
a symmetric polynomial of total degree $2n(n-1)$ and partial degree $2n-2$ in each variable,
subject to the recursion relation obtained by summing Eqs.~\vanish--\recure\ over all $\pi\in CP_n$:
\eqn\recik{\eqalign{ &Z_n(z_1,\ldots,z_{2n})\vert_{z_{i+1}=z_i+1}\cr
&=\left(
\prod_{\scriptstyle k=1\atop\scriptstyle k\ne i,i+1}^{2n} (1+z_{i+1}-z_k)(1+z_k-z_i)
\right)Z_{n-1}(z_1,\ldots,z_{i-1},z_{i+2},\ldots,z_{2n})\cr}}
Together with the initial condition $Z_1(z_1,z_2)=1$,
these properties determine it completely and allow us to prove the

\proclaim Theorem 5. The sum of all entries $Z_n$ has a Pfaffian formulation:
\eqn\brik{ Z_n(z_1,\ldots,z_{2n})=
\pf\left({z_i-z_j\over 1-(z_i-z_j)^2}\right)_{1\leq i,j\leq 2n}
\  \times
\prod_{1\leq i<j\leq 2n} {1-(z_i-z_j)^2\over z_i-z_j}}

Proof. The r.h.s.\ of \brik, which we denote by $K_n$, 
is a symmetric polynomial of the $z_i$,
with total degree $2n(n-1)$ and partial degree $2(n-1)$ in each variable, and such that
when say $z_2\to z_1+1$, the Pfaffian degenerates into $(z_1-z_2)/(1-(z_1-z_2)^2)$ times
that for $z_3,z_4,\ldots,z_{2n}$, while the other quantity reduces to $(1-(z_1-z_2)^2)/(z_1-z_2)$
times
$\prod_{3\leq j\leq 2n}(1+z_2-z_j)(1+z_j-z_1)$ times that for $z_3,z_4,\ldots,z_{2n}$.
As the leading singularity is cancelled, this gives the recursion relation
\eqn\kn{ K_n(z_1,\ldots,z_{2n})\vert_{z_2=z_1+1}=
\left(\prod_{j=3}^{2n} (1+z_2-z_j)(1+z_j-z_1)\right)\   K_{n-1}(z_3,\ldots,z_{2n})}
This is nothing but \recik\ for $i=1$, and we deduce that $Z_n=\alpha K_n$ for some numerical 
factor $\alpha$, fixed to be $1$ by the initial condition $Z_1(z_1,z_2)=1$.

Note that the Pfaffian is naturally a sum over pairings of $2n$ objects, indexed
equivalently by crossing link patterns, and weighted by 
a fermionic sign factor. This decomposes naturally $Z_n$ into an alternating sum over
link patterns. In particular taking all $z_i$ to be large, we obtain the leading degree
contribution to $Z_n$, $Z_n^{max}=\Delta(z) \pf(1/(z_i-z_j))_{1\leq i<j\leq 2n}$, 
which matches exactly the sum of the leading terms
\maxdeg\ obtained in Lemma 2 above. 

We get an interesting corollary of Theorem 5 by taking all $z_i$ to zero.
To do this, notice that if $Z=\pf(A)/\prod_{i<j}f(z_i-z_j)$, $A_{ij}=f(z_i-z_j)$, 
$f$ an odd function such that
$f'(0)\neq 0$, then when  all $z_i$ tend to $0$, we have:
\eqn\limiz{
Z\to {1\over f'(0)^{n(2n-1)}}\pf\left({1\over i!j!}{\der^i\over\der z^i}{\der^j\over\der w^j} f(z-w)\vert_{z=w=0}
\right)_{0\leq i,j\leq 2n-1}}
With $f(x)=x/(1-x^2)$, this gives
\eqn\limzer{Z_n(0,\ldots,0)=
\pf\Big[(-1)^j{i+j\choose i}\delta_{i+j,1\,[2]}\Big]_{0\leq i,j\leq 2n-1}}
where the Kronecker delta symbol ensures that the matrix element vanishes unless $i+j$ is odd.
We also have the following determinantal expression, immediately following from \limzer:
\eqn\betterdet{ Z_n(0,\ldots,0)=\det\Big[{2i+2j+1\choose 2i}\Big]_{0\leq i,j\leq n-1}}
These numbers play for the crossing $O(1)$ loop model the same role as that played
for the non-crossing loop model by the numbers $A_n$ of alternating sign matrices
of size $n\times n$.
They read 
\eqn\numbers{ 1,7,307,82977,137460201,1392263902567\ldots }
for $n=1,2,3,4,5,6\ldots$
A 
remark is in order. One possible interpretation of the numbers \betterdet,
via the Lindstr\"om--Gessel--Viennot formula \LGV, is that they count the total number
of $n$-tuples of non-intersecting lattice paths subject to the following constraints:
the paths are drawn
on the edges of the square lattice, are non-intersecting,
and may only make steps of $(-1,0)$ (horizontal to the left) or $(0,1)$ (vertical up),
and start at the points $(2i,0)$, while they end up at the points $(0,2i+1)$, $i=0,1,\ldots,n-1$.
In \betterdet, the combinatorial number ${2i+2j+i\choose 2i}$ is simply the number of configurations of a 
single path starting at $(2i,0)$ and ending at $(0,2j+1)$.

Finally, it is easy to prove, using standard integrable or matrix model techniques,
that the numbers \betterdet\ behave for large $n$ as 
$Z_n(0,\ldots,0)\approx \left({\pi\over 2}\right)^{2n^2}$. 
This is to be compared with the standard asymptotics for the number of $n\times n$
alternating sign matrices $A_n\approx \left({3\sqrt{3}\over 4}\right)^{n^2}$.

\newsec{Conclusion}
In this paper, we have investigated the ground state vector $\Psi_n$ of the inhomogeneous $O(1)$ crossing
loop model on a cylinder of perimeter $2n$. The inhomogeneities translate into
a collection of $2n$ spectral parameters $z_1,z_2,\ldots z_{2n}$, in terms of which we have proved
that, when suitably normalized, all entries of $\Psi_n$ are polynomials of total degree $2n(n-1)$.
By further characterizing these entries, using in particular an algorithm for generating
them recursively, we have been able to find compact formulas for their sum or partial sum over
a subset of entries related to permutations of $n$ objects.

\subsec{Comparison with the non-crossing loop case}
Comparing this case to that of the standard (non-crossing) $O(1)$ loop model studied
in \DFZJ, the situation is now much simpler. Indeed, the transitivity of the move generated by $f_i$
on the crossing link patterns has allowed us to describe all the entries of $\Psi_n$ simply
in terms of successive local actions of the gauged divided difference
operators $\Theta_{i}$ \deftheta\ on that corresponding
to the maximally crossing link pattern. In particular, as $\Theta_{i}$ is degree-preserving, we
have had no difficulty in proving that the coefficients of $\Psi_n$ are polynomials of degree $2n(n-1)$,
as opposed to the non-crossing case of \DFZJ, where no such simple operator is available, and where
the degree issue was highly non-trivial, and eventually had to be settled by use of the algebraic
Bethe Ansatz. Other properties, such as recursion relations when neighboring spectral parameters
become related, turn out to be very similar in the two models. 
In particular, in light of the formula \brik\ for the sum of all entries of $\Psi_n$, we have
found an analogous Pfaffian formula for the {\it square}\/ of the Izergin--Korepin/Okada 
determinant $Z_n^{IK}$,
equal to the sum of entries in the non-crossing case, reading:
\eqn\pfatl{ Z^{IK}_n(z_1,z_2,\ldots z_{2n})^2=
\pf\left({z_i-z_j\over z_i^2+z_iz_j+z_j^2}\right)_{1\leq i,j\leq 2n}\ 
\times \prod_{1\leq i<j\leq 2n} {z_i^2+z_iz_j+z_j^2\over z_i-z_j} }
The latter corresponds to the partition function of the inhomogeneous
non-crossing $O(1)$ loop model on a complete (infinite) cylinder of perimeter $2n$.
The analogies between the crossing and non-crossing loop cases lead us to expect the existence
of a much more general treatment of integrable lattice models involving loops, which
would display these common features.  
Actually, the right unifying algebraic setting seems to be the so-called ``double affine Hecke algebras''
\Che\ which typically mix loop-like operators with the action of the symmetric
group on spectral parameters (see also \FK).

\subsec{A positive extension of Schubert polynomials}
An intriguing and new feature of the crossing loop model is the emergence
of a special subset of entries of $\Psi_n$, indexed by permutations. Roughly speaking, the
latter is obtained by projecting out the generators $e_i$, $i\neq n,2n$, via the relations \satib.
We have shown that the action of the operators $\Theta_{i}$ drastically simplifies in this sector
so as to resemble the standard divided difference operators commonly used in Schubert calculus.
In this respect, the conjecture of \BRAU\ relating the ground state of the homogeneous
crossing loop model to the degrees of some varieties certainly loses
much of its mystery. We have actually refined this conjecture so as to include all spectral
parameter dependence, by interpreting a suitably normalized version of the entries of $\Psi_n$
in the permutation sector as the multidegree of the varieties studied in \Kn, and gave a
sketch of proof of it.  
Playing around with changes of variables, we have also found an interesting family of polynomials
corresponding to the permutation sector, which seems to have only non-negative integer coefficients.
These correspond to the specializations $z_{n+1-i}=(t_i-1)/(t_i+1), \ z_{n+i}=0$, for $i=1,2,\ldots,n$.
More precisely, we define the family of polynomials $s_\pi$ via:
\eqn\spidef{ s_\pi(t_1,t_2,\ldots,t_n)=
{2^{-n(n-1)}\prod_{i=1}^n (1+t_i)^{2(n-1)}\over \prod_{1\leq i<j\leq n} (1-t_i+3t_j+t_it_j)}
\Psi_{\pi}\left({t_n-1\over t_n+1},\ldots,{t_1-1\over t_1+1},0,\ldots\right) }
The top member of the family
reads simply $s_{\pi_0}=\prod_{1\leq i\leq n} t_i^{n-i}$, while all other members are obtained by repeated
actions of the operators $\theta_{i}$, $i=1,2,\ldots,n-1$, defined by
\eqn\actet{\theta_{i}s_\pi=\left((1+t_i)(1+t_{i+1})\partial_i -\tau_i\right)}
($\theta_i$ implementing as usual multiplication on the right by the elementary transposition
of the permutation $\hat\pi$ such that $\pi(i)=n+\hat\pi(n+1-i)$).
The lowest degree term in the $s_\pi$, $s_\pi^{min}$, is readily identified with the Schubert 
polynomial indexed by $\hat\pi$.
The higher degree
terms all turn out experimentally to have non-negative integer coefficients. Eq.~\actet\
gives a very efficient way of computing the bidegrees of conjecture 1 \bideg:  indeed, 
the latter are simply recovered by taking $t_i=A/B$ for all $i$ and premultiplying $s_\pi$
by $B^{n(n-1)}$.
This extension of Schubert polynomials
awaits some good algebro-geometric or combinatorial interpretation.

\subsec{Other entries of $\Psi$ and more combinatorics}
This subset of entries of particular interest should not let us forget about the general picture we 
have obtained. It is actually quite suggestive that for instance the change of variables
$z_i=A/(A+B), \ z_{i+n}=B/(A+B)$ for $i=1,2,\ldots,n$, together with the global multiplication
by $(A+B)^{2n(n-1)-(n-2)-mod(n,2)}2^{-n(n-1)}$ leads to homogeneous polynomials of $A,B$ 
for {\it all}\/ entries of $\Psi_n$ (not just those of the permutation sector), 
all with {\it non-negative}\/ integer coefficients.
It would be very interesting to find an 
interpretation for these other entries as well.

More combinatorics are undoubtedly hidden in the $O(1)$ crossing loop model. For instance, the specialization
$z_1=(t-1)/(t+1)$ while all other $z_i$ are taken to zero leads after multiplication
by a global factor $((1+t)/2)^{2(n-1)}$ to entries that are all polynomials of $t$ with 
{\it non-negative}\/ integer coefficients. 
The sum over all these entries produces a ``refinement" of the numbers \numbers\
into polynomials of $t$, reading up to $n=4$:
\eqn\refine{\eqalign{ P_1(t)&=1, \quad P_2(t)=1+5t+t^2, \quad P_3(t)=7+63t+167t^2+63t^3+7t^4,\cr
P_4(t)&=307+3991t+18899t^2+36583t^3+18899t^4+3991t^5+307t^6\cr}}
The coefficients of these polynomials play for the crossing loop case the same role as that 
played by refined alternating sign matrix numbers for the non-crossing one \PDFone. Note that they
share with them the property that at $t=0$ one recovers the total number of the preceeding rank
(with $t=1$).
We have also been able to construct counterparts for the doubly-refined alternating sign
matrix numbers \PDFtwo, by specializing our model to $z_1=(t-1)/(t+1)$ and $z_{n+1}=(u-1)/(u+1)$
while all other $z_i$ are zero, in which case we still find that all entries
of a suitably normalized $\Psi_n$ are polynomials of $t,u$ with non-negative 
integer coefficients. 
All these integers await some combinatorial interpretation.

Such an interpretation should be provided by a vertex-type model with fixed boundary conditions
on the square grid of size $n\times n$, generalizing the six-vertex model with domain wall boundary
conditions of the non-crossing case. The obvious candidates are the $OSP(p| 2m)$ vertex
models of \refs{\MR,\MNR}, with $p-2m=1$, but the main problem is that there are many such models,
whose row-to-row transfer matrix acts on a Hilbert space of dimension $(p+2m)^{2n}$ for a system
of size $n$, and that in order to cover at least the $(2n-1)!!$ dimensions of the Hilbert space
of crossing link patterns, both $p$ and $m$ should be at least of the order of the size $n$. This
is the worst scenario that could occur, as the number of spin degrees of freedom on each edge
of the lattice must itself grow with the size of the system. The multiplicity of formulations
of the vertex model of the non-crossing case as: alternating sign matrices, fully-packed loops,
osculating paths, etc leaves much room for imagining a generalization adapted
to the crossing loop case, but this is yet to be found.

\medskip

\leftline{\bf Acknowledgments}
P.D.F.\ thanks B.~Nienhuis, V.~Pasquier and J.-B.~Zuber for discussions.
P.Z.-J.\ thanks F.~Hivert and M.~Kazarian for informative exchange on Schubert polynomials, and
M.~Tsfasman for discussions.
Special thanks to A.~Knutson for his explanations 
on multidegrees, and other valuable comments.

\vfill\eject
\catcode`\@=11 
\def\mini#1{\null\,\vtop{\normalbaselines\m@th%
    \ialign{$##$\hfil&&$##$\hfil\crcr%
      #1}}}
\catcode`\@=12 

\appendix{A}{Entries of $\Psi_2$}
The entries of $\Psi_2$ read:
$$\eqalign{
\Psi
\raise-1.4cm\hbox{\epsfxsize=2cm\epsfbox{diag2-1.eps}}
&=(1+z_1-z_2)(1+z_2-z_3)(1+z_3-z_4)(1+z_4-z_1)\cr
\Psi
\raise-1.4cm\hbox{\epsfxsize=2cm\epsfbox{diag2-2.eps}}
&=\mini{(1+z_2-z_3)(1+z_4-z_1)\cr
\times ((1+z_3-z_1)(2-z_4+z_1)+(1+z_1-z_2)(1+z_1-z_3))\cr}\cr
\Psi
\raise-1.4cm\hbox{\epsfxsize=2cm\epsfbox{diag2-3.eps}}
&=\mini{(1+z_1-z_2)(1+z_3-z_4)\cr
\times ((1+z_4-z_2)(2-z_1+z_2)+(1+z_2-z_3)(1+z_2-z_4))\cr}\cr
}$$
\vfill\eject

\appendix{B}{Entries of $\Psi_3$ from the action of the $\Theta_{i}$}
We show here how to obtain each entry of $\Psi_3$ by repeated
use of Eq.~\translat.

The coefficient $\Psi_{\pi_0}$ is given by Eq.~\valpiob.
It reads
\def\a#1#2{(1+z_{#1}-z_{#2})}
$$
\Psi
\raise-1.4cm\hbox{\epsfxsize=2cm\epsfbox{diag3-8.eps}}
=\mini{\a{1}{2}\a{1}{3}\a{2}{3}\a{2}{4}\cr
\times\a{3}{4}\a{3}{5}\a{4}{5}\a{4}{6}\cr
\times\a{5}{1}\a{6}{1}\a{6}{2}\a{5}{6}\cr}
$$
It has total degree $12$. The other coefficients are given by
$$
\openup2\jot
\displaylines{
\Psi
\raise-1.4cm\hbox{\epsfxsize=2cm\epsfbox{diag3-5.eps}}
=\Theta_3 \Psi_{\pi_0} 
\quad
\Psi
\raise-1.4cm\hbox{\epsfxsize=2cm\epsfbox{diag3-11.eps}}
=\Theta_1\Psi_{\pi_0}
\quad
\Psi
\raise-1.4cm\hbox{\epsfxsize=2cm\epsfbox{diag3-9.eps}}
=\Theta_2\Psi_{\pi_0}
\cr
\Psi
\raise-1.4cm\hbox{\epsfxsize=2cm\epsfbox{diag3-14.eps}}
=\Theta_1\Theta_2\Psi_{\pi_0}
\quad
\Psi
\raise-1.4cm\hbox{\epsfxsize=2cm\epsfbox{diag3-2.eps}}
=\Theta_2\Theta_3\Psi_{\pi_0}
\quad
\Psi
\raise-1.4cm\hbox{\epsfxsize=2cm\epsfbox{diag3-10.eps}}
=\Theta_1\Theta_3\Psi_{\pi_0}
\cr
\Psi
\raise-1.4cm\hbox{\epsfxsize=2cm\epsfbox{diag3-12.eps}}
=\Theta_2\Theta_1\Psi_{\pi_0}
\quad
\Psi
\raise-1.4cm\hbox{\epsfxsize=2cm\epsfbox{diag3-6.eps}}
=\Theta_5\Theta_3\Psi_{\pi_0}
\quad
\Psi
\raise-1.4cm\hbox{\epsfxsize=2cm\epsfbox{diag3-4.eps}}
=\Theta_4\Theta_3\Psi_{\pi_0}
\cr
\Psi
\raise-1.4cm\hbox{\epsfxsize=2cm\epsfbox{diag3-15.eps}}
=\Theta_2\Theta_1\Theta_2\Psi_{\pi_0}
\quad
\Psi
\raise-1.4cm\hbox{\epsfxsize=2cm\epsfbox{diag3-3.eps}}
=\Theta_3\Theta_2\Theta_3\Psi_{\pi_0}
\quad
\Psi
\raise-1.4cm\hbox{\epsfxsize=2cm\epsfbox{diag3-7.eps}}
=\Theta_1\Theta_4\Theta_3\Psi_{\pi_0}
\cr
\Psi
\raise-1.4cm\hbox{\epsfxsize=2cm\epsfbox{diag3-1.eps}}
=\Theta_6\Theta_2\Theta_1\Psi_{\pi_0}
\quad
\Psi
\raise-1.4cm\hbox{\epsfxsize=2cm\epsfbox{diag3-13.eps}}
=\Theta_2\Theta_5\Theta_3\Psi_{\pi_0}
\cr
}
$$

Unfortunately lack of space does not permit us to produce them explicitly here.

Note that we could have restricted ourselves to the computation
of the $\Psi_{\pi}$, taking for $\pi$ one representative in each orbit under rotation,
and extended the result to the remainder of the orbit by use of the cyclic covariance property
\cyclicov. 

Although we have been able to bypass \otrans\ by only using \translat, we may have applied
it to determine the last two entries above
$$\openup2\jot
\eqalign{
\Psi
\raise-1.4cm\hbox{\epsfxsize=2cm\epsfbox{diag3-1.eps}}
&=(\Delta_6\Theta_2\Theta_1\Theta_2- \Theta_2\Theta_1-\Theta_3-1)\Psi_{\pi_0}\cr
\Psi
\raise-1.4cm\hbox{\epsfxsize=2cm\epsfbox{diag3-13.eps}}
&=(\Delta_3\Theta_2\Theta_1\Theta_2-\Theta_1\Theta_2-\Theta_3-1)\Psi_{\pi_0}\cr
}$$
As already mentioned, both $\Delta_6$ and $\Delta_3$ could in principle increase the degree
by one unit, but they both act on the entry $\Theta_2\Theta_1\Theta_2\Psi_{\pi_0}$,
whose maximal degree ($12$) contribution reads
$$\Psi^{max}
\!\!\!\!\raise-1.4cm\hbox{\epsfxsize=2cm\epsfbox{diag3-15.eps}}
=
\mini{
(z_1-z_2)(z_1-z_3)(z_1-z_4)(z_1-z_5)(z_2-z_3)(z_2-z_4)\cr
\times (z_2-z_6)(z_3-z_5)(z_3-z_6)(z_4-z_5)(z_4-z_6)(z_5-z_6)\cr}
$$
according to Lemma 2, Eq.~\maxdeg.
This is clearly invariant both under $z_6\leftrightarrow z_1$ and $z_3\leftrightarrow z_4$, 
and therefore the degree is preserved by both operators.
\vfill\eject

\appendix{C}{Bidegree of the affine homogeneous variety $\aff{\pi}$ for $n=3,4$}
The specialized entries $\Psi_{\pi}$, conjectured to be bidegrees of the $\aff{\pi}$,
are listed below in decreasing number of crossings of the corresponding link pattern.

The $6$ bidegrees at $n=3$:
$$\openup2\jot
\eqalign{ d
\raise-1.4cm\hbox{\epsfxsize=2cm\epsfbox{diag3-8.eps}}
&=A^3B^3\cr
d
\raise-1.4cm\hbox{\epsfxsize=2cm\epsfbox{diag3-9.eps}}
&=A^2B^2(A^2+AB+B^2)\cr
d
\raise-1.4cm\hbox{\epsfxsize=2cm\epsfbox{diag3-11.eps}}
&=A^2B^2(A^2+AB+B^2)\cr
d
\raise-1.4cm\hbox{\epsfxsize=2cm\epsfbox{diag3-12.eps}}
&=AB(A^4+2A^3B+4A^2B^2+4AB^3+2B^4)\cr
d
\raise-1.4cm\hbox{\epsfxsize=2cm\epsfbox{diag3-14.eps}}
&=AB(2A^4+4A^3B+4A^2B^2+2AB^3+B^4)\cr
d
\raise-1.4cm\hbox{\epsfxsize=2cm\epsfbox{diag3-15.eps}}
&=A^6+3A^5B+7A^4B^2+9A^3B^3+7A^2B^4+3AB^5+B^6\cr}
$$
\vfill\eject

The $24$ bidegrees for $n=4$:
$$\openup2\jot
\eqalign{
d
\raise-1.6cm\hbox{\epsfxsize=2cm\epsfbox{diag4perm-1.eps}}
&=A^6 B^6\cr
d
\raise-1.6cm\hbox{\epsfxsize=2cm\epsfbox{diag4perm-7.eps}}
&=d
\raise-1.6cm\hbox{\epsfxsize=2cm\epsfbox{diag4perm-3.eps}}
=d
\raise-1.6cm\hbox{\epsfxsize=2cm\epsfbox{diag4perm-2.eps}}
=A^5B^5 (A^2 + A B + B^2)\cr 
d
\raise-1.6cm\hbox{\epsfxsize=2cm\epsfbox{diag4perm-8.eps}}
&=A^4 B^4 (A^2 + A B + B^2)^2\cr
d
\raise-1.6cm\hbox{\epsfxsize=2cm\epsfbox{diag4perm-9.eps}}
&=d
\raise-1.6cm\hbox{\epsfxsize=2cm\epsfbox{diag4perm-4.eps}}
=A^4 B^4 (A^4 + 2 A^3 B + 4 A^2 B^2 + 4 A B^3 +2 B^4)\cr
d
\raise-1.6cm\hbox{\epsfxsize=2cm\epsfbox{diag4perm-13.eps}}
&=d
\raise-1.6cm\hbox{\epsfxsize=2cm\epsfbox{diag4perm-5.eps}}
=A^4 B^4 (2 A^4 + 4 A^3 B + 4 A^2 B^2 + 2 A B^3 + B^4)\cr 
d
\raise-1.6cm\hbox{\epsfxsize=2cm\epsfbox{diag4perm-6.eps}}
&=d
\raise-1.6cm\hbox{\epsfxsize=2cm\epsfbox{diag4perm-15.eps}}
=\mini{A^3 B^3 (&A^6 + 3 A^5 B + 7 A^4 B^2 + 9 A^3 B^3 + \cr
&7 A^2 B^4 + 3 A B^5 + B^6)\cr}\cr 
d
\raise-1.6cm\hbox{\epsfxsize=2cm\epsfbox{diag4perm-10.eps}}
&=A^3 B^3 (A^6 + 3 A^5 B + 9 A^4 B^2 + 17 A^3 B^3 + 
21 A^2 B^4 + 15 A B^5 + 5 B^6)\cr 
d
\raise-1.6cm\hbox{\epsfxsize=2cm\epsfbox{diag4perm-19.eps}}
&=A^3 B^3 (5 A^6 + 15 A^5 B + 21 A^4 B^2 + 
17 A^3 B^3 + 9 A^2 B^4 + 3 A B^5 + B^6)\cr 
}
$$
$$\openup2\jot
\eqalign{
d
\raise-1.6cm\hbox{\epsfxsize=2cm\epsfbox{diag4perm-11.eps}}
&=d
\raise-1.6cm\hbox{\epsfxsize=2cm\epsfbox{diag4perm-14.eps}}
=\mini{A^3 B^3 (A^2 + A B + B^2)\cr
(2 A^4 + 4 A^3 B + 5 A^2 B^2 + 4 A B^3 + 2 B^4)\cr }
\cr
d
\raise-1.6cm\hbox{\epsfxsize=2cm\epsfbox{diag4perm-12.eps}}
&=d
\raise-1.6cm\hbox{\epsfxsize=2cm\epsfbox{diag4perm-16.eps}}
=
\mini{
A^2 B^2 (&A^8 + 4 A^7 B + 13 A^6 B^2 + 28 A^5 B^3 + 42 A^4 B^4 \cr
&+ 42 A^3 B^5+ 28 A^2 B^6 + 12 A B^7 + 3 B^8)\cr}
\cr
d
\raise-1.6cm\hbox{\epsfxsize=2cm\epsfbox{diag4perm-20.eps}}
&=d
\raise-1.6cm\hbox{\epsfxsize=2cm\epsfbox{diag4perm-21.eps}}
=
\mini{
A^2 B^2 (&3 A^8 + 12 A^7 B + 28 A^6 B^2 + 
42 A^5 B^3 + 42 A^4 B^4\cr
&+ 28 A^3 B^5 + 13 A^2 B^6 + 4 A B^7 + B^8)\cr}
\cr
d
\raise-1.6cm\hbox{\epsfxsize=2cm\epsfbox{diag4perm-17.eps}}
&=
\mini{
A^2 B^2 (&3 A^8 + 12 A^7 B + 27 A^6 B^2 + 
40 A^5 B^3 + 45 A^4 B^4 + 40 A^3 B^5\cr
&+ 27 A^2 B^6 + 12 A B^7 + 3 B^8)\cr}
\cr
d
\raise-1.6cm\hbox{\epsfxsize=2cm\epsfbox{diag4perm-18.eps}}
&=
\mini{
A B (&A^{10} + 5 A^9 B + 19 A^8 B^2 + 47 A^7 B^3 + 
81 A^6 B^4 + 101 A^5 B^5 + 97 A^4 B^6 \cr
&+ 73 A^3 B^7+ 41 A^2 B^8 + 15 A B^9 + 3 B^{10})\cr}
\cr
d
\raise-1.6cm\hbox{\epsfxsize=2cm\epsfbox{diag4perm-23.eps}}
&=
\mini{
A B (&3 A^{10} + 15 A^9 B + 41 A^8 B^2 + 73 A^7 B^3 + 
97 A^6 B^4 + 101 A^5 B^5 \cr
&+ 81 A^4 B^6 + 47 A^3 B^7+ 19 A^2 B^8 + 5 A B^9 + B^{10})\cr}
\cr
d
\raise-1.6cm\hbox{\epsfxsize=2cm\epsfbox{diag4perm-22.eps}}
&=
\mini{
A B (&2 A^{10} + 10 A^9 B + 34 A^8 B^2 + 82 A^7 B^3 + 
141 A^6 B^4 + 169 A^5 B^5 \cr &+ 141 A^4 B^6
+ 82 A^3 B^7+ 34 A^2 B^8 + 10 A B^9 + 2 B^{10})\cr} 
\cr
d
\raise-1.6cm\hbox{\epsfxsize=2cm\epsfbox{diag4perm-24.eps}}
&=
\mini{
A^{12} + 6 A^{11} B + 25 A^{10} B^2 + 70 A^9 B^3 + 
141 A^8 B^4 + 210 A^7 B^5 + 239 A^6 B^6 \cr
+ 210 A^5 B^7+ 141 A^4 B^8 + 70 A^3 B^9 + 25 A^2 B^{10} + 6 A B^{11} + B^{12}\cr}
\cr}
$$

\vfill\eject

\listrefs

\end

For $n=3$, we use the labeling of crossing link patterns of Fig.~\linpa.
We only write explicit formulas for one representant in each of the orbits of $\Psi_{\pi}$ under
rotation. The other members of the orbit are easily obtained by cyclic covariance \cyclicov. For
instance, the five other members of the orbit of the link pattern $5$ of Fig.~\linpa\
are obtained by successive rotations, leading successively to the link patterns
$6,7,8,9,10$.
We have
\eqn\psithree{\eqalign{
\Psi_1&= a_{1,2}a_{1,3}a_{2,3}a_{2,4}a_{3,4}a_{3,5}a_{4,5}a_{4,6}a_{5,1}a_{6,1}a_{6,2}a_{5,6}\cr
\Psi_2&= a_{1,2}a_{2,3}a_{2,4}a_{3,4}a_{3,5}a_{4,5}a_{5,1}a_{6,1}a_{6,2}a_{5,6} P_2\cr
\Psi_5&= a_{1,2}a_{1,3}a_{2,3}a_{3,4}a_{4,5}a_{4,6}a_{5,6} P_5\cr
\Psi_{11}&=a_{1,2}a_{1,3}a_{2,3}a_{4,5}a_{4,6}a_{5,6} P_6\cr
\Psi_{14}&=a_{2,3}a_{4,5}a_{6,1}P_9\cr}}
where $a_{i,j}=1+z_i-z_j$, and the polynomials $P_i$ of total degree $i$ read
$$\eqalign{
P_2&=3 + z_1 - z_3 - z_1 z_3 + z_4 + z_3 z_4 - z_6 + z_1 z_6 - z_4 z_6\cr
P_5&=13 - 10 z_1 + {z_1}^2 - 2 z_2 + 4 z_1 z_2 - 2 {z_1}^2 z_2 - 3 {z_2}^2 - 2 z_1 {z_2}^2 + 
{z_1}^2 {z_2}^2 + 7 z_3 - 2 z_1 z_3 - {z_1}^2 z_3 + 2 z_2 z_3 \cr &- 4 z_1 z_2 z_3+ 
2 {z_1}^2 z_2 z_3 - {z_2}^2 z_3 - 2 z_1 {z_2}^2 z_3 - {z_1}^2 {z_2}^2 z_3 - 7 z_4 + 
5 z_1 z_4 + 3 {z_2}^2 z_4 - z_1 {z_2}^2 z_4 - 5 z_3 z_4  \cr &+ 3 z_1 z_3 z_4+ 
{z_2}^2 z_3 z_4 + z_1 {z_2}^2 z_3 z_4 + 2 z_5 - z_1 z_5 + {z_1}^2 z_5 + z_2 z_5 + 
4 z_1 z_2 z_5 - {z_1}^2 z_2 z_5 + {z_2}^2 z_5 \cr &+ z_1 {z_2}^2 z_5 + 3 z_1 z_3 z_5- 
{z_1}^2 z_3 z_5 + 3 z_2 z_3 z_5 + {z_1}^2 z_2 z_3 z_5 + {z_2}^2 z_3 z_5 + 
z_1 {z_2}^2 z_3 z_5 + 2 z_4 z_5 - 3 z_1 z_4 z_5 \cr &- 3 z_2 z_4 z_5+ 
z_1 z_2 z_4 z_5 - {z_2}^2 z_4 z_5 - z_1 z_3 z_4 z_5 - z_2 z_3 z_4 z_5 - 
z_1 z_2 z_3 z_4 z_5 - {z_2}^2 z_3 z_4 z_5 - 3 {z_5}^2 + z_1 {z_5}^2 \cr &- z_2 {z_5}^2- 
z_1 z_2 {z_5}^2 - 3 z_3 {z_5}^2 + z_1 z_3 {z_5}^2 - z_2 z_3 {z_5}^2 - 
z_1 z_2 z_3 {z_5}^2 + z_4 {z_5}^2 + z_2 z_4 {z_5}^2 + z_3 z_4 {z_5}^2 \cr &+ 
z_2 z_3 z_4 {z_5}^2 + 10 z_6 - 8 z_1 z_6 + 2 {z_1}^2 z_6 - z_2 z_6 + 8 z_1 z_2 z_6 - 
3 {z_1}^2 z_2 z_6 - {z_2}^2 z_6 + {z_1}^2 {z_2}^2 z_6 + 5 z_3 z_6 \cr &- {z_1}^2 z_3 z_6+ 
3 z_2 z_3 z_6 + {z_1}^2 z_2 z_3 z_6 - 2 z_4 z_6 - 3 z_2 z_4 z_6 + 
z_1 z_2 z_4 z_6 + {z_2}^2 z_4 z_6 - z_1 {z_2}^2 z_4 z_6 - 3 z_3 z_4 z_6 \cr &+ 
z_1 z_3 z_4 z_6 - z_2 z_3 z_4 z_6 - z_1 z_2 z_3 z_4 z_6 + 4 z_5 z_6 - 
8 z_1 z_5 z_6 + 2 {z_1}^2 z_5 z_6 - 4 z_2 z_5 z_6 + z_1 z_2 z_5 z_6 \cr &- 
{z_1}^2 z_2 z_5 z_6 - {z_2}^2 z_5 z_6 - z_1 {z_2}^2 z_5 z_6 + z_1 z_3 z_5 z_6 - 
{z_1}^2 z_3 z_5 z_6 + z_2 z_3 z_5 z_6 - z_1 z_2 z_3 z_5 z_6 + 4 z_4 z_5 z_6 \cr &+ 
z_1 z_2 z_4 z_5 z_6 + {z_2}^2 z_4 z_5 z_6 + z_1 z_3 z_4 z_5 z_6 + 
z_2 z_3 z_4 z_5 z_6 + 2 {z_5}^2 z_6 + z_2 {z_5}^2 z_6 + z_1 z_2 {z_5}^2 z_6 \cr &- 
z_3 {z_5}^2 z_6 + z_1 z_3 {z_5}^2 z_6 - 2 z_4 {z_5}^2 z_6 - z_2 z_4 {z_5}^2 z_6 - 
z_3 z_4 {z_5}^2 z_6 + {z_6}^2 - 2 z_1 {z_6}^2 + {z_1}^2 {z_6}^2 - z_2 {z_6}^2 \cr &+ 
2 z_1 z_2 {z_6}^2 - {z_1}^2 z_2 {z_6}^2 + z_4 {z_6}^2 - z_1 z_4 {z_6}^2 - 
z_2 z_4 {z_6}^2 + z_1 z_2 z_4 {z_6}^2 + 2 z_5 {z_6}^2 - 3 z_1 z_5 {z_6}^2 \cr &+ 
{z_1}^2 z_5 {z_6}^2 - z_2 z_5 {z_6}^2 + z_1 z_2 z_5 {z_6}^2 + 2 z_4 z_5 {z_6}^2 - 
z_1 z_4 z_5 {z_6}^2 - z_2 z_4 z_5 {z_6}^2 + {z_5}^2 {z_6}^2 - z_1 {z_5}^2 {z_6}^2 \cr &+ 
z_4 {z_5}^2 {z_6}^2\cr
P_6&=31 - 22 z_1 + 3 {z_1}^2 + 12 z_1 z_2 - 4 {z_1}^2 z_2 - 7 {z_2}^2 + 2 z_1 {z_2}^2 + 
{z_1}^2 {z_2}^2 + 22 z_3 - 4 z_1 z_3 - 2 {z_1}^2 z_3 + 12 z_2 z_3 \cr 
&\!\!\!\!+ 4 {z_1}^2 z_2 z_3- 
2 {z_2}^2 z_3 + 4 z_1 {z_2}^2 z_3 - 2 {z_1}^2 {z_2}^2 z_3 + 3 {z_3}^2 + 2 z_1 {z_3}^2 - 
{z_1}^2 {z_3}^2 + 4 z_2 {z_3}^2 + 4 z_1 z_2 {z_3}^2 + {z_2}^2 {z_3}^2 \cr 
&\!\!+ 
2 z_1 {z_2}^2 {z_3}^2 + {z_1}^2 {z_2}^2 {z_3}^2 - 22 z_4 + 13 z_1 z_4 - {z_1}^2 z_4 - 
3 z_2 z_4 - 6 z_1 z_2 z_4 + {z_1}^2 z_2 z_4 + 5 {z_2}^2 z_4 - 3 z_1 {z_2}^2 z_4 \cr &- 
24 z_3 z_4 + 8 z_1 z_3 z_4 - 8 z_2 z_3 z_4 - 8 z_1 z_2 z_3 z_4 - 6 {z_3}^2 z_4 - 
z_1 {z_3}^2 z_4 + {z_1}^2 {z_3}^2 z_4 - 5 z_2 {z_3}^2 z_4 - 2 z_1 z_2 {z_3}^2 z_4 \cr &- 
{z_1}^2 z_2 {z_3}^2 z_4 - {z_2}^2 {z_3}^2 z_4 - z_1 {z_2}^2 {z_3}^2 z_4 + 3 {z_4}^2 - 
z_1 {z_4}^2 + z_2 {z_4}^2 + z_1 z_2 {z_4}^2 + 6 z_3 {z_4}^2 - 2 z_1 z_3 {z_4}^2 \cr &+ 
2 z_2 z_3 {z_4}^2 + 2 z_1 z_2 z_3 {z_4}^2 + 3 {z_3}^2 {z_4}^2 - z_1 {z_3}^2 {z_4}^2 + 
z_2 {z_3}^2 {z_4}^2 + z_1 z_2 {z_3}^2 {z_4}^2 - 3 z_1 z_5 + {z_1}^2 z_5 - 4 z_2 z_5 \cr
&+2 z_1 z_2 z_5 - 2 {z_1}^2 z_2 z_5 - 3 z_1 {z_2}^2 z_5 + {z_1}^2 {z_2}^2 z_5 - 
3 z_3 z_5 - {z_1}^2 z_3 z_5 - 2 z_2 z_3 z_5 - 8 z_1 z_2 z_3 z_5 \cr &+ 
2 {z_1}^2 z_2 z_3 z_5 - 3 {z_2}^2 z_3 z_5 - {z_1}^2 {z_2}^2 z_3 z_5 - {z_3}^2 z_5 - 
z_1 {z_3}^2 z_5 - 2 z_2 {z_3}^2 z_5 - 2 z_1 z_2 {z_3}^2 z_5 - {z_2}^2 {z_3}^2 z_5 \cr &- 
z_1 {z_2}^2 {z_3}^2 z_5 + 12 z_4 z_5 - 6 z_1 z_4 z_5 + {z_1}^2 z_4 z_5 + 
2 z_2 z_4 z_5 + 7 z_1 z_2 z_4 z_5 - {z_1}^2 z_2 z_4 z_5 + z_1 {z_2}^2 z_4 z_5 \cr &+ 
8 z_3 z_4 z_5 + 3 z_1 z_3 z_4 z_5 - {z_1}^2 z_3 z_4 z_5 + 11 z_2 z_3 z_4 z_5 + 
{z_1}^2 z_2 z_3 z_4 z_5 + {z_2}^2 z_3 z_4 z_5 + z_1 {z_2}^2 z_3 z_4 z_5 \cr 
&+2 {z_3}^2 z_4 z_5 + z_1 {z_3}^2 z_4 z_5 + 3 z_2 {z_3}^2 z_4 z_5 + 
z_1 z_2 {z_3}^2 z_4 z_5 + {z_2}^2 {z_3}^2 z_4 z_5 - 4 {z_4}^2 z_5 + z_1 {z_4}^2 z_5 \cr}$$
\vfill\eject
$$\eqalign{&- 
2 z_2 {z_4}^2 z_5 - z_1 z_2 {z_4}^2 z_5 - 5 z_3 {z_4}^2 z_5 + z_1 z_3 {z_4}^2 z_5 - 
3 z_2 z_3 {z_4}^2 z_5 - z_1 z_2 z_3 {z_4}^2 z_5 - {z_3}^2 {z_4}^2 z_5 \cr &- 
z_2 {z_3}^2 {z_4}^2 z_5 - 7 {z_5}^2 + 5 z_1 {z_5}^2 + 3 {z_2}^2 {z_5}^2 - 
z_1 {z_2}^2 {z_5}^2 - 5 z_3 {z_5}^2 + 3 z_1 z_3 {z_5}^2 + {z_2}^2 z_3 {z_5}^2 \cr &+ 
z_1 {z_2}^2 z_3 {z_5}^2 + 2 z_4 {z_5}^2 - 3 z_1 z_4 {z_5}^2 - 3 z_2 z_4 {z_5}^2 + 
z_1 z_2 z_4 {z_5}^2 - {z_2}^2 z_4 {z_5}^2 - z_1 z_3 z_4 {z_5}^2 \cr &- 
z_2 z_3 z_4 {z_5}^2 - z_1 z_2 z_3 z_4 {z_5}^2 - {z_2}^2 z_3 z_4 {z_5}^2 + 
{z_4}^2 {z_5}^2 + z_2 {z_4}^2 {z_5}^2 + z_3 {z_4}^2 {z_5}^2 + z_2 z_3 {z_4}^2 {z_5}^2 \cr &+ 
22 z_6 - 24 z_1 z_6 + 6 {z_1}^2 z_6 - 3 z_2 z_6 + 8 z_1 z_2 z_6 - 5 {z_1}^2 z_2 z_6 - 
5 {z_2}^2 z_6 + {z_1}^2 {z_2}^2 z_6 + 13 z_3 z_6 - 8 z_1 z_3 z_6 \cr &- {z_1}^2 z_3 z_6 + 
6 z_2 z_3 z_6 - 8 z_1 z_2 z_3 z_6 + 2 {z_1}^2 z_2 z_3 z_6 - 3 {z_2}^2 z_3 z_6 - 
{z_1}^2 {z_2}^2 z_3 z_6 + {z_3}^2 z_6 - {z_1}^2 {z_3}^2 z_6 + z_2 {z_3}^2 z_6 \cr &- 
{z_1}^2 z_2 {z_3}^2 z_6 - 4 z_4 z_6 + 8 z_1 z_4 z_6 - 2 {z_1}^2 z_4 z_6 + 
3 z_1 z_2 z_4 z_6 + {z_1}^2 z_2 z_4 z_6 + 3 {z_2}^2 z_4 z_6 - z_1 {z_2}^2 z_4 z_6 \cr &- 
8 z_3 z_4 z_6 + 11 z_1 z_3 z_4 z_6 - {z_1}^2 z_3 z_4 z_6 + 3 z_2 z_3 z_4 z_6 + 
{z_1}^2 z_2 z_3 z_4 z_6 + {z_2}^2 z_3 z_4 z_6 + z_1 {z_2}^2 z_3 z_4 z_6 \cr &- 
2 {z_3}^2 z_4 z_6 + z_1 {z_3}^2 z_4 z_6 + {z_1}^2 {z_3}^2 z_4 z_6 - 
z_2 {z_3}^2 z_4 z_6 + z_1 z_2 {z_3}^2 z_4 z_6 - 2 {z_4}^2 z_6 - z_2 {z_4}^2 z_6 \cr &- 
z_1 z_2 {z_4}^2 z_6 - z_3 {z_4}^2 z_6 - z_1 z_3 {z_4}^2 z_6 - z_2 z_3 {z_4}^2 z_6 - 
z_1 z_2 z_3 {z_4}^2 z_6 + {z_3}^2 {z_4}^2 z_6 - z_1 {z_3}^2 {z_4}^2 z_6 + 12 z_5 z_6 \cr &- 
8 z_1 z_5 z_6 + 2 {z_1}^2 z_5 z_6 - 2 z_2 z_5 z_6 + 11 z_1 z_2 z_5 z_6 - 
3 {z_1}^2 z_2 z_5 z_6 - z_1 {z_2}^2 z_5 z_6 + {z_1}^2 {z_2}^2 z_5 z_6 \cr &+ 
6 z_3 z_5 z_6 + 3 z_1 z_3 z_5 z_6 - {z_1}^2 z_3 z_5 z_6 + 7 z_2 z_3 z_5 z_6 + 
{z_1}^2 z_2 z_3 z_5 z_6 - {z_2}^2 z_3 z_5 z_6 + z_1 {z_2}^2 z_3 z_5 z_6 \cr &+ 
{z_3}^2 z_5 z_6 + z_1 {z_3}^2 z_5 z_6 + z_2 {z_3}^2 z_5 z_6 + 
z_1 z_2 {z_3}^2 z_5 z_6 - 8 z_1 z_4 z_5 z_6 + 2 {z_1}^2 z_4 z_5 z_6 \cr &- 
8 z_2 z_4 z_5 z_6 - {z_1}^2 z_2 z_4 z_5 z_6 - z_1 {z_2}^2 z_4 z_5 z_6 - 
8 z_3 z_4 z_5 z_6 - {z_1}^2 z_3 z_4 z_5 z_6 - 2 z_1 z_2 z_3 z_4 z_5 z_6 \cr &- 
{z_2}^2 z_3 z_4 z_5 z_6 - 2 {z_3}^2 z_4 z_5 z_6 - z_1 {z_3}^2 z_4 z_5 z_6 - 
z_2 {z_3}^2 z_4 z_5 z_6 + 4 {z_4}^2 z_5 z_6 + 2 z_2 {z_4}^2 z_5 z_6 \cr &+ 
z_1 z_2 {z_4}^2 z_5 z_6 + 2 z_3 {z_4}^2 z_5 z_6 + z_1 z_3 {z_4}^2 z_5 z_6 + 
z_2 z_3 {z_4}^2 z_5 z_6 + {z_3}^2 {z_4}^2 z_5 z_6 - 2 {z_5}^2 z_6 \cr &- 
3 z_2 {z_5}^2 z_6 + z_1 z_2 {z_5}^2 z_6 + {z_2}^2 {z_5}^2 z_6 - 
z_1 {z_2}^2 {z_5}^2 z_6 - 3 z_3 {z_5}^2 z_6 + z_1 z_3 {z_5}^2 z_6 \cr &- 
z_2 z_3 {z_5}^2 z_6 - z_1 z_2 z_3 {z_5}^2 z_6 + 4 z_4 {z_5}^2 z_6 + 
z_1 z_2 z_4 {z_5}^2 z_6 + {z_2}^2 z_4 {z_5}^2 z_6 + z_1 z_3 z_4 {z_5}^2 z_6 \cr &+ 
z_2 z_3 z_4 {z_5}^2 z_6 - 2 {z_4}^2 {z_5}^2 z_6 - z_2 {z_4}^2 {z_5}^2 z_6 - 
z_3 {z_4}^2 {z_5}^2 z_6 + 3 {z_6}^2 - 6 z_1 {z_6}^2 + 3 {z_1}^2 {z_6}^2 - z_2 {z_6}^2 \cr &+ 
2 z_1 z_2 {z_6}^2 - {z_1}^2 z_2 {z_6}^2 + z_3 {z_6}^2 - 2 z_1 z_3 {z_6}^2 + 
{z_1}^2 z_3 {z_6}^2 + z_2 z_3 {z_6}^2 - 2 z_1 z_2 z_3 {z_6}^2 \cr &+ 
{z_1}^2 z_2 z_3 {z_6}^2 + 2 z_4 {z_6}^2 - z_1 z_4 {z_6}^2 - {z_1}^2 z_4 {z_6}^2 -
z_2 z_4 {z_6}^2 + z_1 z_2 z_4 {z_6}^2 + z_1 z_3 z_4 {z_6}^2 \cr &- 
{z_1}^2 z_3 z_4 {z_6}^2 + z_2 z_3 z_4 {z_6}^2 - z_1 z_2 z_3 z_4 {z_6}^2 - 
{z_4}^2 {z_6}^2 + z_1 {z_4}^2 {z_6}^2 - z_3 {z_4}^2 {z_6}^2 + z_1 z_3 {z_4}^2 {z_6}^2 \cr &+ 
4 z_5 {z_6}^2 - 5 z_1 z_5 {z_6}^2 + {z_1}^2 z_5 {z_6}^2 - 2 z_2 z_5 {z_6}^2 + 
3 z_1 z_2 z_5 {z_6}^2 - {z_1}^2 z_2 z_5 {z_6}^2 + z_3 z_5 {z_6}^2 \cr &- 
z_1 z_3 z_5 {z_6}^2 + z_2 z_3 z_5 {z_6}^2 - z_1 z_2 z_3 z_5 {z_6}^2 + 
4 z_4 z_5 {z_6}^2 - 2 z_1 z_4 z_5 {z_6}^2 + {z_1}^2 z_4 z_5 {z_6}^2 \cr &- 
2 z_2 z_4 z_5 {z_6}^2 + z_1 z_2 z_4 z_5 {z_6}^2 + z_1 z_3 z_4 z_5 {z_6}^2 + 
z_2 z_3 z_4 z_5 {z_6}^2 - z_1 {z_4}^2 z_5 {z_6}^2 - z_3 {z_4}^2 z_5 {z_6}^2 \cr &+ 
{z_5}^2 {z_6}^2 - z_1 {z_5}^2 {z_6}^2 - z_2 {z_5}^2 {z_6}^2 + z_1 z_2 {z_5}^2 {z_6}^2 + 
2 z_4 {z_5}^2 {z_6}^2 - z_1 z_4 {z_5}^2 {z_6}^2 - z_2 z_4 {z_5}^2 {z_6}^2 \cr &+ 
{z_4}^2 {z_5}^2 {z_6}^2\cr}$$
\vfill\eject

$P_9$ would occupy 13 pages in the present format .....  

$$\eqalign{P_9&=63 + 19 z_1 - 27 {z_1}^2 - 7 {z_1}^3 - 19 z_2 + 5 z_1 z_2 + 15 {z_1}^2 z_2 + 7 {z_1}^3 z_2 - 
  27 {z_2}^2 - 15 z_1 {z_2}^2 + 7 {z_1}^2 {z_2}^2 \cr
&\!\!\!\!\!\!+ 3 {z_1}^3 {z_2}^2 + 7 {z_2}^3+ 
  7 z_1 {z_2}^3 - 3 {z_1}^2 {z_2}^3 - 3 {z_1}^3 {z_2}^3 + 19 z_3 + 20 z_1 z_3 - 
  {z_1}^2 z_3 + 2 {z_1}^3 z_3 - 3 z_2 z_3 + 30 z_1 z_2 z_3 \cr 
&\!\!\!\!\!\!+ 3 {z_1}^2 z_2 z_3 - 
  2 {z_1}^3 z_2 z_3 + {z_2}^2 z_3 - 8 z_1 {z_2}^2 z_3 - 7 {z_1}^2 {z_2}^2 z_3 - 
  2 {z_1}^3 {z_2}^2 z_3 - 9 {z_2}^3 z_3 - 2 z_1 {z_2}^3 z_3 + 5 {z_1}^2 {z_2}^3 z_3 \cr &+ 
  2 {z_1}^3 {z_2}^3 z_3 - 27 {z_3}^2 - 5 z_1 {z_3}^2 + 11 {z_1}^2 {z_3}^2 + 
  5 {z_1}^3 {z_3}^2 - z_2 {z_3}^2 + 17 z_1 z_2 {z_3}^2 - 3 {z_1}^2 z_2 {z_3}^2 \cr &- 
  5 {z_1}^3 z_2 {z_3}^2 + 7 {z_2}^2 {z_3}^2 + 9 z_1 {z_2}^2 {z_3}^2 + 
  {z_1}^2 {z_2}^2 {z_3}^2 - {z_1}^3 {z_2}^2 {z_3}^2 - 3 {z_2}^3 {z_3}^2 \cr &- 
  5 z_1 {z_2}^3 {z_3}^2 - {z_1}^2 {z_2}^3 {z_3}^2 + {z_1}^3 {z_2}^3 {z_3}^2 - 7 {z_3}^3 - 
  2 z_1 {z_3}^3 + {z_1}^2 {z_3}^3 - 9 z_2 {z_3}^3 - 4 z_1 z_2 {z_3}^3 \cr &+ 
  {z_1}^2 z_2 {z_3}^3 + 3 {z_2}^2 {z_3}^3 - 2 z_1 {z_2}^2 {z_3}^3 - {z_1}^2 {z_2}^2 {z_3}^3 + 
  5 {z_2}^3 {z_3}^3 - {z_1}^2 {z_2}^3 {z_3}^3 - 19 z_4 + 12 z_1 z_4 + 13 {z_1}^2 z_4 \cr &+ 
  2 {z_1}^3 z_4 + 20 z_2 z_4 - z_1 z_2 z_4 - 18 {z_1}^2 z_2 z_4 - 5 {z_1}^3 z_2 z_4 + 
  5 {z_2}^2 z_4 - 12 z_1 {z_2}^2 z_4 - 3 {z_1}^2 {z_2}^2 z_4 + 2 {z_1}^3 {z_2}^2 z_4 \cr &- 
  2 {z_2}^3 z_4 + z_1 {z_2}^3 z_4 + 4 {z_1}^2 {z_2}^3 z_4 + {z_1}^3 {z_2}^3 z_4 + 
  5 z_3 z_4 + z_1 z_3 z_4 - 8 {z_1}^2 z_3 z_4 - 2 {z_1}^3 z_3 z_4 - 30 z_2 z_3 z_4 \cr &+ 
  9 z_1 z_2 z_3 z_4 + 19 {z_1}^2 z_2 z_3 z_4 + 4 {z_1}^3 z_2 z_3 z_4 + 
  17 {z_2}^2 z_3 z_4 - 3 z_1 {z_2}^2 z_3 z_4 - 8 {z_1}^2 {z_2}^2 z_3 z_4 \cr &- 
  2 {z_1}^3 {z_2}^2 z_3 z_4 + 4 {z_2}^3 z_3 z_4 - 3 z_1 {z_2}^3 z_3 z_4 - 
  3 {z_1}^2 {z_2}^3 z_3 z_4 + 15 {z_3}^2 z_4 - 12 z_1 {z_3}^2 z_4 \cr &- 
  11 {z_1}^2 {z_3}^2 z_4 - 8 z_2 {z_3}^2 z_4 + 3 z_1 z_2 {z_3}^2 z_4 + 
  8 {z_1}^2 z_2 {z_3}^2 z_4 + {z_1}^3 z_2 {z_3}^2 z_4 - 9 {z_2}^2 {z_3}^2 z_4 \cr &+ 
  8 z_1 {z_2}^2 {z_3}^2 z_4 + 9 {z_1}^2 {z_2}^2 {z_3}^2 z_4 - 2 {z_2}^3 {z_3}^2 z_4 + 
  z_1 {z_2}^3 {z_3}^2 z_4 - 2 {z_1}^2 {z_2}^3 {z_3}^2 z_4 - {z_1}^3 {z_2}^3 {z_3}^2 z_4 \cr &+ 
  7 {z_3}^3 z_4 - z_1 {z_3}^3 z_4 - 2 {z_1}^2 {z_3}^3 z_4 + 2 z_2 {z_3}^3 z_4 - 
  3 z_1 z_2 {z_3}^3 z_4 - {z_1}^2 z_2 {z_3}^3 z_4 - 5 {z_2}^2 {z_3}^3 z_4 \cr &- 
  z_1 {z_2}^2 {z_3}^3 z_4 + 2 {z_1}^2 {z_2}^2 {z_3}^3 z_4 + z_1 {z_2}^3 {z_3}^3 z_4 + 
  {z_1}^2 {z_2}^3 {z_3}^3 z_4 - 27 {z_4}^2 - 13 z_1 {z_4}^2 + 7 {z_1}^2 {z_4}^2 \cr &+ 
  {z_1}^3 {z_4}^2 + z_2 {z_4}^2 - 8 z_1 z_2 {z_4}^2 - {z_1}^2 z_2 {z_4}^2 + 
  11 {z_2}^2 {z_4}^2 + 11 z_1 {z_2}^2 {z_4}^2 - {z_1}^2 {z_2}^2 {z_4}^2 \cr &- 
  {z_1}^3 {z_2}^2 {z_4}^2 - {z_2}^3 {z_4}^2 - 2 z_1 {z_2}^3 {z_4}^2 - 
  {z_1}^2 {z_2}^3 {z_4}^2 - 15 z_3 {z_4}^2 - 18 z_1 z_3 {z_4}^2 + {z_1}^2 z_3 {z_4}^2 \cr &+ 
  3 z_2 z_3 {z_4}^2 - 19 z_1 z_2 z_3 {z_4}^2 - 4 {z_1}^2 z_2 z_3 {z_4}^2 + 
  3 {z_2}^2 z_3 {z_4}^2 + 8 z_1 {z_2}^2 z_3 {z_4}^2 + 3 {z_1}^2 {z_2}^2 z_3 {z_4}^2 \cr &+ 
  {z_2}^3 z_3 {z_4}^2 + z_1 {z_2}^3 z_3 {z_4}^2 + 7 {z_3}^2 {z_4}^2 + 
  3 z_1 {z_3}^2 {z_4}^2 - {z_1}^2 {z_3}^2 {z_4}^2 - {z_1}^3 {z_3}^2 {z_4}^2 \cr &+ 
  7 z_2 {z_3}^2 {z_4}^2 - 8 z_1 z_2 {z_3}^2 {z_4}^2 - 3 {z_1}^2 z_2 {z_3}^2 {z_4}^2 + 
  {z_2}^2 {z_3}^2 {z_4}^2 - 9 z_1 {z_2}^2 {z_3}^2 {z_4}^2 - {z_1}^2 {z_2}^2 {z_3}^2 {z_4}^2 \cr &+ 
  {z_1}^3 {z_2}^2 {z_3}^2 {z_4}^2 + {z_2}^3 {z_3}^2 {z_4}^2 + 2 z_1 {z_2}^3 {z_3}^2 {z_4}^2 + 
  {z_1}^2 {z_2}^3 {z_3}^2 {z_4}^2 + 3 {z_3}^3 {z_4}^2 + 4 z_1 {z_3}^3 {z_4}^2 \cr &+ 
  {z_1}^2 {z_3}^3 {z_4}^2 + 5 z_2 {z_3}^3 {z_4}^2 + 3 z_1 z_2 {z_3}^3 {z_4}^2 + 
  {z_2}^2 {z_3}^3 {z_4}^2 - 2 z_1 {z_2}^2 {z_3}^3 {z_4}^2 - {z_1}^2 {z_2}^2 {z_3}^3 {z_4}^2 \cr &- 
  {z_2}^3 {z_3}^3 {z_4}^2 - z_1 {z_2}^3 {z_3}^3 {z_4}^2 + 7 {z_4}^3 + 2 z_1 {z_4}^3 - 
  {z_1}^2 {z_4}^3 + 2 z_2 {z_4}^3 + 2 z_1 z_2 {z_4}^3 - 5 {z_2}^2 {z_4}^3 \cr &+ 
  {z_1}^2 {z_2}^2 {z_4}^3 + 7 z_3 {z_4}^3 + 5 z_1 z_3 {z_4}^3 + 2 z_2 z_3 {z_4}^3 + 
  4 z_1 z_2 z_3 {z_4}^3 - 5 {z_2}^2 z_3 {z_4}^3 - z_1 {z_2}^2 z_3 {z_4}^3 \cr &- 
  3 {z_3}^2 {z_4}^3 + 2 z_1 {z_3}^2 {z_4}^3 + {z_1}^2 {z_3}^2 {z_4}^3 - 
  2 z_2 {z_3}^2 {z_4}^3 + 2 z_1 z_2 {z_3}^2 {z_4}^3 + {z_2}^2 {z_3}^2 {z_4}^3 \cr &- 
  {z_1}^2 {z_2}^2 {z_3}^2 {z_4}^3 - 3 {z_3}^3 {z_4}^3 - z_1 {z_3}^3 {z_4}^3 - 
  2 z_2 {z_3}^3 {z_4}^3 + {z_2}^2 {z_3}^3 {z_4}^3 + z_1 {z_2}^2 {z_3}^3 {z_4}^3 + 19 z_5 \cr}$$
  \vfill\eject
  $$\eqalign{&+ 
  20 z_1 z_5 - 5 {z_1}^2 z_5 - 2 {z_1}^3 z_5 + 12 z_2 z_5 + z_1 z_2 z_5 - 
  12 {z_1}^2 z_2 z_5 - {z_1}^3 z_2 z_5 - 13 {z_2}^2 z_5 - 18 z_1 {z_2}^2 z_5 \cr &+ 
  3 {z_1}^2 {z_2}^2 z_5 + 4 {z_1}^3 {z_2}^2 z_5 + 2 {z_2}^3 z_5 + 5 z_1 {z_2}^3 z_5 + 
  2 {z_1}^2 {z_2}^3 z_5 - {z_1}^3 {z_2}^3 z_5 + 20 z_3 z_5 + 10 z_1 z_3 z_5 \cr &- 
  13 {z_1}^2 z_3 z_5 - {z_1}^3 z_3 z_5 - 29 z_2 z_3 z_5 + 26 z_1 z_2 z_3 z_5 + 
  16 {z_1}^2 z_2 z_3 z_5 + 3 {z_1}^3 z_2 z_3 z_5 + 14 {z_2}^2 z_3 z_5 \cr &- 
  4 z_1 {z_2}^2 z_3 z_5 - 11 {z_1}^2 {z_2}^2 z_3 z_5 - 3 {z_1}^3 {z_2}^2 z_3 z_5 - 
  {z_2}^3 z_3 z_5 - 4 z_1 {z_2}^3 z_3 z_5 + {z_1}^3 {z_2}^3 z_3 z_5 - {z_3}^2 z_5 \cr &- 
  13 z_1 {z_3}^2 z_5 - 5 {z_1}^2 {z_3}^2 z_5 + 3 {z_1}^3 {z_3}^2 z_5 - 
  10 z_2 {z_3}^2 z_5 + 16 z_1 z_2 {z_3}^2 z_5 + 4 {z_1}^2 z_2 {z_3}^2 z_5 \cr &- 
  2 {z_1}^3 z_2 {z_3}^2 z_5 - 5 {z_2}^2 {z_3}^2 z_5 + 15 z_1 {z_2}^2 {z_3}^2 z_5 + 
  7 {z_1}^2 {z_2}^2 {z_3}^2 z_5 - {z_1}^3 {z_2}^2 {z_3}^2 z_5 - 4 {z_2}^3 {z_3}^2 z_5 \cr &- 
  2 z_1 {z_2}^3 {z_3}^2 z_5 - 2 {z_1}^2 {z_2}^3 {z_3}^2 z_5 + 2 {z_3}^3 z_5 - 
  z_1 {z_3}^3 z_5 - {z_1}^2 {z_3}^3 z_5 - 5 z_2 {z_3}^3 z_5 - 3 z_1 z_2 {z_3}^3 z_5 \cr &- 
  4 {z_2}^2 {z_3}^3 z_5 - z_1 {z_2}^2 {z_3}^3 z_5 + {z_1}^2 {z_2}^2 {z_3}^3 z_5 + 
  3 {z_2}^3 {z_3}^3 z_5 + z_1 {z_2}^3 {z_3}^3 z_5 - 3 z_4 z_5 - 29 z_1 z_4 z_5 \cr &- 
  8 {z_1}^2 z_4 z_5 + 4 {z_1}^3 z_4 z_5 + 29 z_2 z_4 z_5 + 41 z_1 z_2 z_4 z_5 - 
  4 {z_1}^3 z_2 z_4 z_5 - 8 {z_2}^2 z_4 z_5 + 3 {z_1}^2 {z_2}^2 z_4 z_5 \cr &+ 
  {z_1}^3 {z_2}^2 z_4 z_5 - 4 {z_2}^3 z_4 z_5 - 4 z_1 {z_2}^3 z_4 z_5 - 
  {z_1}^2 {z_2}^3 z_4 z_5 - {z_1}^3 {z_2}^3 z_4 z_5 + 30 z_3 z_4 z_5 \cr &+ 
  26 z_1 z_3 z_4 z_5 - 9 {z_1}^2 z_3 z_4 z_5 - 4 {z_1}^3 z_3 z_4 z_5 - 
  59 z_2 z_3 z_4 z_5 - 19 z_1 z_2 z_3 z_4 z_5 + 10 {z_1}^2 z_2 z_3 z_4 z_5 \cr &+ 
  5 {z_1}^3 z_2 z_3 z_4 z_5 + 19 {z_2}^2 z_3 z_4 z_5 + 4 z_1 {z_2}^2 z_3 z_4 z_5 - 
  6 {z_1}^2 {z_2}^2 z_3 z_4 z_5 - 2 {z_1}^3 {z_2}^2 z_3 z_4 z_5 \cr &+ 
  8 {z_2}^3 z_3 z_4 z_5 + 3 z_1 {z_2}^3 z_3 z_4 z_5 + {z_1}^2 {z_2}^3 z_3 z_4 z_5 + 
  {z_1}^3 {z_2}^3 z_3 z_4 z_5 + 3 {z_3}^2 z_4 z_5 + 16 z_1 {z_3}^2 z_4 z_5 \cr &+ 
  3 {z_1}^2 {z_3}^2 z_4 z_5 - 9 z_2 {z_3}^2 z_4 z_5 - 11 z_1 z_2 {z_3}^2 z_4 z_5 - 
  5 {z_1}^2 z_2 {z_3}^2 z_4 z_5 - {z_1}^3 z_2 {z_3}^2 z_4 z_5 \cr &- 
  4 {z_2}^2 {z_3}^2 z_4 z_5 - 3 z_1 {z_2}^2 {z_3}^2 z_4 z_5 + 
  4 {z_1}^2 {z_2}^2 {z_3}^2 z_4 z_5 + {z_1}^3 {z_2}^2 {z_3}^2 z_4 z_5 \cr &- 
  4 {z_2}^3 {z_3}^2 z_4 z_5 + 2 z_1 {z_2}^3 {z_3}^2 z_4 z_5 - 2 {z_3}^3 z_4 z_5 + 
  3 z_1 {z_3}^3 z_4 z_5 + 2 {z_1}^2 {z_3}^3 z_4 z_5 + 7 z_2 {z_3}^3 z_4 z_5 \cr &+ 
  z_1 z_2 {z_3}^3 z_4 z_5 - {z_1}^2 z_2 {z_3}^3 z_4 z_5 - 3 {z_2}^2 {z_3}^3 z_4 z_5 - 
  5 z_1 {z_2}^2 {z_3}^3 z_4 z_5 - {z_1}^2 {z_2}^2 {z_3}^3 z_4 z_5 \cr &- 
  z_1 {z_2}^3 {z_3}^3 z_4 z_5 + {z_4}^2 z_5 + 14 z_1 {z_4}^2 z_5 + 
  7 {z_1}^2 {z_4}^2 z_5 - 2 {z_1}^3 {z_4}^2 z_5 - 10 z_2 {z_4}^2 z_5 \cr &- 
  16 z_1 z_2 {z_4}^2 z_5 - {z_1}^2 z_2 {z_4}^2 z_5 + {z_1}^3 z_2 {z_4}^2 z_5 - 
  {z_2}^2 {z_4}^2 z_5 - 3 z_1 {z_2}^2 {z_4}^2 z_5 - {z_1}^2 {z_2}^2 {z_4}^2 z_5 \cr &+ 
  {z_1}^3 {z_2}^2 {z_4}^2 z_5 + 2 {z_2}^3 {z_4}^2 z_5 + 3 z_1 {z_2}^3 {z_4}^2 z_5 + 
  {z_1}^2 {z_2}^3 {z_4}^2 z_5 - 8 z_3 {z_4}^2 z_5 - 4 z_1 z_3 {z_4}^2 z_5 \cr &+ 
  7 {z_1}^2 z_3 {z_4}^2 z_5 + {z_1}^3 z_3 {z_4}^2 z_5 + 9 z_2 z_3 {z_4}^2 z_5 - 
  14 z_1 z_2 z_3 {z_4}^2 z_5 - 4 {z_1}^2 z_2 z_3 {z_4}^2 z_5 \cr &- 
  2 {z_2}^2 z_3 {z_4}^2 z_5 + 7 z_1 {z_2}^2 z_3 {z_4}^2 z_5 + 
  2 {z_1}^2 {z_2}^2 z_3 {z_4}^2 z_5 - {z_1}^3 {z_2}^2 z_3 {z_4}^2 z_5 \cr &- 
  3 {z_2}^3 z_3 {z_4}^2 z_5 - 3 z_1 {z_2}^3 z_3 {z_4}^2 z_5 - 
  {z_1}^2 {z_2}^3 z_3 {z_4}^2 z_5 - 7 {z_3}^2 {z_4}^2 z_5 - 11 z_1 {z_3}^2 {z_4}^2 z_5 \cr &- 
  {z_1}^2 {z_3}^2 {z_4}^2 z_5 + {z_1}^3 {z_3}^2 {z_4}^2 z_5 + 8 z_2 {z_3}^2 {z_4}^2 z_5 + 
  3 z_1 z_2 {z_3}^2 {z_4}^2 z_5 - {z_1}^3 z_2 {z_3}^2 {z_4}^2 z_5 \cr &+ 
  7 {z_2}^2 {z_3}^2 {z_4}^2 z_5 - 2 z_1 {z_2}^2 {z_3}^2 {z_4}^2 z_5 - 
  {z_1}^2 {z_2}^2 {z_3}^2 {z_4}^2 z_5 - 2 {z_3}^3 {z_4}^2 z_5 \cr &- 
  3 z_1 {z_3}^3 {z_4}^2 z_5 - {z_1}^2 {z_3}^3 {z_4}^2 z_5 + z_2 {z_3}^3 {z_4}^2 z_5 + 
  3 z_1 z_2 {z_3}^3 {z_4}^2 z_5 + {z_1}^2 z_2 {z_3}^3 {z_4}^2 z_5 \cr }$$
  \vfill\eject
  $$\eqalign{&+ 
  4 {z_2}^2 {z_3}^3 {z_4}^2 z_5 + 2 z_1 {z_2}^2 {z_3}^3 {z_4}^2 z_5 + 
  {z_2}^3 {z_3}^3 {z_4}^2 z_5 - 9 {z_4}^3 z_5 - z_1 {z_4}^3 z_5 + 2 {z_1}^2 {z_4}^3 z_5 \cr &+ 
  5 z_2 {z_4}^3 z_5 - {z_1}^2 z_2 {z_4}^3 z_5 + 2 {z_2}^2 {z_4}^3 z_5 - 
  z_1 {z_2}^2 {z_4}^3 z_5 - {z_1}^2 {z_2}^2 {z_4}^3 z_5 - 2 z_3 {z_4}^3 z_5 \cr &- 
  4 z_1 z_3 {z_4}^3 z_5 - {z_1}^2 z_3 {z_4}^3 z_5 + 7 z_2 z_3 {z_4}^3 z_5 + 
  3 z_1 z_2 z_3 {z_4}^3 z_5 - 3 {z_2}^2 z_3 {z_4}^3 z_5 \cr &+ 
  z_1 {z_2}^2 z_3 {z_4}^3 z_5 + {z_1}^2 {z_2}^2 z_3 {z_4}^3 z_5 + 5 {z_3}^2 {z_4}^3 z_5 - 
  {z_1}^2 {z_3}^2 {z_4}^3 z_5 - z_2 {z_3}^2 {z_4}^3 z_5 + 2 z_1 z_2 {z_3}^2 {z_4}^3 z_5 \cr &+ 
  {z_1}^2 z_2 {z_3}^2 {z_4}^3 z_5 - 2 {z_2}^2 {z_3}^2 {z_4}^3 z_5 + 
  2 {z_3}^3 {z_4}^3 z_5 + z_1 {z_3}^3 {z_4}^3 z_5 - 3 z_2 {z_3}^3 {z_4}^3 z_5 \cr &- 
  z_1 z_2 {z_3}^3 {z_4}^3 z_5 - {z_2}^2 {z_3}^3 {z_4}^3 z_5 - 27 {z_5}^2 - z_1 {z_5}^2 + 
  11 {z_1}^2 {z_5}^2 + {z_1}^3 {z_5}^2 + 13 z_2 {z_5}^2 - 8 z_1 z_2 {z_5}^2 \cr &- 
  11 {z_1}^2 z_2 {z_5}^2 - 2 {z_1}^3 z_2 {z_5}^2 + 7 {z_2}^2 {z_5}^2 + z_1 {z_2}^2 {z_5}^2 - 
  {z_1}^2 {z_2}^2 {z_5}^2 + {z_1}^3 {z_2}^2 {z_5}^2 - {z_2}^3 {z_5}^2 \cr &+ 
  {z_1}^2 {z_2}^3 {z_5}^2 - 5 z_3 {z_5}^2 - 13 z_1 z_3 {z_5}^2 - 5 {z_1}^2 z_3 {z_5}^2 - 
  {z_1}^3 z_3 {z_5}^2 - 8 z_2 z_3 {z_5}^2 - 9 z_1 z_2 z_3 {z_5}^2 \cr &+ 
  3 {z_1}^2 z_2 z_3 {z_5}^2 + 2 {z_1}^3 z_2 z_3 {z_5}^2 + 7 {z_2}^2 z_3 {z_5}^2 + 
  7 z_1 {z_2}^2 z_3 {z_5}^2 - {z_1}^2 {z_2}^2 z_3 {z_5}^2 - {z_1}^3 {z_2}^2 z_3 {z_5}^2 \cr &+ 
  2 {z_2}^3 z_3 {z_5}^2 - z_1 {z_2}^3 z_3 {z_5}^2 - {z_1}^2 {z_2}^3 z_3 {z_5}^2 + 
  11 {z_3}^2 {z_5}^2 - 5 z_1 {z_3}^2 {z_5}^2 - 6 {z_1}^2 {z_3}^2 {z_5}^2 \cr &+ 
  z_2 {z_3}^2 {z_5}^2 - z_1 z_2 {z_3}^2 {z_5}^2 - 3 {z_2}^2 {z_3}^2 {z_5}^2 + 
  z_1 {z_2}^2 {z_3}^2 {z_5}^2 + 2 {z_1}^2 {z_2}^2 {z_3}^2 {z_5}^2 - {z_2}^3 {z_3}^2 {z_5}^2 \cr &+ 
  z_1 {z_2}^3 {z_3}^2 {z_5}^2 + 5 {z_3}^3 {z_5}^2 + 3 z_1 {z_3}^3 {z_5}^2 + 
  2 z_2 {z_3}^3 {z_5}^2 + 2 z_1 z_2 {z_3}^3 {z_5}^2 - 3 {z_2}^2 {z_3}^3 {z_5}^2 \cr &- 
  z_1 {z_2}^2 {z_3}^3 {z_5}^2 - z_4 {z_5}^2 - 10 z_1 z_4 {z_5}^2 + {z_1}^2 z_4 {z_5}^2 + 
  2 {z_1}^3 z_4 {z_5}^2 + 14 z_2 z_4 {z_5}^2 + 16 z_1 z_2 z_4 {z_5}^2 \cr &- 
  3 {z_1}^2 z_2 z_4 {z_5}^2 - 3 {z_1}^3 z_2 z_4 {z_5}^2 - 7 {z_2}^2 z_4 {z_5}^2 - 
  z_1 {z_2}^2 z_4 {z_5}^2 + {z_1}^2 {z_2}^2 z_4 {z_5}^2 + {z_1}^3 {z_2}^2 z_4 {z_5}^2 \cr &- 
  2 {z_2}^3 z_4 {z_5}^2 - z_1 {z_2}^3 z_4 {z_5}^2 + {z_1}^2 {z_2}^3 z_4 {z_5}^2 + 
  17 z_3 z_4 {z_5}^2 + 16 z_1 z_3 z_4 {z_5}^2 - {z_1}^2 z_3 z_4 {z_5}^2 \cr &- 
  2 {z_1}^3 z_3 z_4 {z_5}^2 - 19 z_2 z_3 z_4 {z_5}^2 - 16 z_1 z_2 z_3 z_4 {z_5}^2 + 
  {z_1}^2 z_2 z_3 z_4 {z_5}^2 + 3 {z_1}^3 z_2 z_3 z_4 {z_5}^2 \cr &+ 
  4 z_1 {z_2}^2 z_3 z_4 {z_5}^2 - {z_1}^2 {z_2}^2 z_3 z_4 {z_5}^2 - 
  {z_1}^3 {z_2}^2 z_3 z_4 {z_5}^2 + 4 {z_2}^3 z_3 z_4 {z_5}^2 \cr &- 
  {z_1}^2 {z_2}^3 z_3 z_4 {z_5}^2 - 3 {z_3}^2 z_4 {z_5}^2 + 4 z_1 {z_3}^2 z_4 {z_5}^2 - 
  2 z_2 {z_3}^2 z_4 {z_5}^2 - 3 z_1 z_2 {z_3}^2 z_4 {z_5}^2 \cr &- 
  2 {z_1}^2 z_2 {z_3}^2 z_4 {z_5}^2 + 3 {z_2}^2 {z_3}^2 z_4 {z_5}^2 - 
  2 {z_2}^3 {z_3}^2 z_4 {z_5}^2 + z_1 {z_2}^3 {z_3}^2 z_4 {z_5}^2 \cr &- 
  5 {z_3}^3 z_4 {z_5}^2 - 2 z_1 {z_3}^3 z_4 {z_5}^2 + 3 z_2 {z_3}^3 z_4 {z_5}^2 + 
  3 z_1 z_2 {z_3}^3 z_4 {z_5}^2 + z_1 {z_2}^2 {z_3}^3 z_4 {z_5}^2 + 7 {z_4}^2 {z_5}^2 \cr &- 
  5 z_1 {z_4}^2 {z_5}^2 - 3 {z_1}^2 {z_4}^2 {z_5}^2 + {z_1}^3 {z_4}^2 {z_5}^2 + 
  5 z_2 {z_4}^2 {z_5}^2 + 19 z_1 z_2 {z_4}^2 {z_5}^2 + 5 {z_1}^2 z_2 {z_4}^2 {z_5}^2 \cr &- 
  {z_1}^3 z_2 {z_4}^2 {z_5}^2 - 3 {z_2}^2 {z_4}^2 {z_5}^2 - 5 z_1 {z_2}^2 {z_4}^2 {z_5}^2 - 
  2 {z_1}^2 {z_2}^2 {z_4}^2 {z_5}^2 - {z_2}^3 {z_4}^2 {z_5}^2 - z_1 {z_2}^3 {z_4}^2 {z_5}^2 \cr &+ 
  9 z_3 {z_4}^2 {z_5}^2 + 15 z_1 z_3 {z_4}^2 {z_5}^2 + {z_1}^2 z_3 {z_4}^2 {z_5}^2 - 
  {z_1}^3 z_3 {z_4}^2 {z_5}^2 - 4 z_2 z_3 {z_4}^2 {z_5}^2 \cr &- 
  4 z_1 z_2 z_3 {z_4}^2 {z_5}^2 - {z_1}^2 z_2 z_3 {z_4}^2 {z_5}^2 + 
  {z_1}^3 z_2 z_3 {z_4}^2 {z_5}^2 - 3 {z_2}^2 z_3 {z_4}^2 {z_5}^2 \cr &+ 
  2 z_1 {z_2}^2 z_3 {z_4}^2 {z_5}^2 + 2 {z_1}^2 {z_2}^2 z_3 {z_4}^2 {z_5}^2 + 
  2 {z_2}^3 z_3 {z_4}^2 {z_5}^2 + z_1 {z_2}^3 z_3 {z_4}^2 {z_5}^2 \cr &+ 
  {z_3}^2 {z_4}^2 {z_5}^2 + 7 z_1 {z_3}^2 {z_4}^2 {z_5}^2 + 
  2 {z_1}^2 {z_3}^2 {z_4}^2 {z_5}^2 - 7 z_2 {z_3}^2 {z_4}^2 {z_5}^2 \cr }$$
  \vfill\eject
  $$\eqalign{&- 
  2 z_1 z_2 {z_3}^2 {z_4}^2 {z_5}^2 - {z_2}^2 {z_3}^2 {z_4}^2 {z_5}^2 - 
  z_1 {z_2}^2 {z_3}^2 {z_4}^2 {z_5}^2 - {z_2}^3 {z_3}^2 {z_4}^2 {z_5}^2 \cr &- 
  {z_3}^3 {z_4}^2 {z_5}^2 - z_1 {z_3}^3 {z_4}^2 {z_5}^2 - 2 z_2 {z_3}^3 {z_4}^2 {z_5}^2 - 
  z_1 z_2 {z_3}^3 {z_4}^2 {z_5}^2 - {z_2}^2 {z_3}^3 {z_4}^2 {z_5}^2 - 3 {z_4}^3 {z_5}^2 \cr &- 
  4 z_1 {z_4}^3 {z_5}^2 - {z_1}^2 {z_4}^3 {z_5}^2 - 4 z_2 {z_4}^3 {z_5}^2 - 
  z_1 z_2 {z_4}^3 {z_5}^2 + {z_1}^2 z_2 {z_4}^3 {z_5}^2 + 3 {z_2}^2 {z_4}^3 {z_5}^2 \cr &+ 
  z_1 {z_2}^2 {z_4}^3 {z_5}^2 - 5 z_3 {z_4}^3 {z_5}^2 - 2 z_1 z_3 {z_4}^3 {z_5}^2 + 
  {z_1}^2 z_3 {z_4}^3 {z_5}^2 + 3 z_2 z_3 {z_4}^3 {z_5}^2 \cr &- 
  3 z_1 z_2 z_3 {z_4}^3 {z_5}^2 - {z_1}^2 z_2 z_3 {z_4}^3 {z_5}^2 - 
  z_1 {z_2}^2 z_3 {z_4}^3 {z_5}^2 - {z_3}^2 {z_4}^3 {z_5}^2 \cr &- 
  2 z_1 {z_3}^2 {z_4}^3 {z_5}^2 + 4 z_2 {z_3}^2 {z_4}^3 {z_5}^2 + 
  {z_2}^2 {z_3}^2 {z_4}^3 {z_5}^2 + {z_3}^3 {z_4}^3 {z_5}^2 + z_2 {z_3}^3 {z_4}^3 {z_5}^2 \cr &- 
  7 {z_5}^3 + 2 z_1 {z_5}^3 + 5 {z_1}^2 {z_5}^3 + 2 z_2 {z_5}^3 - 2 z_1 z_2 {z_5}^3 + 
  {z_2}^2 {z_5}^3 - {z_1}^2 {z_2}^2 {z_5}^3 - 2 z_3 {z_5}^3 - z_1 z_3 {z_5}^3 \cr &+ 
  3 {z_1}^2 z_3 {z_5}^3 + 4 z_2 z_3 {z_5}^3 - 4 z_1 z_2 z_3 {z_5}^3 - 
  2 {z_2}^2 z_3 {z_5}^3 + z_1 {z_2}^2 z_3 {z_5}^3 + {z_1}^2 {z_2}^2 z_3 {z_5}^3 \cr &+ 
  {z_3}^2 {z_5}^3 - z_1 {z_3}^2 {z_5}^3 + 2 z_2 {z_3}^2 {z_5}^3 - 
  2 z_1 z_2 {z_3}^2 {z_5}^3 + {z_2}^2 {z_3}^2 {z_5}^3 - z_1 {z_2}^2 {z_3}^2 {z_5}^3 \cr &- 
  9 z_4 {z_5}^3 - 5 z_1 z_4 {z_5}^3 + 2 {z_1}^2 z_4 {z_5}^3 + z_2 z_4 {z_5}^3 + 
  {z_1}^2 z_2 z_4 {z_5}^3 + 2 {z_2}^2 z_4 {z_5}^3 + z_1 {z_2}^2 z_4 {z_5}^3 \cr &- 
  {z_1}^2 {z_2}^2 z_4 {z_5}^3 - 4 z_3 z_4 {z_5}^3 - 3 z_1 z_3 z_4 {z_5}^3 + 
  2 {z_1}^2 z_3 z_4 {z_5}^3 + 8 z_2 z_3 z_4 {z_5}^3 - 3 z_1 z_2 z_3 z_4 {z_5}^3 \cr &- 
  {z_1}^2 z_2 z_3 z_4 {z_5}^3 - 4 {z_2}^2 z_3 z_4 {z_5}^3 + 
  {z_1}^2 {z_2}^2 z_3 z_4 {z_5}^3 + {z_3}^2 z_4 {z_5}^3 + 3 z_2 {z_3}^2 z_4 {z_5}^3 \cr &- 
  z_1 z_2 {z_3}^2 z_4 {z_5}^3 + 2 {z_2}^2 {z_3}^2 z_4 {z_5}^3 - 
  z_1 {z_2}^2 {z_3}^2 z_4 {z_5}^3 + 3 {z_4}^2 {z_5}^3 - 4 z_1 {z_4}^2 {z_5}^3 \cr &- 
  3 {z_1}^2 {z_4}^2 {z_5}^3 - 4 z_2 {z_4}^2 {z_5}^3 + z_1 z_2 {z_4}^2 {z_5}^3 + 
  {z_1}^2 z_2 {z_4}^2 {z_5}^3 + {z_2}^2 {z_4}^2 {z_5}^3 + z_1 {z_2}^2 {z_4}^2 {z_5}^3 \cr &- 
  2 z_3 {z_4}^2 {z_5}^3 - z_1 z_3 {z_4}^2 {z_5}^3 - {z_1}^2 z_3 {z_4}^2 {z_5}^3 + 
  4 z_2 z_3 {z_4}^2 {z_5}^3 + 2 z_1 z_2 z_3 {z_4}^2 {z_5}^3 \cr &- 
  {z_1}^2 z_2 z_3 {z_4}^2 {z_5}^3 - 2 {z_2}^2 z_3 {z_4}^2 {z_5}^3 - 
  z_1 {z_2}^2 z_3 {z_4}^2 {z_5}^3 - {z_3}^2 {z_4}^2 {z_5}^3 + z_1 {z_3}^2 {z_4}^2 {z_5}^3 \cr &+ 
  z_1 z_2 {z_3}^2 {z_4}^2 {z_5}^3 + {z_2}^2 {z_3}^2 {z_4}^2 {z_5}^3 + 5 {z_4}^3 {z_5}^3 + 
  3 z_1 {z_4}^3 {z_5}^3 - 3 z_2 {z_4}^3 {z_5}^3 - z_1 z_2 {z_4}^3 {z_5}^3 \cr &+ 
  z_1 z_3 {z_4}^3 {z_5}^3 + z_1 z_2 z_3 {z_4}^3 {z_5}^3 - {z_3}^2 {z_4}^3 {z_5}^3 - 
  z_2 {z_3}^2 {z_4}^3 {z_5}^3 - 19 z_6 - 3 z_1 z_6 - {z_1}^2 z_6 - 9 {z_1}^3 z_6 \cr &+ 
  20 z_2 z_6 - 30 z_1 z_2 z_6 - 8 {z_1}^2 z_2 z_6 + 2 {z_1}^3 z_2 z_6 + {z_2}^2 z_6 + 
  3 z_1 {z_2}^2 z_6 + 7 {z_1}^2 {z_2}^2 z_6 + 5 {z_1}^3 {z_2}^2 z_6 + 2 {z_2}^3 z_6 \cr &+ 
  2 z_1 {z_2}^3 z_6 - 2 {z_1}^2 {z_2}^3 z_6 - 2 {z_1}^3 {z_2}^3 z_6 + 12 z_3 z_6 - 
  29 z_1 z_3 z_6 - 10 {z_1}^2 z_3 z_6 - 5 {z_1}^3 z_3 z_6 + 29 z_2 z_3 z_6 \cr &- 
  59 z_1 z_2 z_3 z_6 - 9 {z_1}^2 z_2 z_3 z_6 + 7 {z_1}^3 z_2 z_3 z_6 - 
  10 {z_2}^2 z_3 z_6 + 9 z_1 {z_2}^2 z_3 z_6 + 8 {z_1}^2 {z_2}^2 z_3 z_6 \cr &+ 
  {z_1}^3 {z_2}^2 z_3 z_6 + 5 {z_2}^3 z_3 z_6 + 7 z_1 {z_2}^3 z_3 z_6 - 
  {z_1}^2 {z_2}^3 z_3 z_6 - 3 {z_1}^3 {z_2}^3 z_3 z_6 + 13 {z_3}^2 z_6 \cr &- 
  8 z_1 {z_3}^2 z_6 + {z_1}^2 {z_3}^2 z_6 + 2 {z_1}^3 {z_3}^2 z_6 + 14 z_2 {z_3}^2 z_6 - 
  19 z_1 z_2 {z_3}^2 z_6 - 2 {z_1}^2 z_2 {z_3}^2 z_6 + 3 {z_1}^3 z_2 {z_3}^2 z_6  \cr &+ 
  5 {z_2}^2 {z_3}^2 z_6 - 4 z_1 {z_2}^2 {z_3}^2 z_6 - 7 {z_1}^2 {z_2}^2 {z_3}^2 z_6 - 
  2 {z_1}^3 {z_2}^2 {z_3}^2 z_6 - 4 {z_2}^3 {z_3}^2 z_6 + 3 z_1 {z_2}^3 {z_3}^2 z_6  \cr &+ 
  4 {z_1}^2 {z_2}^3 {z_3}^2 z_6 + {z_1}^3 {z_2}^3 {z_3}^2 z_6 + 2 {z_3}^3 z_6 +
  4 z_1 {z_3}^3 z_6 + 2 {z_1}^2 {z_3}^3 z_6 + z_2 {z_3}^3 z_6 + 8 z_1 z_2 {z_3}^3 z_6  \cr &+ 
  3 {z_1}^2 z_2 {z_3}^3 z_6 - 4 {z_2}^2 {z_3}^3 z_6 + 4 z_1 {z_2}^2 {z_3}^3 z_6 - 
  3 {z_2}^3 {z_3}^3 z_6 - {z_1}^2 {z_2}^3 {z_3}^3 z_6 + 20 z_4 z_6 + 29 z_1 z_4 z_6  \cr }$$
  \vfill\eject
  $$\eqalign{&+ 
  14 {z_1}^2 z_4 z_6 + {z_1}^3 z_4 z_6 - 10 z_2 z_4 z_6 + 26 z_1 z_2 z_4 z_6 + 
  4 {z_1}^2 z_2 z_4 z_6 - 4 {z_1}^3 z_2 z_4 z_6 - 13 {z_2}^2 z_4 z_6  \cr &- 
  16 z_1 {z_2}^2 z_4 z_6 - 11 {z_1}^2 {z_2}^2 z_4 z_6 + {z_2}^3 z_4 z_6 + 
  3 z_1 {z_2}^3 z_4 z_6 + 3 {z_1}^2 {z_2}^3 z_4 z_6 + {z_1}^3 {z_2}^3 z_4 z_6  \cr &- 
  z_3 z_4 z_6 + 41 z_1 z_3 z_4 z_6 + 16 {z_1}^2 z_3 z_4 z_6 + 26 z_2 z_3 z_4 z_6 + 
  19 z_1 z_2 z_3 z_4 z_6 - 14 {z_1}^2 z_2 z_3 z_4 z_6 - 3 {z_1}^3 z_2 z_3 z_4 z_6  \cr &- 
  16 {z_2}^2 z_3 z_4 z_6 - 11 z_1 {z_2}^2 z_3 z_4 z_6 - 
  3 {z_1}^2 {z_2}^2 z_3 z_4 z_6 + 2 {z_1}^3 {z_2}^2 z_3 z_4 z_6  \cr &- 
  3 {z_2}^3 z_3 z_4 z_6 - z_1 {z_2}^3 z_3 z_4 z_6 + 3 {z_1}^2 {z_2}^3 z_3 z_4 z_6 + 
  {z_1}^3 {z_2}^3 z_3 z_4 z_6 - 18 {z_3}^2 z_4 z_6 - 3 {z_1}^2 {z_3}^2 z_4 z_6  \cr &+ 
  {z_1}^3 {z_3}^2 z_4 z_6 + 4 z_2 {z_3}^2 z_4 z_6 + 4 z_1 z_2 {z_3}^2 z_4 z_6 - 
  7 {z_1}^2 z_2 {z_3}^2 z_4 z_6 + {z_1}^3 z_2 {z_3}^2 z_4 z_6  \cr &+ 
  15 {z_2}^2 {z_3}^2 z_4 z_6 + 3 z_1 {z_2}^2 {z_3}^2 z_4 z_6 - 
  2 {z_1}^2 {z_2}^2 {z_3}^2 z_4 z_6 + {z_2}^3 {z_3}^2 z_4 z_6  \cr &- 
  5 z_1 {z_2}^3 {z_3}^2 z_4 z_6 - 2 {z_1}^2 {z_2}^3 {z_3}^2 z_4 z_6 - 
  5 {z_3}^3 z_4 z_6 - 4 z_1 {z_3}^3 z_4 z_6 - 3 {z_1}^2 {z_3}^3 z_4 z_6  \cr &- 
  4 z_2 {z_3}^3 z_4 z_6 - 3 z_1 z_2 {z_3}^3 z_4 z_6 - 3 {z_1}^2 z_2 {z_3}^3 z_4 z_6 + 
  2 {z_2}^2 {z_3}^3 z_4 z_6 + 2 z_1 {z_2}^2 {z_3}^3 z_4 z_6 + {z_2}^3 {z_3}^3 z_4 z_6  \cr &+ 
  z_1 {z_2}^3 {z_3}^3 z_4 z_6 + 5 {z_4}^2 z_6 - 8 z_1 {z_4}^2 z_6 - 
  7 {z_1}^2 {z_4}^2 z_6 + 2 {z_1}^3 {z_4}^2 z_6 - 13 z_2 {z_4}^2 z_6  \cr &+ 
  9 z_1 z_2 {z_4}^2 z_6 + 7 {z_1}^2 z_2 {z_4}^2 z_6 + {z_1}^3 z_2 {z_4}^2 z_6 + 
  5 {z_2}^2 {z_4}^2 z_6 + 3 z_1 {z_2}^2 {z_4}^2 z_6 + {z_1}^2 {z_2}^2 {z_4}^2 z_6  \cr &- 
  {z_1}^3 {z_2}^2 {z_4}^2 z_6 - {z_2}^3 {z_4}^2 z_6 - 2 z_1 {z_2}^3 {z_4}^2 z_6 - 
  {z_1}^2 {z_2}^3 {z_4}^2 z_6 - 12 z_3 {z_4}^2 z_6 - {z_1}^2 z_3 {z_4}^2 z_6  \cr &+ 
  {z_1}^3 z_3 {z_4}^2 z_6 - 16 z_2 z_3 {z_4}^2 z_6 + 10 z_1 z_2 z_3 {z_4}^2 z_6 + 
  4 {z_1}^2 z_2 z_3 {z_4}^2 z_6 + 4 {z_2}^2 z_3 {z_4}^2 z_6  \cr &+ 
  5 z_1 {z_2}^2 z_3 {z_4}^2 z_6 - {z_1}^3 {z_2}^2 z_3 {z_4}^2 z_6 - 
  z_1 {z_2}^3 z_3 {z_4}^2 z_6 - {z_1}^2 {z_2}^3 z_3 {z_4}^2 z_6 - 3 {z_3}^2 {z_4}^2 z_6  \cr &+ 
  3 z_1 {z_3}^2 {z_4}^2 z_6 + {z_1}^2 {z_3}^2 {z_4}^2 z_6 - {z_1}^3 {z_3}^2 {z_4}^2 z_6 - 
  11 z_2 {z_3}^2 {z_4}^2 z_6 + 6 z_1 z_2 {z_3}^2 {z_4}^2 z_6  \cr &+ 
  2 {z_1}^2 z_2 {z_3}^2 {z_4}^2 z_6 - {z_1}^3 z_2 {z_3}^2 {z_4}^2 z_6 - 
  7 {z_2}^2 {z_3}^2 {z_4}^2 z_6 + 4 z_1 {z_2}^2 {z_3}^2 {z_4}^2 z_6  \cr &+ 
  {z_1}^2 {z_2}^2 {z_3}^2 {z_4}^2 z_6 + {z_2}^3 {z_3}^2 {z_4}^2 z_6 + 
  z_1 {z_2}^3 {z_3}^2 {z_4}^2 z_6 + 2 {z_3}^3 {z_4}^2 z_6 + z_1 {z_3}^3 {z_4}^2 z_6  \cr &+ 
  {z_1}^2 {z_3}^3 {z_4}^2 z_6 + z_1 z_2 {z_3}^3 {z_4}^2 z_6 + 
  {z_1}^2 z_2 {z_3}^3 {z_4}^2 z_6 - 2 {z_2}^2 {z_3}^3 {z_4}^2 z_6 - 2 {z_4}^3 z_6  \cr &- 
  4 z_1 {z_4}^3 z_6 - 2 {z_1}^2 {z_4}^3 z_6 + z_2 {z_4}^3 z_6 - 4 z_1 z_2 {z_4}^3 z_6 - 
  {z_1}^2 z_2 {z_4}^3 z_6 + 3 {z_2}^2 {z_4}^3 z_6 + {z_1}^2 {z_2}^2 {z_4}^3 z_6  \cr &+ 
  z_3 {z_4}^3 z_6 - 4 z_1 z_3 {z_4}^3 z_6 - {z_1}^2 z_3 {z_4}^3 z_6 + 
  3 z_2 z_3 {z_4}^3 z_6 - 5 z_1 z_2 z_3 {z_4}^3 z_6 + 2 {z_2}^2 z_3 {z_4}^3 z_6  \cr &- 
  z_1 {z_2}^2 z_3 {z_4}^3 z_6 + {z_1}^2 {z_2}^2 z_3 {z_4}^3 z_6 + 4 {z_3}^2 {z_4}^3 z_6 - 
  z_1 {z_3}^2 {z_4}^3 z_6 + {z_1}^2 {z_3}^2 {z_4}^3 z_6 + 3 z_2 {z_3}^2 {z_4}^3 z_6  \cr &- 
  2 z_1 z_2 {z_3}^2 {z_4}^3 z_6 + {z_1}^2 z_2 {z_3}^2 {z_4}^3 z_6 - 
  {z_2}^2 {z_3}^2 {z_4}^3 z_6 - z_1 {z_2}^2 {z_3}^2 {z_4}^3 z_6 + {z_3}^3 {z_4}^3 z_6  \cr &- 
  z_1 {z_3}^3 {z_4}^3 z_6 + z_2 {z_3}^3 {z_4}^3 z_6 - z_1 z_2 {z_3}^3 {z_4}^3 z_6 + 
  5 z_5 z_6 + 30 z_1 z_5 z_6 + 17 {z_1}^2 z_5 z_6 - 4 {z_1}^3 z_5 z_6 - z_2 z_5 z_6 \cr &+ 
  9 z_1 z_2 z_5 z_6 + 3 {z_1}^2 z_2 z_5 z_6 - 3 {z_1}^3 z_2 z_5 z_6 - 
  8 {z_2}^2 z_5 z_6 - 19 z_1 {z_2}^2 z_5 z_6 - 8 {z_1}^2 {z_2}^2 z_5 z_6 \cr &+ 
  3 {z_1}^3 {z_2}^2 z_5 z_6 + 2 {z_2}^3 z_5 z_6 + 4 z_1 {z_2}^3 z_5 z_6  + 
  2 {z_1}^2 {z_2}^3 z_5 z_6 + z_3 z_5 z_6 + 26 z_1 z_3 z_5 z_6 \cr &+ 
  16 {z_1}^2 z_3 z_5 z_6 - 3 {z_1}^3 z_3 z_5 z_6 + 41 z_2 z_3 z_5 z_6- 
  19 z_1 z_2 z_3 z_5 z_6 - 11 {z_1}^2 z_2 z_3 z_5 z_6 + {z_1}^3 z_2 z_3 z_5 z_6 \cr}$$
  \vfill\eject
  $$\eqalign{&- 
  16 {z_2}^2 z_3 z_5 z_6 - 14 z_1 {z_2}^2 z_3 z_5 z_6+ 3 {z_1}^2 {z_2}^2 z_3 z_5 z_6 
  + 3 {z_1}^3 {z_2}^2 z_3 z_5 z_6 +3 z_1{z_2}^3 z_3 z_5 z_6 +2 {z_1}^2 {z_2}^3 z_3 z_5 z_6  \cr &- 
  {z_1}^3 {z_2}^3 z_3 z_5 z_6 - 8 {z_3}^2 z_5 z_6 - 9 z_1 {z_3}^2 z_5 z_6 - 
  {z_1}^2 {z_3}^2 z_5 z_6 + 2 {z_1}^3 {z_3}^2 z_5 z_6 + 16 z_2 {z_3}^2 z_5 z_6  \cr &- 
  16 z_1 z_2 {z_3}^2 z_5 z_6 - 3 {z_1}^2 z_2 {z_3}^2 z_5 z_6 + 
  3 {z_1}^3 z_2 {z_3}^2 z_5 z_6 + 19 {z_2}^2 {z_3}^2 z_5 z_6  \cr &- 
  4 z_1 {z_2}^2 {z_3}^2 z_5 z_6 - 2 {z_1}^2 {z_2}^2 {z_3}^2 z_5 z_6 - 
  {z_1}^3 {z_2}^2 {z_3}^2 z_5 z_6 - {z_2}^3 {z_3}^2 z_5 z_6  \cr &- 
  3 z_1 {z_2}^3 {z_3}^2 z_5 z_6 - 2 {z_3}^3 z_5 z_6 - 4 z_1 {z_3}^3 z_5 z_6 - 
  2 {z_1}^2 {z_3}^3 z_5 z_6 - 3 z_1 z_2 {z_3}^3 z_5 z_6  \cr &- 
  {z_1}^2 z_2 {z_3}^3 z_5 z_6 + {z_2}^2 {z_3}^3 z_5 z_6 + 
  2 z_1 {z_2}^2 {z_3}^3 z_5 z_6 + {z_1}^2 {z_2}^2 {z_3}^3 z_5 z_6  \cr &- 
  {z_2}^3 {z_3}^3 z_5 z_6 + z_1 {z_2}^3 {z_3}^3 z_5 z_6 - 30 z_4 z_5 z_6 - 
  59 z_1 z_4 z_5 z_6 - 19 {z_1}^2 z_4 z_5 z_6 + 8 {z_1}^3 z_4 z_5 z_6  \cr &+ 
  26 z_2 z_4 z_5 z_6 + 19 z_1 z_2 z_4 z_5 z_6 + 4 {z_1}^2 z_2 z_4 z_5 z_6 - 
  3 {z_1}^3 z_2 z_4 z_5 z_6 + 9 {z_2}^2 z_4 z_5 z_6 + 10 z_1 {z_2}^2 z_4 z_5 z_6  \cr &+ 
  6 {z_1}^2 {z_2}^2 z_4 z_5 z_6 + {z_1}^3 {z_2}^2 z_4 z_5 z_6 - 
  4 {z_2}^3 z_4 z_5 z_6 - 5 z_1 {z_2}^3 z_4 z_5 z_6 - 2 {z_1}^2 {z_2}^3 z_4 z_5 z_6  \cr &- 
  {z_1}^3 {z_2}^3 z_4 z_5 z_6 + 9 z_3 z_4 z_5 z_6 - 19 z_1 z_3 z_4 z_5 z_6 - 
  16 {z_1}^2 z_3 z_4 z_5 z_6 - 3 {z_1}^3 z_3 z_4 z_5 z_6 + 19 z_2 z_3 z_4 z_5 z_6  \cr &+ 
  54 z_1 z_2 z_3 z_4 z_5 z_6 + 13 {z_1}^2 z_2 z_3 z_4 z_5 z_6 - 
  16 {z_2}^2 z_3 z_4 z_5 z_6 - 13 z_1 {z_2}^2 z_3 z_4 z_5 z_6  \cr &- 
  7 {z_1}^2 {z_2}^2 z_3 z_4 z_5 z_6 - {z_1}^3 {z_2}^2 z_3 z_4 z_5 z_6 + 
  3 {z_2}^3 z_3 z_4 z_5 z_6 + {z_1}^2 {z_2}^3 z_3 z_4 z_5 z_6  \cr &+ 
  19 {z_3}^2 z_4 z_5 z_6 + 10 z_1 {z_3}^2 z_4 z_5 z_6 + {z_1}^2 {z_3}^2 z_4 z_5 z_6 - 
  {z_1}^3 {z_3}^2 z_4 z_5 z_6 - 14 z_2 {z_3}^2 z_4 z_5 z_6  \cr &+ 
  13 z_1 z_2 {z_3}^2 z_4 z_5 z_6 + 2 {z_1}^2 z_2 {z_3}^2 z_4 z_5 z_6 + 
  4 {z_2}^2 {z_3}^2 z_4 z_5 z_6 + 5 z_1 {z_2}^2 {z_3}^2 z_4 z_5 z_6  \cr &+ 
  2 {z_1}^2 {z_2}^2 {z_3}^2 z_4 z_5 z_6 + 2 {z_2}^3 {z_3}^2 z_4 z_5 z_6 + 
  z_1 {z_2}^3 {z_3}^2 z_4 z_5 z_6 + 4 {z_3}^3 z_4 z_5 z_6  \cr &+ 
  5 z_1 {z_3}^3 z_4 z_5 z_6 + 3 {z_1}^2 {z_3}^3 z_4 z_5 z_6 - 
  3 z_2 {z_3}^3 z_4 z_5 z_6 - 3 {z_2}^2 {z_3}^3 z_4 z_5 z_6  \cr &- 
  z_1 {z_2}^2 {z_3}^3 z_4 z_5 z_6 - {z_2}^3 {z_3}^3 z_4 z_5 z_6 + 17 {z_4}^2 z_5 z_6 + 
  19 z_1 {z_4}^2 z_5 z_6 - 4 {z_1}^3 {z_4}^2 z_5 z_6 - 16 z_2 {z_4}^2 z_5 z_6  \cr &- 
  16 z_1 z_2 {z_4}^2 z_5 z_6 - 4 {z_1}^2 z_2 {z_4}^2 z_5 z_6 - 
  {z_2}^2 {z_4}^2 z_5 z_6 - z_1 {z_2}^2 {z_4}^2 z_5 z_6  \cr &- 
  {z_1}^2 {z_2}^2 {z_4}^2 z_5 z_6 + {z_1}^3 {z_2}^2 {z_4}^2 z_5 z_6 + 
  2 {z_2}^3 {z_4}^2 z_5 z_6 + 3 z_1 {z_2}^3 {z_4}^2 z_5 z_6  \cr &+ 
  {z_1}^2 {z_2}^3 {z_4}^2 z_5 z_6 - 3 z_3 {z_4}^2 z_5 z_6 + 
  4 z_1 z_3 {z_4}^2 z_5 z_6 + 4 {z_1}^2 z_3 {z_4}^2 z_5 z_6  \cr &- 
  11 z_2 z_3 {z_4}^2 z_5 z_6 - 13 z_1 z_2 z_3 {z_4}^2 z_5 z_6 - 
  2 {z_1}^2 z_2 z_3 {z_4}^2 z_5 z_6 + 2 {z_1}^3 z_2 z_3 {z_4}^2 z_5 z_6  \cr &+ 
  3 {z_2}^2 z_3 {z_4}^2 z_5 z_6 + 2 z_1 {z_2}^2 z_3 {z_4}^2 z_5 z_6 + 
  {z_1}^2 {z_2}^2 z_3 {z_4}^2 z_5 z_6 - {z_2}^3 z_3 {z_4}^2 z_5 z_6  \cr &- 
  8 {z_3}^2 {z_4}^2 z_5 z_6 - 6 z_1 {z_3}^2 {z_4}^2 z_5 z_6 - 
  {z_1}^2 {z_3}^2 {z_4}^2 z_5 z_6 + {z_1}^3 {z_3}^2 {z_4}^2 z_5 z_6  \cr &- 
  3 z_2 {z_3}^2 {z_4}^2 z_5 z_6 - 7 z_1 z_2 {z_3}^2 {z_4}^2 z_5 z_6 - 
  {z_1}^2 z_2 {z_3}^2 {z_4}^2 z_5 z_6 - 2 {z_2}^2 {z_3}^2 {z_4}^2 z_5 z_6  \cr &- 
  2 z_1 {z_2}^2 {z_3}^2 {z_4}^2 z_5 z_6 - {z_2}^3 {z_3}^2 {z_4}^2 z_5 z_6 - 
  2 {z_3}^3 {z_4}^2 z_5 z_6 - 2 z_1 {z_3}^3 {z_4}^2 z_5 z_6  \cr &- 
  {z_1}^2 {z_3}^3 {z_4}^2 z_5 z_6 + 2 z_2 {z_3}^3 {z_4}^2 z_5 z_6 - 
  z_1 z_2 {z_3}^3 {z_4}^2 z_5 z_6 + 4 {z_4}^3 z_5 z_6 + 8 z_1 {z_4}^3 z_5 z_6  \cr }$$
  \vfill\eject
$$\eqalign{&+ 
  4 {z_1}^2 {z_4}^3 z_5 z_6 - 3 z_2 {z_4}^3 z_5 z_6 + 3 z_1 z_2 {z_4}^3 z_5 z_6 - 
  2 {z_2}^2 {z_4}^3 z_5 z_6 - z_1 {z_2}^2 {z_4}^3 z_5 z_6  \cr &- 
  {z_1}^2 {z_2}^2 {z_4}^3 z_5 z_6 - 3 z_3 {z_4}^3 z_5 z_6 + 
  3 z_1 z_3 {z_4}^3 z_5 z_6 - z_2 z_3 {z_4}^3 z_5 z_6  \cr &- 
  2 {z_1}^2 z_2 z_3 {z_4}^3 z_5 z_6 + 3 {z_2}^2 z_3 {z_4}^3 z_5 z_6 - 
  3 {z_3}^2 {z_4}^3 z_5 z_6 + z_1 {z_3}^2 {z_4}^3 z_5 z_6  \cr &- 
  {z_1}^2 {z_3}^2 {z_4}^3 z_5 z_6 + 3 z_2 {z_3}^2 {z_4}^3 z_5 z_6 + 
  z_1 z_2 {z_3}^2 {z_4}^3 z_5 z_6 + {z_2}^2 {z_3}^2 {z_4}^3 z_5 z_6  \cr &+ 
  z_1 {z_3}^3 {z_4}^3 z_5 z_6 + z_2 {z_3}^3 {z_4}^3 z_5 z_6 + 15 {z_5}^2 z_6 + 
  3 z_1 {z_5}^2 z_6 - 3 {z_1}^2 {z_5}^2 z_6 + {z_1}^3 {z_5}^2 z_6 - 18 z_2 {z_5}^2 z_6  \cr &+ 
  19 z_1 z_2 {z_5}^2 z_6 + 8 {z_1}^2 z_2 {z_5}^2 z_6 - {z_1}^3 z_2 {z_5}^2 z_6 - 
  {z_2}^2 {z_5}^2 z_6 - 4 z_1 {z_2}^2 {z_5}^2 z_6 - 3 {z_1}^2 {z_2}^2 {z_5}^2 z_6  \cr &- 
  12 z_3 {z_5}^2 z_6 + 16 z_1 z_3 {z_5}^2 z_6 + 4 {z_1}^2 z_3 {z_5}^2 z_6 + 
  10 z_1 z_2 z_3 {z_5}^2 z_6 - 5 {z_1}^2 z_2 z_3 {z_5}^2 z_6  \cr &- 
  {z_1}^3 z_2 z_3 {z_5}^2 z_6 - {z_2}^2 z_3 {z_5}^2 z_6 - 
  4 z_1 {z_2}^2 z_3 {z_5}^2 z_6 + {z_1}^3 {z_2}^2 z_3 {z_5}^2 z_6  \cr &- 
  {z_2}^3 z_3 {z_5}^2 z_6 + {z_1}^2 {z_2}^3 z_3 {z_5}^2 z_6 - 11 {z_3}^2 {z_5}^2 z_6 + 
  3 z_1 {z_3}^2 {z_5}^2 z_6 - 3 z_2 {z_3}^2 {z_5}^2 z_6 + z_1 z_2 {z_3}^2 {z_5}^2 z_6  \cr &- 
  2 {z_1}^2 z_2 {z_3}^2 {z_5}^2 z_6 + 5 {z_2}^2 {z_3}^2 {z_5}^2 z_6 - 
  z_1 {z_2}^2 {z_3}^2 {z_5}^2 z_6 + {z_2}^3 {z_3}^2 {z_5}^2 z_6  \cr &- 
  z_1 {z_2}^3 {z_3}^2 {z_5}^2 z_6 + z_2 {z_3}^3 {z_5}^2 z_6 - 
  z_1 z_2 {z_3}^3 {z_5}^2 z_6 + {z_2}^2 {z_3}^3 {z_5}^2 z_6  \cr &- 
  z_1 {z_2}^2 {z_3}^3 {z_5}^2 z_6 - 8 z_4 {z_5}^2 z_6 - 9 z_1 z_4 {z_5}^2 z_6 - 
  2 {z_1}^2 z_4 {z_5}^2 z_6 + 3 {z_1}^3 z_4 {z_5}^2 z_6 + 4 z_2 z_4 {z_5}^2 z_6  \cr &- 
  14 z_1 z_2 z_4 {z_5}^2 z_6 - 7 {z_1}^2 z_2 z_4 {z_5}^2 z_6 - 
  3 {z_1}^3 z_2 z_4 {z_5}^2 z_6 + 7 {z_2}^2 z_4 {z_5}^2 z_6  \cr &+ 
  4 z_1 {z_2}^2 z_4 {z_5}^2 z_6 + 2 {z_1}^2 {z_2}^2 z_4 {z_5}^2 z_6 + 
  {z_1}^3 {z_2}^2 z_4 {z_5}^2 z_6 - {z_2}^3 z_4 {z_5}^2 z_6  \cr &+ 
  {z_1}^2 {z_2}^3 z_4 {z_5}^2 z_6 + 3 z_3 z_4 {z_5}^2 z_6 - 
  11 z_1 z_3 z_4 {z_5}^2 z_6 - 3 {z_1}^2 z_3 z_4 {z_5}^2 z_6  \cr &- 
  {z_1}^3 z_3 z_4 {z_5}^2 z_6 + 4 z_2 z_3 z_4 {z_5}^2 z_6 + 
  13 z_1 z_2 z_3 z_4 {z_5}^2 z_6 + 2 {z_1}^2 z_2 z_3 z_4 {z_5}^2 z_6  \cr &- 
  4 {z_2}^2 z_3 z_4 {z_5}^2 z_6 - 2 z_1 {z_2}^2 z_3 z_4 {z_5}^2 z_6 - 
  {z_1}^2 {z_2}^2 z_3 z_4 {z_5}^2 z_6 - 2 z_1 {z_2}^3 z_3 z_4 {z_5}^2 z_6  \cr &+ 
  8 {z_3}^2 z_4 {z_5}^2 z_6 - 5 z_1 {z_3}^2 z_4 {z_5}^2 z_6 - 
  2 {z_1}^2 {z_3}^2 z_4 {z_5}^2 z_6 - 7 z_2 {z_3}^2 z_4 {z_5}^2 z_6  \cr &+ 
  2 z_1 z_2 {z_3}^2 z_4 {z_5}^2 z_6 + 2 {z_2}^2 {z_3}^2 z_4 {z_5}^2 z_6 - 
  z_1 {z_2}^2 {z_3}^2 z_4 {z_5}^2 z_6 + {z_2}^3 {z_3}^2 z_4 {z_5}^2 z_6  \cr &+ 
  {z_3}^3 z_4 {z_5}^2 z_6 - z_1 {z_3}^3 z_4 {z_5}^2 z_6 + z_2 {z_3}^3 z_4 {z_5}^2 z_6 + 
  {z_2}^2 {z_3}^3 z_4 {z_5}^2 z_6 - 9 {z_4}^2 {z_5}^2 z_6 - 4 z_1 {z_4}^2 {z_5}^2 z_6  \cr &+ 
  3 {z_1}^2 {z_4}^2 {z_5}^2 z_6 + 2 {z_1}^3 {z_4}^2 {z_5}^2 z_6 + 
  15 z_2 {z_4}^2 {z_5}^2 z_6 + 4 z_1 z_2 {z_4}^2 {z_5}^2 z_6  \cr &+ 
  2 {z_1}^2 z_2 {z_4}^2 {z_5}^2 z_6 - {z_1}^3 z_2 {z_4}^2 {z_5}^2 z_6 - 
  {z_2}^2 {z_4}^2 {z_5}^2 z_6 - z_1 {z_2}^2 {z_4}^2 {z_5}^2 z_6  \cr &- 
  2 {z_1}^2 {z_2}^2 {z_4}^2 {z_5}^2 z_6 - {z_2}^3 {z_4}^2 {z_5}^2 z_6 - 
  z_1 {z_2}^3 {z_4}^2 {z_5}^2 z_6 + 8 z_3 {z_4}^2 {z_5}^2 z_6  \cr &- 
  3 z_1 z_3 {z_4}^2 {z_5}^2 z_6 - {z_1}^3 z_3 {z_4}^2 {z_5}^2 z_6 + 
  3 z_2 z_3 {z_4}^2 {z_5}^2 z_6 + 5 z_1 z_2 z_3 {z_4}^2 {z_5}^2 z_6  \cr &- 
  {z_1}^2 z_2 z_3 {z_4}^2 {z_5}^2 z_6 + z_1 {z_2}^2 z_3 {z_4}^2 {z_5}^2 z_6 + 
  {z_2}^3 z_3 {z_4}^2 {z_5}^2 z_6 + 9 {z_3}^2 {z_4}^2 {z_5}^2 z_6  \cr &+ 
  4 z_1 {z_3}^2 {z_4}^2 {z_5}^2 z_6 - 2 z_2 {z_3}^2 {z_4}^2 {z_5}^2 z_6 + 
  2 z_1 z_2 {z_3}^2 {z_4}^2 {z_5}^2 z_6 + {z_2}^2 {z_3}^2 {z_4}^2 {z_5}^2 z_6  \cr }$$
\vfill\eject
$$\eqalign{&+ 
  z_1 {z_3}^3 {z_4}^2 {z_5}^2 z_6 - 2 {z_4}^3 {z_5}^2 z_6 - 4 z_1 {z_4}^3 {z_5}^2 z_6 - 
  2 {z_1}^2 {z_4}^3 {z_5}^2 z_6 + z_2 {z_4}^3 {z_5}^2 z_6  \cr &+ 
  2 z_1 z_2 {z_4}^3 {z_5}^2 z_6 + {z_1}^2 z_2 {z_4}^3 {z_5}^2 z_6 - 
  {z_2}^2 {z_4}^3 {z_5}^2 z_6 + z_1 {z_2}^2 {z_4}^3 {z_5}^2 z_6  \cr &+ 
  z_3 {z_4}^3 {z_5}^2 z_6 + 2 z_1 z_3 {z_4}^3 {z_5}^2 z_6 + 
  {z_1}^2 z_3 {z_4}^3 {z_5}^2 z_6 - 5 z_2 z_3 {z_4}^3 {z_5}^2 z_6  \cr &+ 
  z_1 z_2 z_3 {z_4}^3 {z_5}^2 z_6 - {z_2}^2 z_3 {z_4}^3 {z_5}^2 z_6 - 
  2 {z_3}^2 {z_4}^3 {z_5}^2 z_6 - 2 z_2 {z_3}^2 {z_4}^3 {z_5}^2 z_6  \cr &- 
  {z_3}^3 {z_4}^3 {z_5}^2 z_6 + 7 {z_5}^3 z_6 - 2 z_1 {z_5}^3 z_6 - 
  5 {z_1}^2 {z_5}^3 z_6 - 5 z_2 {z_5}^3 z_6 + 4 z_1 z_2 {z_5}^3 z_6  \cr &+ 
  {z_1}^2 z_2 {z_5}^3 z_6 - z_3 {z_5}^3 z_6 + 3 z_1 z_3 {z_5}^3 z_6 - 
  2 {z_1}^2 z_3 {z_5}^3 z_6 - 4 z_2 z_3 {z_5}^3 z_6 + 5 z_1 z_2 z_3 {z_5}^3 z_6  \cr &- 
  {z_1}^2 z_2 z_3 {z_5}^3 z_6 + {z_2}^2 z_3 {z_5}^3 z_6 - 
  {z_1}^2 {z_2}^2 z_3 {z_5}^3 z_6 - 2 {z_3}^2 {z_5}^3 z_6 + 2 z_1 {z_3}^2 {z_5}^3 z_6  \cr &- 
  3 z_2 {z_3}^2 {z_5}^3 z_6 + 3 z_1 z_2 {z_3}^2 {z_5}^3 z_6 - 
  {z_2}^2 {z_3}^2 {z_5}^3 z_6 + z_1 {z_2}^2 {z_3}^2 {z_5}^3 z_6 + 2 z_4 {z_5}^3 z_6  \cr &+ 
  7 z_1 z_4 {z_5}^3 z_6 + 3 {z_1}^2 z_4 {z_5}^3 z_6 - 4 z_2 z_4 {z_5}^3 z_6 - 
  3 z_1 z_2 z_4 {z_5}^3 z_6 + {z_1}^2 z_2 z_4 {z_5}^3 z_6 + {z_2}^2 z_4 {z_5}^3 z_6  \cr &- 
  {z_1}^2 {z_2}^2 z_4 {z_5}^3 z_6 - 3 z_3 z_4 {z_5}^3 z_6 + z_1 z_3 z_4 {z_5}^3 z_6 + 
  3 {z_1}^2 z_3 z_4 {z_5}^3 z_6 - 3 z_2 z_3 z_4 {z_5}^3 z_6  \cr &+ 
  2 z_1 {z_2}^2 z_3 z_4 {z_5}^3 z_6 - {z_3}^2 z_4 {z_5}^3 z_6 - 
  z_1 {z_3}^2 z_4 {z_5}^3 z_6 - 3 z_2 {z_3}^2 z_4 {z_5}^3 z_6  \cr &- 
  {z_2}^2 {z_3}^2 z_4 {z_5}^3 z_6 - 5 {z_4}^2 {z_5}^3 z_6 - 3 z_1 {z_4}^2 {z_5}^3 z_6 + 
  2 z_2 {z_4}^2 {z_5}^3 z_6 - 3 z_1 z_2 {z_4}^2 {z_5}^3 z_6  \cr &+ 
  {z_1}^2 z_2 {z_4}^2 {z_5}^3 z_6 + {z_2}^2 {z_4}^2 {z_5}^3 z_6 + 
  z_1 {z_2}^2 {z_4}^2 {z_5}^3 z_6 - z_3 {z_4}^2 {z_5}^3 z_6  \cr &- 
  5 z_1 z_3 {z_4}^2 {z_5}^3 z_6 + {z_1}^2 z_3 {z_4}^2 {z_5}^3 z_6 + 
  2 z_2 z_3 {z_4}^2 {z_5}^3 z_6 - z_1 z_2 z_3 {z_4}^2 {z_5}^3 z_6  \cr &- 
  {z_2}^2 z_3 {z_4}^2 {z_5}^3 z_6 + 2 {z_3}^2 {z_4}^2 {z_5}^3 z_6 - 
  z_1 {z_3}^2 {z_4}^2 {z_5}^3 z_6 + z_2 {z_4}^3 {z_5}^3 z_6  \cr &- 
  z_1 z_2 {z_4}^3 {z_5}^3 z_6 + z_3 {z_4}^3 {z_5}^3 z_6 - z_1 z_3 {z_4}^3 {z_5}^3 z_6 + 
  z_2 z_3 {z_4}^3 {z_5}^3 z_6 + {z_3}^2 {z_4}^3 {z_5}^3 z_6 - 27 {z_6}^2 + z_1 {z_6}^2  \cr &+ 
  7 {z_1}^2 {z_6}^2 + 3 {z_1}^3 {z_6}^2 + 5 z_2 {z_6}^2 + 17 z_1 z_2 {z_6}^2 - 
  9 {z_1}^2 z_2 {z_6}^2 - 5 {z_1}^3 z_2 {z_6}^2 + 11 {z_2}^2 {z_6}^2  \cr &+ 
  3 z_1 {z_2}^2 {z_6}^2 + {z_1}^2 {z_2}^2 {z_6}^2 + {z_1}^3 {z_2}^2 {z_6}^2 - 
  5 {z_2}^3 {z_6}^2 - 5 z_1 {z_2}^3 {z_6}^2 + {z_1}^2 {z_2}^3 {z_6}^2  \cr &+ 
  {z_1}^3 {z_2}^3 {z_6}^2 - 13 z_3 {z_6}^2 + 14 z_1 z_3 {z_6}^2 - 5 {z_1}^2 z_3 {z_6}^2 - 
  4 {z_1}^3 z_3 {z_6}^2 - 8 z_2 z_3 {z_6}^2 + 19 z_1 z_2 z_3 {z_6}^2  \cr &- 
  4 {z_1}^2 z_2 z_3 {z_6}^2 - 3 {z_1}^3 z_2 z_3 {z_6}^2 - {z_2}^2 z_3 {z_6}^2 - 
  2 z_1 {z_2}^2 z_3 {z_6}^2 + 7 {z_1}^2 {z_2}^2 z_3 {z_6}^2  \cr &+ 
  4 {z_1}^3 {z_2}^2 z_3 {z_6}^2 + 2 {z_2}^3 z_3 {z_6}^2 - 3 z_1 {z_2}^3 z_3 {z_6}^2 - 
  2 {z_1}^2 {z_2}^3 z_3 {z_6}^2 - {z_1}^3 {z_2}^3 z_3 {z_6}^2 + 7 {z_3}^2 {z_6}^2  \cr &+ 
  7 z_1 {z_3}^2 {z_6}^2 - 3 {z_1}^2 {z_3}^2 {z_6}^2 - 3 {z_1}^3 {z_3}^2 {z_6}^2 - 
  7 z_2 {z_3}^2 {z_6}^2 + 3 {z_1}^2 z_2 {z_3}^2 {z_6}^2 - 3 {z_2}^2 {z_3}^2 {z_6}^2  \cr &- 
  3 z_1 {z_2}^2 {z_3}^2 {z_6}^2 - {z_1}^2 {z_2}^2 {z_3}^2 {z_6}^2 - 
  {z_1}^3 {z_2}^2 {z_3}^2 {z_6}^2 + 3 {z_2}^3 {z_3}^2 {z_6}^2  \cr &+ 
  {z_1}^2 {z_2}^3 {z_3}^2 {z_6}^2 + {z_3}^3 {z_6}^2 - 2 z_1 {z_3}^3 {z_6}^2 + 
  {z_1}^2 {z_3}^3 {z_6}^2 + 2 z_2 {z_3}^3 {z_6}^2 - 4 z_1 z_2 {z_3}^3 {z_6}^2  \cr &+ 
  2 {z_1}^2 z_2 {z_3}^3 {z_6}^2 + {z_2}^2 {z_3}^3 {z_6}^2 - 2 z_1 {z_2}^2 {z_3}^3 {z_6}^2 + 
  {z_1}^2 {z_2}^2 {z_3}^3 {z_6}^2 + z_4 {z_6}^2 - 10 z_1 z_4 {z_6}^2  \cr &+ 
  5 {z_1}^2 z_4 {z_6}^2 - 4 {z_1}^3 z_4 {z_6}^2 - 13 z_2 z_4 {z_6}^2 - 
  16 z_1 z_2 z_4 {z_6}^2 + 15 {z_1}^2 z_2 z_4 {z_6}^2 + 2 {z_1}^3 z_2 z_4 {z_6}^2  \cr }$$
\vfill\eject
$$\eqalign{&+ 
  5 {z_2}^2 z_4 {z_6}^2 + 4 z_1 {z_2}^2 z_4 {z_6}^2 - 7 {z_1}^2 {z_2}^2 z_4 {z_6}^2 - 
  2 {z_1}^3 {z_2}^2 z_4 {z_6}^2 + 3 {z_2}^3 z_4 {z_6}^2 + 2 z_1 {z_2}^3 z_4 {z_6}^2  \cr &- 
  {z_1}^2 {z_2}^3 z_4 {z_6}^2 - 8 z_3 z_4 {z_6}^2 - 16 z_1 z_3 z_4 {z_6}^2 + 
  19 {z_1}^2 z_3 z_4 {z_6}^2 + {z_1}^3 z_3 z_4 {z_6}^2 + 9 z_2 z_3 z_4 {z_6}^2  \cr &- 
  16 z_1 z_2 z_3 z_4 {z_6}^2 + 4 {z_1}^2 z_2 z_3 z_4 {z_6}^2 - 
  3 {z_1}^3 z_2 z_3 z_4 {z_6}^2 - {z_2}^2 z_3 z_4 {z_6}^2  \cr &+ 
  3 z_1 {z_2}^2 z_3 z_4 {z_6}^2 - 2 {z_1}^2 {z_2}^2 z_3 z_4 {z_6}^2 - 
  2 {z_2}^3 z_3 z_4 {z_6}^2 + 3 z_1 {z_2}^3 z_3 z_4 {z_6}^2  \cr &+ 
  {z_1}^2 {z_2}^3 z_3 z_4 {z_6}^2 - {z_3}^2 z_4 {z_6}^2 - z_1 {z_3}^2 z_4 {z_6}^2 + 
  5 {z_1}^2 {z_3}^2 z_4 {z_6}^2 + {z_1}^3 {z_3}^2 z_4 {z_6}^2  \cr &+ 
  7 z_2 {z_3}^2 z_4 {z_6}^2 - 4 z_1 z_2 {z_3}^2 z_4 {z_6}^2 + 
  2 {z_1}^2 z_2 {z_3}^2 z_4 {z_6}^2 + {z_1}^3 z_2 {z_3}^2 z_4 {z_6}^2  \cr &- 
  {z_2}^2 {z_3}^2 z_4 {z_6}^2 + {z_1}^2 {z_2}^2 {z_3}^2 z_4 {z_6}^2 - 
  {z_2}^3 {z_3}^2 z_4 {z_6}^2 - z_1 {z_2}^3 {z_3}^2 z_4 {z_6}^2  \cr &+ 
  z_1 {z_3}^3 z_4 {z_6}^2 - {z_1}^2 {z_3}^3 z_4 {z_6}^2 + z_2 {z_3}^3 z_4 {z_6}^2 - 
  {z_1}^2 z_2 {z_3}^3 z_4 {z_6}^2 + {z_2}^2 {z_3}^3 z_4 {z_6}^2  \cr &- 
  z_1 {z_2}^2 {z_3}^3 z_4 {z_6}^2 + 11 {z_4}^2 {z_6}^2 - z_1 {z_4}^2 {z_6}^2 - 
  3 {z_1}^2 {z_4}^2 {z_6}^2 + {z_1}^3 {z_4}^2 {z_6}^2 + 5 z_2 {z_4}^2 {z_6}^2  \cr &- 
  z_1 z_2 {z_4}^2 {z_6}^2 - {z_1}^2 z_2 {z_4}^2 {z_6}^2 + {z_1}^3 z_2 {z_4}^2 {z_6}^2 - 
  6 {z_2}^2 {z_4}^2 {z_6}^2 + 2 {z_1}^2 {z_2}^2 {z_4}^2 {z_6}^2 + 11 z_3 {z_4}^2 {z_6}^2  \cr &- 
  3 z_1 z_3 {z_4}^2 {z_6}^2 - 5 {z_1}^2 z_3 {z_4}^2 {z_6}^2 + 
  {z_1}^3 z_3 {z_4}^2 {z_6}^2 + 3 z_2 z_3 {z_4}^2 {z_6}^2 - z_1 z_2 z_3 {z_4}^2 {z_6}^2  \cr &- 
  {z_1}^2 z_2 z_3 {z_4}^2 {z_6}^2 + {z_1}^3 z_2 z_3 {z_4}^2 {z_6}^2 - 
  2 z_1 {z_2}^2 z_3 {z_4}^2 {z_6}^2 - {z_3}^2 {z_4}^2 {z_6}^2  \cr &- 
  z_1 {z_3}^2 {z_4}^2 {z_6}^2 - 2 {z_1}^2 {z_3}^2 {z_4}^2 {z_6}^2 + 
  z_2 {z_3}^2 {z_4}^2 {z_6}^2 - z_1 z_2 {z_3}^2 {z_4}^2 {z_6}^2  \cr &- 
  2 {z_1}^2 z_2 {z_3}^2 {z_4}^2 {z_6}^2 + 2 {z_2}^2 {z_3}^2 {z_4}^2 {z_6}^2 - 
  {z_3}^3 {z_4}^2 {z_6}^2 + z_1 {z_3}^3 {z_4}^2 {z_6}^2 - z_2 {z_3}^3 {z_4}^2 {z_6}^2  \cr &+ 
  z_1 z_2 {z_3}^3 {z_4}^2 {z_6}^2 - {z_4}^3 {z_6}^2 + 2 z_1 {z_4}^3 {z_6}^2 - 
  {z_1}^2 {z_4}^3 {z_6}^2 - z_2 {z_4}^3 {z_6}^2 + 2 z_1 z_2 {z_4}^3 {z_6}^2  \cr &- 
  {z_1}^2 z_2 {z_4}^3 {z_6}^2 - 2 z_3 {z_4}^3 {z_6}^2 + 3 z_1 z_3 {z_4}^3 {z_6}^2 - 
  {z_1}^2 z_3 {z_4}^3 {z_6}^2 - 2 z_2 z_3 {z_4}^3 {z_6}^2  \cr &+ 
  3 z_1 z_2 z_3 {z_4}^3 {z_6}^2 - {z_1}^2 z_2 z_3 {z_4}^3 {z_6}^2 - 
  {z_3}^2 {z_4}^3 {z_6}^2 + z_1 {z_3}^2 {z_4}^3 {z_6}^2 - z_2 {z_3}^2 {z_4}^3 {z_6}^2  \cr &+ 
  z_1 z_2 {z_3}^2 {z_4}^3 {z_6}^2 - 15 z_5 {z_6}^2 - 8 z_1 z_5 {z_6}^2 + 
  9 {z_1}^2 z_5 {z_6}^2 - 2 {z_1}^3 z_5 {z_6}^2 - 12 z_2 z_5 {z_6}^2  \cr &- 
  3 z_1 z_2 z_5 {z_6}^2 + 8 {z_1}^2 z_2 z_5 {z_6}^2 - {z_1}^3 z_2 z_5 {z_6}^2 +
  11 {z_2}^2 z_5 {z_6}^2 + 8 z_1 {z_2}^2 z_5 {z_6}^2 - 9 {z_1}^2 {z_2}^2 z_5 {z_6}^2  \cr &- 
  2 {z_1}^3 {z_2}^2 z_5 {z_6}^2 - z_1 {z_2}^3 z_5 {z_6}^2 + {z_1}^3 {z_2}^3 z_5 {z_6}^2 - 
  18 z_3 z_5 {z_6}^2 - 4 z_1 z_3 z_5 {z_6}^2 + 15 {z_1}^2 z_3 z_5 {z_6}^2  \cr &- 
  {z_1}^3 z_3 z_5 {z_6}^2 + 4 z_1 z_2 z_3 z_5 {z_6}^2 - 
  3 {z_1}^2 z_2 z_3 z_5 {z_6}^2 - 5 {z_1}^3 z_2 z_3 z_5 {z_6}^2  \cr &- 
  3 {z_2}^2 z_3 z_5 {z_6}^2 + 7 z_1 {z_2}^2 z_3 z_5 {z_6}^2 - 
  2 {z_1}^2 {z_2}^2 z_3 z_5 {z_6}^2 + 2 {z_1}^3 {z_2}^2 z_3 z_5 {z_6}^2  \cr &- 
  {z_2}^3 z_3 z_5 {z_6}^2 + z_1 {z_2}^3 z_3 z_5 {z_6}^2 + {z_3}^2 z_5 {z_6}^2 + 
  7 z_1 {z_3}^2 z_5 {z_6}^2 + {z_1}^2 {z_3}^2 z_5 {z_6}^2 - {z_1}^3 {z_3}^2 z_5 {z_6}^2  \cr &- 
  z_2 {z_3}^2 z_5 {z_6}^2 + 4 z_1 z_2 {z_3}^2 z_5 {z_6}^2 + 
  {z_1}^3 z_2 {z_3}^2 z_5 {z_6}^2 - 5 {z_2}^2 {z_3}^2 z_5 {z_6}^2  \cr &+ 
  2 z_1 {z_2}^2 {z_3}^2 z_5 {z_6}^2 - {z_1}^2 {z_2}^2 {z_3}^2 z_5 {z_6}^2 + 
  {z_2}^3 {z_3}^2 z_5 {z_6}^2 - z_1 {z_2}^3 {z_3}^2 z_5 {z_6}^2  \cr &+ 
  z_1 {z_3}^3 z_5 {z_6}^2 - {z_1}^2 {z_3}^3 z_5 {z_6}^2 + z_2 {z_3}^3 z_5 {z_6}^2 - 
  {z_1}^2 z_2 {z_3}^3 z_5 {z_6}^2 + {z_2}^2 {z_3}^3 z_5 {z_6}^2  \cr }$$
\vfill\eject
$$\eqalign{&- 
  z_1 {z_2}^2 {z_3}^3 z_5 {z_6}^2 + 3 z_4 z_5 {z_6}^2 + 9 z_1 z_4 z_5 {z_6}^2 - 
  4 {z_1}^2 z_4 z_5 {z_6}^2 + 4 {z_1}^3 z_4 z_5 {z_6}^2 - 16 z_2 z_4 z_5 {z_6}^2  \cr &- 
  11 z_1 z_2 z_4 z_5 {z_6}^2 + 3 {z_1}^2 z_2 z_4 z_5 {z_6}^2 + 
  2 {z_1}^3 z_2 z_4 z_5 {z_6}^2 + 3 {z_2}^2 z_4 z_5 {z_6}^2  \cr &+ 
  5 z_1 {z_2}^2 z_4 z_5 {z_6}^2 + 4 {z_1}^2 {z_2}^2 z_4 z_5 {z_6}^2 - 
  z_1 {z_2}^3 z_4 z_5 {z_6}^2 - {z_1}^2 {z_2}^3 z_4 z_5 {z_6}^2  \cr &- 
  19 z_3 z_4 z_5 {z_6}^2 - 14 z_1 z_3 z_4 z_5 {z_6}^2 - 
  4 {z_1}^2 z_3 z_4 z_5 {z_6}^2 + 2 {z_1}^3 z_3 z_4 z_5 {z_6}^2  \cr &+ 
  10 z_2 z_3 z_4 z_5 {z_6}^2 - 13 z_1 z_2 z_3 z_4 z_5 {z_6}^2 + 
  5 {z_1}^2 z_2 z_3 z_4 z_5 {z_6}^2 - {z_1}^3 z_2 z_3 z_4 z_5 {z_6}^2  \cr &- 
  {z_2}^2 z_3 z_4 z_5 {z_6}^2 + 2 z_1 {z_2}^2 z_3 z_4 z_5 {z_6}^2 - 
  2 {z_1}^2 {z_2}^2 z_3 z_4 z_5 {z_6}^2 - {z_2}^3 z_3 z_4 z_5 {z_6}^2  \cr &- 
  4 {z_3}^2 z_4 z_5 {z_6}^2 - 4 z_1 {z_3}^2 z_4 z_5 {z_6}^2 - 
  {z_1}^2 {z_3}^2 z_4 z_5 {z_6}^2 - {z_1}^3 {z_3}^2 z_4 z_5 {z_6}^2  \cr &+ 
  4 z_2 {z_3}^2 z_4 z_5 {z_6}^2 - 2 z_1 z_2 {z_3}^2 z_4 z_5 {z_6}^2 - 
  {z_1}^2 z_2 {z_3}^2 z_4 z_5 {z_6}^2 - {z_2}^2 {z_3}^2 z_4 z_5 {z_6}^2  \cr &+ 
  z_1 {z_2}^2 {z_3}^2 z_4 z_5 {z_6}^2 + {z_2}^3 {z_3}^2 z_4 z_5 {z_6}^2 + 
  {z_1}^2 {z_3}^3 z_4 z_5 {z_6}^2 + 2 z_1 z_2 {z_3}^3 z_4 z_5 {z_6}^2  \cr &+ 
  {z_2}^2 {z_3}^3 z_4 z_5 {z_6}^2 + 3 {z_4}^2 z_5 {z_6}^2 - 2 z_1 {z_4}^2 z_5 {z_6}^2 - 
  3 {z_1}^2 {z_4}^2 z_5 {z_6}^2 - 2 {z_1}^3 {z_4}^2 z_5 {z_6}^2  \cr &+ 
  4 z_2 {z_4}^2 z_5 {z_6}^2 + 3 z_1 z_2 {z_4}^2 z_5 {z_6}^2 - 
  {z_1}^3 z_2 {z_4}^2 z_5 {z_6}^2 - 2 z_1 {z_2}^2 {z_4}^2 z_5 {z_6}^2  \cr &+ 
  8 z_3 {z_4}^2 z_5 {z_6}^2 + 7 z_1 z_3 {z_4}^2 z_5 {z_6}^2 + 
  2 {z_1}^2 z_3 {z_4}^2 z_5 {z_6}^2 - {z_1}^3 z_3 {z_4}^2 z_5 {z_6}^2  \cr &+ 
  5 z_2 z_3 {z_4}^2 z_5 {z_6}^2 + 2 z_1 z_2 z_3 {z_4}^2 z_5 {z_6}^2 + 
  {z_1}^2 z_2 z_3 {z_4}^2 z_5 {z_6}^2 - 2 {z_2}^2 z_3 {z_4}^2 z_5 {z_6}^2  \cr &+ 
  3 {z_3}^2 {z_4}^2 z_5 {z_6}^2 + 2 z_1 {z_3}^2 {z_4}^2 z_5 {z_6}^2 + 
  2 {z_1}^2 {z_3}^2 {z_4}^2 z_5 {z_6}^2 + z_1 z_2 {z_3}^2 {z_4}^2 z_5 {z_6}^2  \cr &- 
  z_1 {z_3}^3 {z_4}^2 z_5 {z_6}^2 - z_2 {z_3}^3 {z_4}^2 z_5 {z_6}^2 + 
  {z_4}^3 z_5 {z_6}^2 - 3 z_1 {z_4}^3 z_5 {z_6}^2 + 2 {z_1}^2 {z_4}^3 z_5 {z_6}^2  \cr &- 
  z_1 z_2 {z_4}^3 z_5 {z_6}^2 + {z_1}^2 z_2 {z_4}^3 z_5 {z_6}^2 + 
  z_3 {z_4}^3 z_5 {z_6}^2 - 3 z_1 z_3 {z_4}^3 z_5 {z_6}^2  \cr &+ 
  {z_1}^2 z_3 {z_4}^3 z_5 {z_6}^2 - z_2 z_3 {z_4}^3 z_5 {z_6}^2 - 
  z_1 {z_3}^2 {z_4}^3 z_5 {z_6}^2 - z_2 {z_3}^2 {z_4}^3 z_5 {z_6}^2 + 7 {z_5}^2 {z_6}^2  \cr &- 
  7 z_1 {z_5}^2 {z_6}^2 + {z_1}^2 {z_5}^2 {z_6}^2 - {z_1}^3 {z_5}^2 {z_6}^2 - 
  3 z_2 {z_5}^2 {z_6}^2 - 8 z_1 z_2 {z_5}^2 {z_6}^2 + 9 {z_1}^2 z_2 {z_5}^2 {z_6}^2  \cr &+ 
  2 {z_1}^3 z_2 {z_5}^2 {z_6}^2 - {z_2}^2 {z_5}^2 {z_6}^2 + 3 z_1 {z_2}^2 {z_5}^2 {z_6}^2 - 
  {z_1}^2 {z_2}^2 {z_5}^2 {z_6}^2 - {z_1}^3 {z_2}^2 {z_5}^2 {z_6}^2  \cr &+ 
  {z_2}^3 {z_5}^2 {z_6}^2 - {z_1}^2 {z_2}^3 {z_5}^2 {z_6}^2 + 3 z_3 {z_5}^2 {z_6}^2 - 
  11 z_1 z_3 {z_5}^2 {z_6}^2 + 7 {z_1}^2 z_3 {z_5}^2 {z_6}^2  \cr &+ 
  {z_1}^3 z_3 {z_5}^2 {z_6}^2 + 3 z_2 z_3 {z_5}^2 {z_6}^2 - 
  6 z_1 z_2 z_3 {z_5}^2 {z_6}^2 + 4 {z_1}^2 z_2 z_3 {z_5}^2 {z_6}^2  \cr &- 
  {z_1}^3 z_2 z_3 {z_5}^2 {z_6}^2 - {z_2}^2 z_3 {z_5}^2 {z_6}^2 + 
  2 z_1 {z_2}^2 z_3 {z_5}^2 {z_6}^2 - {z_1}^2 {z_2}^2 z_3 {z_5}^2 {z_6}^2  \cr &- 
  {z_2}^3 z_3 {z_5}^2 {z_6}^2 + z_1 {z_2}^3 z_3 {z_5}^2 {z_6}^2 - {z_3}^2 {z_5}^2 {z_6}^2 - 
  z_1 {z_3}^2 {z_5}^2 {z_6}^2 + 2 {z_1}^2 {z_3}^2 {z_5}^2 {z_6}^2  \cr &+ 
  z_2 {z_3}^2 {z_5}^2 {z_6}^2 - z_1 z_2 {z_3}^2 {z_5}^2 {z_6}^2 - 
  2 {z_2}^2 {z_3}^2 {z_5}^2 {z_6}^2 + 2 z_1 {z_2}^2 {z_3}^2 {z_5}^2 {z_6}^2  \cr &- 
  {z_3}^3 {z_5}^2 {z_6}^2 + z_1 {z_3}^3 {z_5}^2 {z_6}^2 - z_2 {z_3}^3 {z_5}^2 {z_6}^2 + 
  z_1 z_2 {z_3}^3 {z_5}^2 {z_6}^2 + 7 z_4 {z_5}^2 {z_6}^2 + 8 z_1 z_4 {z_5}^2 {z_6}^2  \cr &- 
  7 {z_1}^2 z_4 {z_5}^2 {z_6}^2 - 11 z_2 z_4 {z_5}^2 {z_6}^2 - 
  3 z_1 z_2 z_4 {z_5}^2 {z_6}^2 - 2 {z_1}^2 z_2 z_4 {z_5}^2 {z_6}^2  \cr }$$
\vfill\eject
$$\eqalign{&+ 
  {z_2}^2 z_4 {z_5}^2 {z_6}^2 + {z_1}^2 {z_2}^2 z_4 {z_5}^2 {z_6}^2 + 
  {z_2}^3 z_4 {z_5}^2 {z_6}^2 + z_1 {z_2}^3 z_4 {z_5}^2 {z_6}^2  \cr &- 
  8 z_3 z_4 {z_5}^2 {z_6}^2 + 3 z_1 z_3 z_4 {z_5}^2 {z_6}^2 - 
  2 {z_1}^2 z_3 z_4 {z_5}^2 {z_6}^2 + {z_1}^3 z_3 z_4 {z_5}^2 {z_6}^2  \cr &+ 
  6 z_2 z_3 z_4 {z_5}^2 {z_6}^2 - 7 z_1 z_2 z_3 z_4 {z_5}^2 {z_6}^2 + 
  2 {z_1}^2 z_2 z_3 z_4 {z_5}^2 {z_6}^2 - {z_2}^2 z_3 z_4 {z_5}^2 {z_6}^2  \cr &+ 
  z_1 {z_2}^2 z_3 z_4 {z_5}^2 {z_6}^2 - {z_2}^3 z_3 z_4 {z_5}^2 {z_6}^2 - 
  3 {z_3}^2 z_4 {z_5}^2 {z_6}^2 + 2 z_2 {z_3}^2 z_4 {z_5}^2 {z_6}^2  \cr &- 
  z_1 z_2 {z_3}^2 z_4 {z_5}^2 {z_6}^2 - 2 {z_2}^2 {z_3}^2 z_4 {z_5}^2 {z_6}^2 - 
  z_1 {z_3}^3 z_4 {z_5}^2 {z_6}^2 - z_2 {z_3}^3 z_4 {z_5}^2 {z_6}^2  \cr &+ 
  {z_4}^2 {z_5}^2 {z_6}^2 + 7 z_1 {z_4}^2 {z_5}^2 {z_6}^2 - {z_1}^2 {z_4}^2 {z_5}^2 {z_6}^2 + 
  {z_1}^3 {z_4}^2 {z_5}^2 {z_6}^2 - 7 z_2 {z_4}^2 {z_5}^2 {z_6}^2  \cr &- 
  2 z_1 z_2 {z_4}^2 {z_5}^2 {z_6}^2 + {z_1}^2 z_2 {z_4}^2 {z_5}^2 {z_6}^2 + 
  2 {z_2}^2 {z_4}^2 {z_5}^2 {z_6}^2 - 9 z_3 {z_4}^2 {z_5}^2 {z_6}^2  \cr &- 
  2 z_1 z_3 {z_4}^2 {z_5}^2 {z_6}^2 - {z_1}^2 z_3 {z_4}^2 {z_5}^2 {z_6}^2 + 
  4 z_2 z_3 {z_4}^2 {z_5}^2 {z_6}^2 - 2 z_1 z_2 z_3 {z_4}^2 {z_5}^2 {z_6}^2  \cr &- 
  {z_3}^2 {z_4}^2 {z_5}^2 {z_6}^2 - z_1 {z_3}^2 {z_4}^2 {z_5}^2 {z_6}^2 + 
  z_2 {z_3}^2 {z_4}^2 {z_5}^2 {z_6}^2 + {z_3}^3 {z_4}^2 {z_5}^2 {z_6}^2  \cr &+ 
  {z_4}^3 {z_5}^2 {z_6}^2 - {z_1}^2 {z_4}^3 {z_5}^2 {z_6}^2 + z_2 {z_4}^3 {z_5}^2 {z_6}^2 - 
  z_1 z_2 {z_4}^3 {z_5}^2 {z_6}^2 + 2 z_3 {z_4}^3 {z_5}^2 {z_6}^2  \cr &+ 
  z_2 z_3 {z_4}^3 {z_5}^2 {z_6}^2 + {z_3}^2 {z_4}^3 {z_5}^2 {z_6}^2 + 3 {z_5}^3 {z_6}^2 - 
  2 z_1 {z_5}^3 {z_6}^2 - {z_1}^2 {z_5}^3 {z_6}^2 + 2 z_2 {z_5}^3 {z_6}^2  \cr &- 
  2 z_1 z_2 {z_5}^3 {z_6}^2 - {z_2}^2 {z_5}^3 {z_6}^2 + {z_1}^2 {z_2}^2 {z_5}^3 {z_6}^2 + 
  4 z_3 {z_5}^3 {z_6}^2 - 3 z_1 z_3 {z_5}^3 {z_6}^2 - {z_1}^2 z_3 {z_5}^3 {z_6}^2  \cr &+ 
  z_2 z_3 {z_5}^3 {z_6}^2 - 2 z_1 z_2 z_3 {z_5}^3 {z_6}^2 + 
  {z_1}^2 z_2 z_3 {z_5}^3 {z_6}^2 + {z_2}^2 z_3 {z_5}^3 {z_6}^2  \cr &- 
  z_1 {z_2}^2 z_3 {z_5}^3 {z_6}^2 + {z_3}^2 {z_5}^3 {z_6}^2 - z_1 {z_3}^2 {z_5}^3 {z_6}^2 + 
  z_2 {z_3}^2 {z_5}^3 {z_6}^2 - z_1 z_2 {z_3}^2 {z_5}^3 {z_6}^2 + 5 z_4 {z_5}^3 {z_6}^2  \cr &+ 
  z_1 z_4 {z_5}^3 {z_6}^2 - 2 {z_1}^2 z_4 {z_5}^3 {z_6}^2 + 
  2 z_1 z_2 z_4 {z_5}^3 {z_6}^2 - {z_2}^2 z_4 {z_5}^3 {z_6}^2  \cr &- 
  z_1 {z_2}^2 z_4 {z_5}^3 {z_6}^2 + 3 z_3 z_4 {z_5}^3 {z_6}^2 + 
  3 z_1 z_3 z_4 {z_5}^3 {z_6}^2 - {z_1}^2 z_3 z_4 {z_5}^3 {z_6}^2  \cr &+ 
  z_2 z_3 z_4 {z_5}^3 {z_6}^2 - z_1 z_2 z_3 z_4 {z_5}^3 {z_6}^2 + 
  {z_2}^2 z_3 z_4 {z_5}^3 {z_6}^2 + z_1 {z_3}^2 z_4 {z_5}^3 {z_6}^2  \cr &+ 
  z_2 {z_3}^2 z_4 {z_5}^3 {z_6}^2 + {z_4}^2 {z_5}^3 {z_6}^2 + 
  4 z_1 {z_4}^2 {z_5}^3 {z_6}^2 - {z_1}^2 {z_4}^2 {z_5}^3 {z_6}^2  \cr &- 
  2 z_2 {z_4}^2 {z_5}^3 {z_6}^2 - 2 z_3 {z_4}^2 {z_5}^3 {z_6}^2 + 
  2 z_1 z_3 {z_4}^2 {z_5}^3 {z_6}^2 - {z_3}^2 {z_4}^2 {z_5}^3 {z_6}^2  \cr &- 
  {z_4}^3 {z_5}^3 {z_6}^2 + z_1 {z_4}^3 {z_5}^3 {z_6}^2 - z_3 {z_4}^3 {z_5}^3 {z_6}^2 + 
  7 {z_6}^3 - 9 z_1 {z_6}^3 - 3 {z_1}^2 {z_6}^3 + 5 {z_1}^3 {z_6}^3 - 2 z_2 {z_6}^3  \cr &+ 
  4 z_1 z_2 {z_6}^3 - 2 {z_1}^2 z_2 {z_6}^3 - {z_2}^2 {z_6}^3 + z_1 {z_2}^2 {z_6}^3 + 
  {z_1}^2 {z_2}^2 {z_6}^3 - {z_1}^3 {z_2}^2 {z_6}^3 + 2 z_3 {z_6}^3 - z_1 z_3 {z_6}^3  \cr &- 
  4 {z_1}^2 z_3 {z_6}^3 + 3 {z_1}^3 z_3 {z_6}^3 - 4 z_2 z_3 {z_6}^3 + 
  8 z_1 z_2 z_3 {z_6}^3 - 4 {z_1}^2 z_2 z_3 {z_6}^3 + 2 {z_2}^2 z_3 {z_6}^3  \cr &- 
  3 z_1 {z_2}^2 z_3 {z_6}^3 + {z_1}^3 {z_2}^2 z_3 {z_6}^3 - {z_3}^2 {z_6}^3 + 
  2 z_1 {z_3}^2 {z_6}^3 - {z_1}^2 {z_3}^2 {z_6}^3 - 2 z_2 {z_3}^2 {z_6}^3  \cr &+ 
  4 z_1 z_2 {z_3}^2 {z_6}^3 - 2 {z_1}^2 z_2 {z_3}^2 {z_6}^3 - {z_2}^2 {z_3}^2 {z_6}^3 + 
  2 z_1 {z_2}^2 {z_3}^2 {z_6}^3 - {z_1}^2 {z_2}^2 {z_3}^2 {z_6}^3 + 2 z_4 {z_6}^3  \cr &+ 
  5 z_1 z_4 {z_6}^3 - 4 {z_1}^2 z_4 {z_6}^3 - 3 {z_1}^3 z_4 {z_6}^3 + z_2 z_4 {z_6}^3 - 
  3 z_1 z_2 z_4 {z_6}^3 + {z_1}^2 z_2 z_4 {z_6}^3 + {z_1}^3 z_2 z_4 {z_6}^3  \cr &- 
  {z_2}^2 z_4 {z_6}^3 + {z_1}^2 {z_2}^2 z_4 {z_6}^3 + 2 z_3 z_4 {z_6}^3 - 
  {z_1}^2 z_3 z_4 {z_6}^3 - {z_1}^3 z_3 z_4 {z_6}^3 - 4 z_2 z_3 z_4 {z_6}^3  \cr }$$
\vfill\eject
$$\eqalign{&+ 
  3 z_1 z_2 z_3 z_4 {z_6}^3 + 2 {z_1}^2 z_2 z_3 z_4 {z_6}^3 - 
  {z_1}^3 z_2 z_3 z_4 {z_6}^3 + 2 {z_2}^2 z_3 z_4 {z_6}^3  \cr &- 
  z_1 {z_2}^2 z_3 z_4 {z_6}^3 - {z_1}^2 {z_2}^2 z_3 z_4 {z_6}^3 - 
  z_1 {z_3}^2 z_4 {z_6}^3 + {z_1}^2 {z_3}^2 z_4 {z_6}^3 - z_2 {z_3}^2 z_4 {z_6}^3  \cr &+ 
  {z_1}^2 z_2 {z_3}^2 z_4 {z_6}^3 - {z_2}^2 {z_3}^2 z_4 {z_6}^3 + 
  z_1 {z_2}^2 {z_3}^2 z_4 {z_6}^3 - 5 {z_4}^2 {z_6}^3 + 2 z_1 {z_4}^2 {z_6}^3  \cr &+ 
  3 {z_1}^2 {z_4}^2 {z_6}^3 + 3 z_2 {z_4}^2 {z_6}^3 - 2 z_1 z_2 {z_4}^2 {z_6}^3 - 
  {z_1}^2 z_2 {z_4}^2 {z_6}^3 - z_1 z_3 {z_4}^2 {z_6}^3 + {z_1}^2 z_3 {z_4}^2 {z_6}^3  \cr &- 
  z_1 z_2 z_3 {z_4}^2 {z_6}^3 + {z_1}^2 z_2 z_3 {z_4}^2 {z_6}^3 + 
  {z_3}^2 {z_4}^2 {z_6}^3 - z_1 {z_3}^2 {z_4}^2 {z_6}^3 + z_2 {z_3}^2 {z_4}^2 {z_6}^3  \cr &- 
  z_1 z_2 {z_3}^2 {z_4}^2 {z_6}^3 + 7 z_5 {z_6}^3 - 2 z_1 z_5 {z_6}^3 - 
  5 {z_1}^2 z_5 {z_6}^3 + z_2 z_5 {z_6}^3 - 3 z_1 z_2 z_5 {z_6}^3  \cr &+ 
  {z_1}^2 z_2 z_5 {z_6}^3 + {z_1}^3 z_2 z_5 {z_6}^3 - 2 {z_2}^2 z_5 {z_6}^3 + 
  z_1 {z_2}^2 z_5 {z_6}^3 + 2 {z_1}^2 {z_2}^2 z_5 {z_6}^3 - {z_1}^3 {z_2}^2 z_5 {z_6}^3  \cr &+ 
  5 z_3 z_5 {z_6}^3 - 4 z_1 z_3 z_5 {z_6}^3 - 2 {z_1}^2 z_3 z_5 {z_6}^3 + 
  {z_1}^3 z_3 z_5 {z_6}^3 - 4 z_2 z_3 z_5 {z_6}^3 + 3 z_1 z_2 z_3 z_5 {z_6}^3  \cr &+ 
  2 {z_1}^2 z_2 z_3 z_5 {z_6}^3 - {z_1}^3 z_2 z_3 z_5 {z_6}^3 + 
  3 {z_2}^2 z_3 z_5 {z_6}^3 - 3 z_1 {z_2}^2 z_3 z_5 {z_6}^3 - z_1 {z_3}^2 z_5 {z_6}^3  \cr &+ 
  {z_1}^2 {z_3}^2 z_5 {z_6}^3 - z_2 {z_3}^2 z_5 {z_6}^3 + 
  {z_1}^2 z_2 {z_3}^2 z_5 {z_6}^3 - {z_2}^2 {z_3}^2 z_5 {z_6}^3  \cr &+ 
  z_1 {z_2}^2 {z_3}^2 z_5 {z_6}^3 + 2 z_4 z_5 {z_6}^3 + 7 z_1 z_4 z_5 {z_6}^3 + 
  3 {z_1}^2 z_4 z_5 {z_6}^3 + 3 z_2 z_4 z_5 {z_6}^3 - z_1 z_2 z_4 z_5 {z_6}^3  \cr &- 
  5 {z_1}^2 z_2 z_4 z_5 {z_6}^3 + {z_1}^3 z_2 z_4 z_5 {z_6}^3 - 
  2 {z_2}^2 z_4 z_5 {z_6}^3 - z_1 {z_2}^2 z_4 z_5 {z_6}^3  \cr &+ 
  {z_1}^2 {z_2}^2 z_4 z_5 {z_6}^3 + 4 z_3 z_4 z_5 {z_6}^3 + 
  3 z_1 z_3 z_4 z_5 {z_6}^3 - 3 {z_1}^2 z_3 z_4 z_5 {z_6}^3  \cr &+ 
  {z_1}^3 z_3 z_4 z_5 {z_6}^3 - 5 z_2 z_3 z_4 z_5 {z_6}^3 + 
  {z_1}^2 z_2 z_3 z_4 z_5 {z_6}^3 + 3 {z_2}^2 z_3 z_4 z_5 {z_6}^3  \cr &- 
  {z_1}^2 {z_3}^2 z_4 z_5 {z_6}^3 - 2 z_1 z_2 {z_3}^2 z_4 z_5 {z_6}^3 - 
  {z_2}^2 {z_3}^2 z_4 z_5 {z_6}^3 - 5 {z_4}^2 z_5 {z_6}^3 - 3 z_1 {z_4}^2 z_5 {z_6}^3  \cr &+ 
  2 z_2 {z_4}^2 z_5 {z_6}^3 + 3 z_1 z_2 {z_4}^2 z_5 {z_6}^3 -
  {z_1}^2 z_2 {z_4}^2 z_5 {z_6}^3 - z_3 {z_4}^2 z_5 {z_6}^3  \cr &+ 
  z_1 z_3 {z_4}^2 z_5 {z_6}^3 - {z_1}^2 z_3 {z_4}^2 z_5 {z_6}^3 - 
  z_2 z_3 {z_4}^2 z_5 {z_6}^3 + z_1 {z_3}^2 {z_4}^2 z_5 {z_6}^3  \cr &+ 
  z_2 {z_3}^2 {z_4}^2 z_5 {z_6}^3 - 3 {z_5}^2 {z_6}^3 + 5 z_1 {z_5}^2 {z_6}^3 - 
  {z_1}^2 {z_5}^2 {z_6}^3 - {z_1}^3 {z_5}^2 {z_6}^3 + 4 z_2 {z_5}^2 {z_6}^3  \cr &- 
  3 z_1 z_2 {z_5}^2 {z_6}^3 - 2 {z_1}^2 z_2 {z_5}^2 {z_6}^3 + 
  {z_1}^3 z_2 {z_5}^2 {z_6}^3 - {z_2}^2 {z_5}^2 {z_6}^3 + {z_1}^2 {z_2}^2 {z_5}^2 {z_6}^3  \cr &+ 
  2 z_3 {z_5}^2 {z_6}^3 - 2 {z_1}^2 z_3 {z_5}^2 {z_6}^3 - z_2 z_3 {z_5}^2 {z_6}^3 + 
  z_1 z_2 z_3 {z_5}^2 {z_6}^3 + {z_2}^2 z_3 {z_5}^2 {z_6}^3  \cr &- 
  z_1 {z_2}^2 z_3 {z_5}^2 {z_6}^3 + {z_3}^2 {z_5}^2 {z_6}^3 - z_1 {z_3}^2 {z_5}^2 {z_6}^3 + 
  z_2 {z_3}^2 {z_5}^2 {z_6}^3 - z_1 z_2 {z_3}^2 {z_5}^2 {z_6}^3 - 2 z_4 {z_5}^2 {z_6}^3  \cr &- 
  z_1 z_4 {z_5}^2 {z_6}^3 + 4 {z_1}^2 z_4 {z_5}^2 {z_6}^3 - {z_1}^3 z_4 {z_5}^2 {z_6}^3 + 
  3 z_2 z_4 {z_5}^2 {z_6}^3 + 3 z_1 z_2 z_4 {z_5}^2 {z_6}^3  \cr &- 
  2 {z_1}^2 z_2 z_4 {z_5}^2 {z_6}^3 - {z_2}^2 z_4 {z_5}^2 {z_6}^3 -
  z_1 {z_2}^2 z_4 {z_5}^2 {z_6}^3 + 2 z_3 z_4 {z_5}^2 {z_6}^3  \cr &+ 
  2 z_1 z_3 z_4 {z_5}^2 {z_6}^3 - 2 z_2 z_3 z_4 {z_5}^2 {z_6}^3 + 
  z_1 z_2 z_3 z_4 {z_5}^2 {z_6}^3 + {z_2}^2 z_3 z_4 {z_5}^2 {z_6}^3  \cr &+ 
  z_1 {z_3}^2 z_4 {z_5}^2 {z_6}^3 + z_2 {z_3}^2 z_4 {z_5}^2 {z_6}^3 + 
  {z_4}^2 {z_5}^2 {z_6}^3 - 2 z_1 {z_4}^2 {z_5}^2 {z_6}^3 + {z_1}^2 {z_4}^2 {z_5}^2 {z_6}^3  \cr &- 
  z_2 {z_4}^2 {z_5}^2 {z_6}^3 + z_1 z_2 {z_4}^2 {z_5}^2 {z_6}^3 - 
  z_2 z_3 {z_4}^2 {z_5}^2 {z_6}^3 - {z_3}^2 {z_4}^2 {z_5}^2 {z_6}^3 - 3 {z_5}^3 {z_6}^3  \cr &+ 
  2 z_1 {z_5}^3 {z_6}^3 + {z_1}^2 {z_5}^3 {z_6}^3 + z_2 {z_5}^3 {z_6}^3 - 
  {z_1}^2 z_2 {z_5}^3 {z_6}^3 - z_3 {z_5}^3 {z_6}^3 + z_1 z_3 {z_5}^3 {z_6}^3  \cr }$$
\vfill\eject
$$\eqalign{&- 
  z_2 z_3 {z_5}^3 {z_6}^3 + z_1 z_2 z_3 {z_5}^3 {z_6}^3 - 2 z_4 {z_5}^3 {z_6}^3 - 
  3 z_1 z_4 {z_5}^3 {z_6}^3 + {z_1}^2 z_4 {z_5}^3 {z_6}^3 + z_2 z_4 {z_5}^3 {z_6}^3  \cr &+ 
  z_1 z_2 z_4 {z_5}^3 {z_6}^3 - z_1 z_3 z_4 {z_5}^3 {z_6}^3 - 
  z_2 z_3 z_4 {z_5}^3 {z_6}^3 + {z_4}^2 {z_5}^3 {z_6}^3 - z_1 {z_4}^2 {z_5}^3 {z_6}^3  \cr &+ 
  z_3 {z_4}^2 {z_5}^3 {z_6}^3\cr}$$

\subsec{A candidate for multidegree of the affine variety $\aff{\pi}$}
[cette section meriterait evidemment d'etre affinee... sans jeu de mots.
depend bcp de si on fait du progres ou non cf
ces histoires de polynomes de Schubert  ... OK]

It is interesting to note that an efficient way of computing the bidegrees \bideg\ of
Conjecture 1 consists in performing the change of variables $z_i=(t_i-1)/(t_i+1)$
for $i=1,2,\ldots, n$ while taking $z_i=0$ for $i=n+1,n+2,\ldots,2n$. Using Eq.~\valpiob, we find:
\eqn\pioval{ \Psi_{\pi_0}= 2^{n(n-1)}
{\prod_{1\leq i<j\leq n} (1+3t_i-t_j+t_it_j)\over \prod_{i=1}^n (1+t_i)^{2(n-1)}}
\prod_{i=1}^n t_i^{i-1} }
This suggests to define for $\pi\in P_n$ the ``multidegree" 
\eqn\multideg{ d_\pi(t_1,t_2,\ldots,t_n)=
{\prod_{i=1}^n (1+t_i)^{2(n-1)}\over 2^{n(n-1)}\prod_{1\leq i<j\leq n} (1+3t_i-t_j+t_it_j)} 
\Psi_{\pi}\left({t_1-1\over t_1+1},\ldots,{t_n-1\over t_n+1},0,\ldots,0\right) }
These are polynomials of the $t_i$, obtained by repeated action on $d_{\pi_0}=\prod t_i^{i-1}$
of the operators $\Theta_{i}$, $i=1,2,\ldots,n-1$.
With the above change of variables, $\Theta_{i}$
acts as 
\eqn\newthet{ \theta_{n,i}f(t,u)=(1+t)(1+u){f(u,t)-f(t,u)\over t-u} -f(u,t)}
on the $i$-th and $i+1$-th variables $t,u$ of the function $f$. 
Note that $\theta_{n,i}$ acting on polynomials produces polynomials,
with degree increased by at most $1$.
Moreover, the action of the $\theta_{n,i}$ is readily seen to produce only polynomials with
integer coefficients, starting from $d_{\pi_0}$. This property has a nice consequence on the
homogeneous limit where all $t_i=1$, as it implies that all multidegrees become integers, 
and in particular
$d_{\pi_0}^{hom}=1$ in this limit. Moreover, in this case
the prefactor in \multideg\ reduces to $1$ hence, as by the Perron--Frobenius property all
the entries of $\Psi_n$ are positive, so are the $d_\pi^{hom}=\Psi_{\pi}^{hom}$, $\pi\in P_n$. 
The latter are therefore all non-negative integers,
with $d_{\pi_0}^{hom}=\Psi_{\pi_0}^{hom}=1$. 
We deduce that when all $z_i\to 0$, the entries $\Psi_{\pi}$ become non-negative
integers for all $\pi\in P_n$, the smallest of which being $\Psi_{\pi_0}^{hom}=1$. 
This proves a weaker version of 
the non-negative integrality conjecture of \BRAU\ on the components of the ground state vector of
the homogeneous case, here reduced to the permutation sector (or any of its rotations as well).

At generic values of the $t_i$, we have observed that all coefficients of $d_\pi$ as polynomials of the $t_i$
are non-negative as well, and we conjecture that this is the case in general. Modulo this conjecture,
we therefore expect the multidegree
$d_\pi$ of Eq.~\multideg\ to have some combinatorial meaning in the context of the affine variety
$\aff{\pi}$. 
For illustration, in the case $n=3$, the multidegrees read:
[xxx ... utiliser l'indexation par diagrammes? appendice?]
\eqn\multithree{\eqalign{ d_{(41)(52)(63)}&=t_2t_3^2\cr
d_{(41)(53)(62)}&=t_2t_3(1+t_3+t_2t_3)\cr
d_{(42)(51)(63)}&=t_3^2(1+t_2+t_1t_2)\cr
d_{(43)(51)(62)}&=t_3(1 + t_2 + t_1 t_2 + t_3 + t_1 t_3 + 2t_2 t_3 + 
3 t_1 t_2 t_3 + t_1^2 t_2 t_3 + t_2^2 t_3 + t_1 t_2^2 t_3)\cr
d_{(42)(53)(61)}&=t_2 + t_3 + 3t_2 t_3 + t_1 t_2 t_3 + t_2^2 t_3 + 
t_3^2 + 2 t_2 t_3^2 + t_1 t_2 t_3^2 + t_2^2 t_3^2 + t_1 t_2^2 t_3^2\cr
d_{(43)(52)(61)}&=1 + t_2 + t_1 t_2 + 2 t_3 + t_1 t_3 + 4 t_2 t_3 + 4 t_1 t_2 t_3 
+ t_1^2 t_2 t_3 + t_2^2 t_3 + t_1t_2^2 t_3 \cr
&+ t_3^2 + t_1 t_3^2 + 3 t_2 t_3^2 +4t_1 t_2t_3^2 + t_1^2 t_2 t_3^2 + t_2^2t_3^2 
+ 2t_1t_2^2t_3^2 + t_1^2 t_2^2 t_3^2\cr}}
One may make this look more symmetric by performing the change of variables $t_i=A_i/B_i$ for all $i$
and premultiply by a factor $\prod_{1\leq i\leq n}B_i^{n-1}$.
The bidegree of Sect.~4.1 is then recovered from the multidegree
by taking $t_i=A_i/B_i$, $A_i=A$ and $B_i=B$ for all $i$ and multiplying by an overall factor
$B^{n(n-1)}$, and for $n=3$ \multithree\ leads to the expressions of appendix C.
Finally, the sum rule \permu\ together with the normalization \multideg\ yield a general sum
rule for the multidegrees, namely
\eqn\sumulti{ \sum_{\pi\in P_n} d_\pi(t_1,\ldots,t_n)=\prod_{1\leq i<j\leq n} (1+2t_j+t_it_j) }

[aussi, factorisation du degre?]